\documentclass[10pt]{article}
\usepackage{latexsym,graphicx,multirow}
\usepackage{amssymb}
\usepackage{amsmath}
\usepackage{amscd}
\usepackage{amsthm}
\usepackage[left=1cm,top=2.5cm,right=1.2cm,bottom=1.5cm]{geometry}
\usepackage{hyperref}
\DeclareGraphicsExtensions{.eps}
\DeclareGraphicsExtensions{.jpg}
\usepackage{epstopdf}
\usepackage{}
\usepackage{float}

\begin{document}
	
	\begin{center}
	\large{\bf{Traversable wormholes with logarithmic shape function in $f(R, T)$ gravity}} \\
	\vspace{10mm}
	\normalsize{Archana Dixit$^1$, Chanchal Chawla$^2$, Anirudh Pradhan$^3$ }\\
	\vspace{5mm}
	\normalsize{$^{1,3}$Department of Mathematics, 
	Institute of Applied Sciences and Humanities, GLA University,
		Mathura-281 406, Uttar Pradesh, India}\\
	\vspace{2mm}	
	\normalsize{$^{2}$Department of Distance Education, Punjabi University Patiala-147002, Punjab, India.}\\
	
	\vspace{2mm}
	$^1$E-mail: archana.dixit@gla.ac.in.\\
	$^2$E-mail: c.chawla137@gmail.com\\
	$^3$E-mail: pradhan.anirudh@gmail.com\\

	\end{center}	
	
	\vspace{10mm}
	
%%\date{}
%%\makefile
\begin{abstract}
In the present work, a new form of the logarithmic shape function is proposed for the linear $f(R,T)$ gravity, $f(R,T)=R+2\lambda T$ 
where $\lambda$ is an arbitrary coupling constant, in wormhole geometry. The desired logarithmic shape function accomplishes all necessary 
conditions for traversable and asymptotically flat wormholes. The obtained wormhole solutions are analyzed from the energy conditions 
for different values of $\lambda$. It has been observed that our proposed shape function for the linear form of $f(R,T)$ gravity, 
represents the existence of exotic matter and non-exotic matter. Moreover, for $\lambda=0$ i.e. for the general relativity case, the 
existence of exotic matter for the wormhole geometry has been confirmed. Further, the behaviour of the radial state parameter $\omega_{r}$, 
the tangential state parameter $\omega_{t}$ and the anisotropy parameter $\triangle$ describing the geometry of the universe, has been presented 
for different values of $\lambda$ chosen in $[-100,100]$.
\end{abstract}
\vspace{5mm}
%%%%%%%%%%%%%%%%%%%%%%%%%%%%%%%%% section 1 Introduction %%%%%%%%%%%%%%%%%%%%%%%%%%
\section{Introduction}
Wormholes are a common science fiction tools that serve as a tunnel-like structure lying in the same universe, linking multiple universes 
or widely separated regions. Such geometric models can be regarded as a means of fast interstellar travel, time machines and warp drives. 
In this direction, in the mid-1935s, work led by Einstein and Rosen \cite{ref1}, designed an elementary particle model described by a 
bridge linking two identical sheets: the Einstein-Rosen bridge (ERB) and carried out for a comprehensive study of wormhole solutions. 
Wheeler \cite{ref2} has obtained Kerr wormholes which are the objects of quantum foam, connect different parts of space-time and 
operates at the Planck scale. Thorne and his student Morris \cite{ref3}, understood the structure of wormholes with a throat and two 
mouths and introduced static traversable wormholes and sparked this area. They used general relativity concepts and discovered a 
possible way to travel in time. Wormholes are consistent with the general theory of relativity but it remains to be seen if wormholes 
actually exist. Some scientists postulate wormholes are merely a 4th dimension projection, similar to how a 2D being might only 
perceive a portion of a 3D object.\\

Generally, a wormhole is a tool or a tunnel that connects two distinct regions of spacetime. In the event that the two regions are 
in a similar universe, it is called an intra-universe wormhole, while if the two regions are in two different universes, it is 
called an inter-universal wormhole. A significant component of a wormhole is its traversability. Generally, if a substance can enter 
from one side of the wormhole and exit from the opposite, the wormhole is called traversable. It is exceptional that ongoing 
perceptions have affirmed that the Universe is extraordinary a period of accelerated expansion. Scientific proof of this cosmological 
growth, derived from estimates of supernovae of type Ia (SNe Ia) \cite{ref4}-\cite{ref5} and independent of the cosmic percentage 
microwave base radiation \cite{ref6}-\cite{ref7}, shows that the Universe also contains a sort of negative pressure ``dark energy ". 
The Wilkinson Microwave Anisotropy Sample (WMAP), intended to quantify the CMB anisotropy with significant precision and accuracy, 
has subsequently verified that the Universe consists of around 70 percent of dark energy \cite{ref6}. In literature, there are lots 
of candidates have been proposed for dark energy, which has to be explicit, a positive cosmological constant, the quintessence fields,  
dilation model Chaplygin gas and tachyon models, etc.\\

Wormhole material science is a particular case of embracing the opposite way of thinking of illuminating the Einstein field equation, 
by first developing the spacetime metric, at that point where reducing the stress-energy tensor segments. In this way, it was found 
that these traversable wormholes have a stress-energy tensor that disregards the null energy condition (NEC) \cite{ref8}-\cite{ref9}. 
In fact, they violate all the point-wise energy conditions and arrived at the midpoint of energy conditions, which are basic of the 
singularity hypotheses. It is fascinating to take note that recent observation in cosmology strongly propose that the cosmological 
fluid disregards the strong energy condition (SEC), and gives indications that the NEC might possibly be damaged in a classical 
regime \cite{ref10}-\cite{ref11}. In wormholes physics, the presence of throat is the fundamental feature which satisfies flaring-out 
condition\cite{ref12}. Theoretically, wormholes throats can be formulated in the absence of exotic matter\cite{ref13}.\\

In the current article, we will concentrate on the f(R,T) theory of gravity. This theory has appeared to give a decent option in 
contrast to the cosmological issues, in \cite{ref14}-\cite{ref19}. Other intriguing references have been suggested wormhole 
solutions in \cite{ref20}-\cite{ref23}. This theory attracts more due to its unique feature that is a non-minimal coupling of matter 
and geometry\cite{ref24}. In the recent past, several applications of $f(R,T)$ \cite{ref25}-\cite{ref32} have been reported in the literature.
The most fascinating type of Dark energy is a phantom with ($\omega<-1$ )\cite{ref33}, for which the weak energy condition (WEC) is violated. 
The exotic nature of phantom energy uncovers itself in various abnormal cosmological results. One of them is a major big rip \cite{ref14}. 
Sahoo \cite{ref34} study supports the exotic wormholes. Subsequently, we can say that phantom energy as an essential candidate
for exotic matter. Another intriguing phenomenon is that every single black hole in the apparition universe reduces their masses to 
disappear precisely in the big rip \cite{ref35}. \\

Many cosmologists have considered, wormholes in $f(R,T)$ gravity in various perspectives. Now  we are going to examine the contribution 
of certain cosmologists in $f(R)$ and $f(R,T)$ gravity theory. Harko \cite{ref15} worked on $f(R)$ theory and \cite{ref36}-\cite{ref41} 
by replacing the function $f(R)$ with an arbitrary function $f(R, T)$. The anisotropic cosmological models in $f(R, T)$ gravity with 
$\Lambda(T)$ discussed in \cite{ref42}. Moraes et al. \cite{ref43} contemplated the charged wormhole arrangements in $f(R, T)$  gravity 
and obtained fulfillment of energy conditions. In the same context \cite{ref44} utilized the analytic technique to investigate the wormhole 
solutions in $f(R, T)$ gravity. Bhatti et al. \cite{ref45} considered exponential $f(R, T)$ gravity model and the solutions of wormholes. 
Elizalde and Khurshudyan \cite{ref46} explored the transferable wormhole solutions in $f(R, T)$ gravity by assuming the different sorts 
of energy density. Sharif and Nawazish \cite{ref47} explored wormhole solutions for dust and non dust appropriations by the utilization 
of No-ether symmetry technique in $f(R, T)$ gravity.\\

Several functional forms for $f(R, T)$ were studied in various contexts for the derivation of cosmological dynamics. Zubair\cite{ref48} 
used three types of fluid and explores energy conditions and wormhole solutions. Godani and Samanta \cite{ref49} established a non-linear 
function and investigated the spherical regions for static transverse wormholes. They also worked on comparative analysis in GR of traversable 
WHs with exponential form function of modified gravity of $f(R)$ and $f(R, T)$. Simultaneously, NEC violation prompts the presence of WH 
solutions \cite{ref50}-\cite{ref55}, $36$ which are theoretical sections between two districts of spacetime supported by exotic  matter".\\

In the present model, we have a tendency to specialize in the phantom development representational process accelerated enlargement of 
the universe for WH geometry in adjusted $f(R, T)$ gravity. The enormous speeding up of the universe driven by ghost DE characterized 
through EoS with $\omega< 1$. In this model we select a new shape function to study the transversal solution in f(R,T) gravity  
with specific form of $f(R,T)= R+2\lambda T$ and designed a new shape function $b(r)=r[1-\frac{log(1+r-\gamma)}{log(1+r)}]$. The Sections I, 
brief reviews of wormholes and $f(R, T)$ gravity. Section II discuss $f(R, T)$ model and wormhole. In Section III related to wormhole solutions. 
Our results are discussed in Section IV and the concluded remarks are discussed in last section.

%%%%%%%%%%%%%%%%%%%%%%%%%%%%%%%  SECTION 2  %%%%%%%%%%%%%%%%%%%%%%%%%%%%%%%%%%%%%%%%%%%%%%%%%%%%%%%%%%%%%
\section{The formulation of f(R,T) gravity model}
In the Hilbert-Einstein action the respective field equation of $f(R, T)$ gravity model proposed by Harko et al. \cite{ref15} is 
formulated as follows:

\begin{equation}
\label{eq1}
S=\int{\sqrt{-g}\left( \frac{1}{16\pi G}f(R,T)+L_{m}\right)}d^{4}x,
\end{equation}
where $L_{m}$ is the Lagrangian density of matter source,  $g$ is the determinant of the metric tensor $g_{\mu\nu}$ and  $T$ is the 
trace of the energy-momentum tensor $T_{\mu\nu}$

Here the $T_{\mu\nu}$ from Lagrangian matter is characterized in the structure as
\begin{equation}
\label{eq2}
T_{\mu\nu}=-\frac{2}{\sqrt{-g}}\frac{\delta{\sqrt{-g}L_{m}}}{\delta{g^{\mu\nu}}}
\end{equation}

By varying the action $S$ in eqn. (\ref{eq1}) with respect to $g_{\mu\nu}$, the $f(R, T)$ gravity field equations are acquired as

\begin{equation}
\label{eq3}
f_{R}(R,T)R_{\mu\nu}-\frac{1}{2}f(R,T)g_{\mu\nu}+\left(g_{\mu\nu}\nabla^{\mu}\nabla_{\nu}-\nabla_{\mu}\nabla_{\nu}\right)f_{R}(R,T)=
8\pi T_{\mu\nu}-f_{T}(R,T)T_{\mu\nu}-f_{T}(R,T)\Theta_{\mu\nu}
\end{equation}
where
\begin{equation}
\label{eq4}
\Theta_{\mu\nu}=-2T_{\mu\nu}+g_{\mu\nu}L_{m}-2g^{lm}\frac{\partial^{2}L_{m}}{\partial g^{\mu}\partial g^{lm}}
\end{equation}
Here, $\nabla_{\mu}$ is the co-variant derivative, $f_{R}(R,T)=\frac{\partial f(R,T)}{\partial R}$, $f_{T}(R,T)=
\frac{\partial f(R,T)}{\partial T}$.

The curvature of space is the source term of the energy-momentum tensor. 
Every one of its segments can be viewed as sources of gravity and not simply mass density alone. Right now, the energy-momentum 
tensor for anisotropic fluid is represented as;

\begin{equation}
\label{eq5}
T^{\mu}_{\nu}=(\rho+p_{t})u^{\mu}u_{\nu}-p_{t}g^{\mu}_{\nu}+(p_{r}-p_{t})x^{\mu}x_{\mu}
\end{equation}

Here  $p_{t}$ is the tangential pressure, $\rho$ is the energy density, and $p_{r}$ is the radial pressure. $x_{\mu}$ is the radial 
unit four vector  and $u_{\mu}$  is the four-velocity vector and satisfied with the relation $x_{\mu}x_{\nu}=-1$  and $u^{\mu}u_{\nu}=1$. 
We select the Lagrangian matter as $L {m} = -P$, in which the maximum pressure is $L_{m} =-P$, where $P=\frac{p_{r}+2p_{t}}{3}$. 

In this paper, we consider the linear form of $f(R,T)$ function given by

\begin{equation}
\label{eq6}
f(R,T) = R+2f(T)=R+2\lambda T,
\end{equation}

where $\lambda$ represents an arbitrary constant. The models based on linearly in terms of $R$ or $T$ are also described in the 
literature in the sense of the gravity theories of both $f(R)$ and $f(R, T)$. 

The general $f(R,T)$ gravity field equations
(under the gravitational units $c=G=1$) for the above form (\ref{eq5}) of $f(R,T)$, is rewrite as ;

\begin{equation}
\label{eq7}
G^{\mu}_{\nu}=(8\pi+2\lambda)T^{\mu}_{\nu}+\lambda(\rho-P)
\end{equation}

\section{Wormhole solution}
Let us consider a static spherically symmetric wormhole metric \cite{ref8,ref9} in \textit{Schwarzschild} co-ordinate $(t,r,\theta,\phi)$ as

\begin{equation}
\label{eq8}
ds^{2}=-e^{2f(r)}dt^{2}+\left(1-\frac{b(r)}{r}\right)^{-1}dr^{2}+r^{2}\left(d\theta^{2}+sin^{2}\theta d\phi^{2}\right),
\end{equation}

where $f(r)$ and $b(r)$ are two arbitrary functions of the radial co-ordinate $r$, called redshift function and shape function, individually. 
The radial coordinate $r$ changes from $\gamma$ to infinity for example $\gamma \leq r \leq \infty$, where $\gamma$ is known as the throat 
radius of wormhole.\\

Additionally, the following conditions must be satisfied for the wormhole to be traversable and asymptotically flat by the redshift 
function $f(r)$ and the shape function $b(r)$:

\begin{itemize}
\item Absence of event horizon (required for traversable wormhole) : $f(r)$ should be finite.
\item Throat Condition : $b(\gamma) = \gamma$ and $b(r)$ should be less than $r$ for $r> \gamma$.
\item Flaring out condition : $b^{\prime}(\gamma) <1$ (in general $\frac{b-b^{\prime}r}{b^{2}}>0$), where $\prime$ denotes the 
derivative w.r.t. $r$.
\item Asymptotically Flatness : $\frac{b(r)}{r}\rightarrow 0$ as $r\rightarrow \infty$.
\end{itemize} 
\noindent In the present analysis, for the calculations, we consider a constant redshift function i.e. $f^{\prime}(r)=0$. 
Therefore, the field equations (\ref{eq6}) for the metric (\ref{eq7}) are given as:
\begin{equation}
\label{eq9}
\frac{b^{\prime}}{r^{2}}=\left(8\pi+3\lambda\right)\rho-\frac{1}{3}\lambda p_{r}-\frac{2}{3}\lambda p_{t},
\end{equation}
\begin{equation}
\label{eq10} \frac{b}{r^{3}}=-\left(8\pi+\frac{7}{3}\lambda \right)p_{r}+\lambda\rho-\frac{2}{3}\lambda p_{t},
\end{equation}
\begin{equation}
\label{eq11} \frac{b^{\prime}r-b}{2r^{3}}=-\left(8\pi+\frac{8}{3}\lambda \right)p_{t}+\lambda\rho-\frac{1}{3}\lambda p_{r}.
\end{equation}
Solving the above equations, the explicit solutions for the $\rho$, $p_{t}$ and $p_{r}$ are obtained as:
\begin{equation}
\label{eq12} p_{r}=\frac{b^{\prime} \lambda r-3(2\pi+\lambda)b}{2\left(3\lambda^{2}+62\pi\lambda+24\pi^{2}\right)}
\end{equation}
\begin{equation}
\label{eq13} p_{t}=\frac{\left[3\left(\lambda^{2}+66\pi\lambda+40\pi^{2}\right)b(r)-\left(3\lambda^{2}+70\pi\lambda+24\pi^{2}
\right)b^{\prime}r+2\lambda^{2}bb^{\prime}r\right]}{4(4\pi+\lambda)\left(3\lambda^{2}+62\pi\lambda+24\pi^{2}\right)r^{3}},
\end{equation}

\[
\rho = \frac{1}{8\pi+3\lambda}\Bigl[\frac{b^{\prime}}{r^{2}}+\frac{\lambda}{6(4\pi+\lambda)
\left(3\lambda^{2}+62\pi\lambda+24\pi^{2}\right)r^{3}}\Bigl((4\pi+\lambda)\left(b^{\prime}\lambda r-3(2\pi+\lambda)b\right)r^{3} 
\]
\begin{equation}
\label{eq14}
 +3\left(\lambda^{2} +66\pi\lambda+40\pi^{2}\right)b-\left(3\lambda^{2}+70\pi\lambda+24\pi^{2}\right)b^{\prime}r+2\lambda^{2}bb^{\prime}r\Bigr)\Bigr]
 \end{equation}

The above equations define the matter which threads the wormhole as
a function of the $b(r)$ shape function and the $\lambda$ coupling parameter. 
In particular, for $\lambda=0$, our proposed model represent the scenario in general relativity.
Now, the geometry of wormholes typically depends upon the choice of shape function $b(r)$. In the recent literature, various mathematical 
forms of the shape function such as $b(r)=r_{0}(r/r_{0})^{\frac{-1}{\omega}}$ for phantom fluid models $\omega < -1$ 
\cite{ref56}, $b(r)=\sqrt{r_{0}r}$ for Chaplygin gas wormholes \cite{ref57}, $b(r)=r_{0}^{2}/r$ \cite{ref58} (shape function considered for 
Ellis wormhole),
$b(r)=\sqrt{r_{0}r}$ and $b(r)=r_{0}+\gamma^{2}r_{0}(1-r_{0}/r)$ for $0<\gamma <1$ \cite{ref13}, $b(r)=r_{0}(r/r_{0})^{\gamma}$ for $0 < \gamma <1$ 
and $b(r) =r_{0}log(r+1)/log(r_{0}+1)$ \cite{ref59}, 
$b(r)=r_{0}(r/r_{0})^{n}$ 
and $b(r)=r_{0} (1+\delta^{2}(1-r_{0}/r))$ for $\delta^{2}<1$ \cite{ref60}, 
are used for the investigation of wormhole solutions. For our present analysis on  traversable and asymptotically flat wormholes, we 
propose our new logarithmic form of the shape function, given as:

\begin{equation}
\label{eq15}
b(r)=r\left[1-\frac{log(1+r-\gamma)}{log(1+r)}\right],
\end{equation}

It is clearly shown in Fig. $1$ that our proposed form of the shape function satisfies all the conditions required for the wormholes 
to be traversable and asymptotically flat. With the help of the above shape function, we can find our implicit wormhole solutions that 
represent the non-vacuum solutions of Einstein’s field equations and according to Einstein’s field theory, they may be filled with a 
matter which is different from the normal matter, called the exotic matter. Now, a natural question that comes to our mind: does the 
logarithmic dependence (both directly and inversely) of the shape function $b(r)$ on the radial coordinate $r$, give wormhole solutions 
with or without the presence of exotic matter. Our main motivation to use such new shape function for the linear $f(R,T)=R+2\lambda T$ 
the form is to find the different values of coupling constant $\lambda$ giving different types of wormhole geometry models with and without 
the presence of exotic matter.
%%%%%%%%%%%%%%%%%%%%%%%%%%%%%%%%%%%%%%%%%%%%%%%%%%%%%%%% Figure 1 %%%%%%%%%%%%%%%%%%%%%%%%%%%%%%%%%%%%%%%%%%%%%%%%%%%%%%%%%%%
\begin{figure}[htb]
	\centering
	\includegraphics[width=10cm,height=8cm,angle=0]{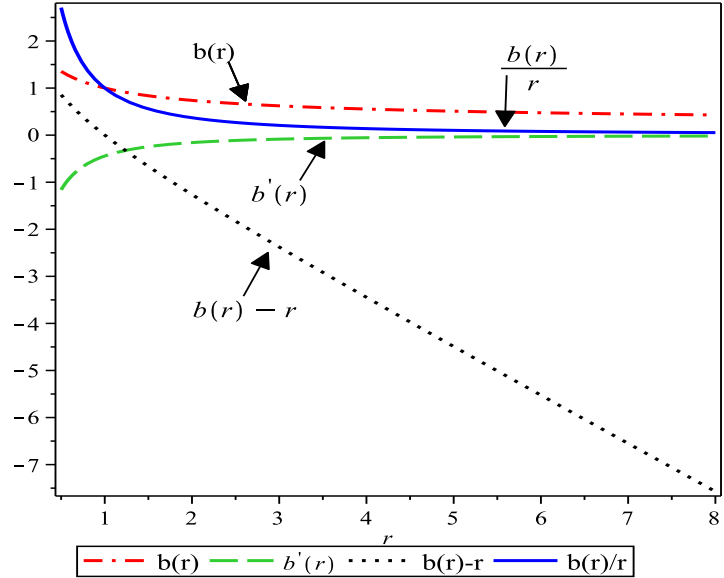}
	\caption{The Plot of $b(r)$, $b^{\prime}(r)$, $b(r)-r$ and $b(r)/r$ vs. Radial Co-ordinate ($r$). Here $\gamma = 1$.}
	\label{pic1}
\end{figure}
%%%%%%%%%%%%%%%%%%%%%%%%%%%%%%%%%%%%%%%%%%%%%%%%%%%%%%%%%%%%%%%%%%%%%%%%%%%%%%%%%%%%%%%%%%%%%%%%%%%%%%%%%%%%%%%%%

Using the shape function (\ref{eq15}) in Eqs. (\ref{eq12})$-$(\ref{eq14}),

 Implicit wormhole solutions were obtained and different combinations of
 energy density, radial and tangential pressures were measured, given as:

\[
p_{r}= \frac{1}{6\lambda^{2}+124\pi\lambda+48\pi^{2}}\Bigl[\Bigl(1-\phi(r_{1})+r\Bigl(-\frac{\phi(r_{1})}{(r_{1}-\gamma)
log(r_{1}-\gamma)}+\frac{\phi(r_{1})}{r_{1}log(r_{1})}\Bigr)\Bigr)\lambda r 
\]
\begin{equation}
\label{eq16}
 -3(2\pi+\lambda)r\left(1-\phi(r_{1})\right)\Bigr],
\end{equation}

\[
 p_{t}= \frac{1}{4}\frac{1}{r^{3}(4\pi+\lambda)\left(3\lambda^{2}+62\pi\lambda+24\pi^{2}\right)}\Bigl[3\left(\lambda^{2}+
66\pi\lambda+40\pi^{2}\right)r\left(1-\phi(r_{1})\right)+\Bigl(-\left(3\lambda^{2}+70\pi\lambda+24\pi^{2}\right)r
\]
\begin{align}
\label{eq17}  
 +2\lambda^{2}r^{2}(1-\phi(r_{1}))\Bigr)\Bigl(1-\phi(r_{1}) +r\Bigl(-\frac{\phi(r_{1})}{(r_{1}-\gamma)log(r_{1}-\gamma)}+
\frac{\phi(r_{1})}{r_{1} log(r_{1})}\Bigr)\Bigr)\Bigr],                                                                                                                                                           
\end{align}

\[
 \rho =  \frac{1}{8\pi+3\lambda}\Bigl[\frac{1-\phi(r_{1})+r\left(-\frac{\phi(r_{1})}{(r_{1}-\gamma)log(r_{1}-\gamma)}+
\frac{\phi(r_{1})}{r_{1}log(r_{1})}\right)}{r^{2}} +\frac{1}{6r^{3}(4\pi+\lambda)(3\lambda^{2}+62\pi \lambda+24\pi^{2})}
\]
\[
 \times \Bigl(3\lambda \Bigl((\lambda^{2}+66\pi\lambda+40\pi^{2})-(2\pi+\lambda)(4\pi+\lambda)r^{3}\Bigr) r(1-\phi(r_{1}))+
\lambda\Bigl(\lambda(4\pi+\lambda)r^{3}-(3\lambda^{2}+70\pi\lambda+24\pi^{2})
\]
\begin{equation}
\label{eq18}
+2\lambda^{2}r(1-\phi(r_{1}))\Bigr) r \Bigl(1-\phi(r_{1})+r\Bigl(-\frac{\phi(r_{1})}{(r_{1}-\gamma)log(r_{1}-\gamma)}+
\frac{\phi(r_{1})}{r_{1}log(r_{1})}\Bigr)\Bigr)\Bigr)\Bigr],
\end{equation}

\[
 \rho +p_{r}= \Bigl[ \frac{1}{8\pi+3\lambda}\Bigl[\frac{1-\phi(r_{1})+r\left(-\frac{\phi(r_{1})}{(r_{1}-\gamma)log(r_{1}-\gamma)}+
\frac{\phi(r_{1})}{r_{1}log(r_{1})}\right)}{r^{2}} +\frac{1}{6r^{3}(4\pi+\lambda)(3\lambda^{2}+62\pi \lambda+24\pi^{2})}
\]
\[
  \times \Bigl(3\lambda \Bigl((\lambda^{2}+66\pi\lambda+40\pi^{2})-(2\pi+\lambda)(4\pi+\lambda)r^{3}\Bigr)r(1-\phi(r_{1}))+
\lambda\Bigl(\lambda(4\pi+\lambda)r^{3}-(3\lambda^{2}+70\pi\lambda+24\pi^{2})
\]
\[
 +2\lambda^{2}r(1-\phi(r_{1}))\Bigr) r \Bigl(1-\phi(r_{1})+r\Bigl(-\frac{\phi(r_{1})}{(r_{1}-\gamma)log(r_{1}-\gamma)}+
\frac{\phi(r_{1})}{r_{1}log(r_{1})}\Bigr)\Bigr)\Bigr)\Bigr]\Bigr] +
\]
\begin{equation}
\label{eq19}
\Bigl[\frac{1}{6\lambda^{2}+124\pi\lambda+48\pi^{2}}\Bigl[\Bigl(1-\phi(r_{1})+r\Bigl(-\frac{\phi(r_{1})}{(r_{1}-\gamma)log(r_{1}-\gamma)}+
\frac{\phi(r_{1})}{r_{1}log(r_{1})}\Bigr)\Bigr)\lambda r 
-3(2\pi+\lambda)r\left(1-\phi(r_{1})\right)\Bigr]\Bigr],
\end{equation}

\[
\rho +p_{t}= \Bigl[\frac{1}{8\pi+3\lambda}\Bigl[\frac{1-\phi(r_{1})+r\left(-\frac{\phi(r_{1})}{(r_{1}-\gamma)log(r_{1}-\gamma)}+
\frac{\phi(r_{1})}{r_{1}log(r_{1})}\right)}{r^{2}} +\frac{1}{6r^{3}(4\pi+\lambda)(3\lambda^{2}+62\pi \lambda+24\pi^{2})}
\]
\[
 \times \Bigl(3\lambda \Bigl((\lambda^{2}+66\pi\lambda+40\pi^{2})-(2\pi+\lambda)(4\pi+\lambda)r^{3}\Bigr)r(1-\phi(r_{1}))+
\lambda\Bigl(\lambda(4\pi+\lambda)r^{3}-(3\lambda^{2}+70\pi\lambda+24\pi^{2}) +
\]
\[
2\lambda^{2}r(1-\phi(r_{1}))\Bigr) r \Bigl(1-\phi(r_{1})+r\Bigl(-\frac{\phi(r_{1})}{(r_{1}-\gamma)log(r_{1}-\gamma)}+
\frac{\phi(r_{1})}{r_{1} log(r_{1})}\Bigr)\Bigr)\Bigr)\Bigr]\Bigr] +
\]
\[
\Bigl[\frac{1}{4}\frac{1}{r^{3}(4\pi+\lambda)\left(3\lambda^{2}+62\pi\lambda+24\pi^{2}\right)}
\Bigl[3\left(\lambda^{2}+66\pi\lambda+40\pi^{2}\right)r\left(1-\phi(r_{1})\right)+\Bigl(-\left(3\lambda^{2}+70\pi\lambda+24\pi^{2}\right)r 
\]
\begin{equation}
\label{eq20}
+2\lambda^{2}r^{2}(1-\phi(r_{1}))\Bigr)\Bigl(1-\phi(r_{1}) +r\Bigl(-\frac{\phi(r_{1})}{(r_{1}-\gamma)log(r_{1}-\gamma)}+
\frac{\phi(r_{1})}{r_{1}log(r_{1})}\Bigr)\Bigr)\Bigr]\Bigr],
\end{equation}

\[
\rho- |p_{r}| = \Bigl[ \frac{1}{8\pi+3\lambda}\Bigl[\frac{1-\phi(r_{1})+r\left(-\frac{\phi(r_{1})}{(r_{1}-\gamma)log(r_{1}-\gamma)}+
\frac{\phi(r_{1})}{r_{1}log(r_{1})}\right)}{r^{2}} +\frac{1}{6r^{3}(4\pi+\lambda)(3\lambda^{2}+62\pi \lambda+24\pi^{2})}
\]
\[
 \times \Bigl(3\lambda \Bigl((\lambda^{2}+66\pi\lambda+40\pi^{2})-(2\pi+\lambda)(4\pi+\lambda)r^{3}\Bigr)r(1-\phi(r_{1}))+
\lambda\Bigl(\lambda(4\pi+\lambda)r^{3}-(3\lambda^{2}+70\pi\lambda+24\pi^{2}) +
\]
\[
2\lambda^{2}r(1-\phi(r_{1}))\Bigr) r \Bigl(1-\phi(r_{1})+r\Bigl(-\frac{\phi(r_{1})}{(r_{1}-\gamma)log(r_{1}-\gamma)}+
\frac{\phi(r_{1})}{r_{1}log(r_{1})}\Bigr)\Bigr)\Bigr)\Bigr]\Bigr] -
\]
\begin{equation}
\label{eq21} 
\Bigl|\frac{1}{6\lambda^{2}+124\pi\lambda+48\pi^{2}}\Bigl[\Bigl(1-\phi(r_{1})+
r\Bigl(-\frac{\phi(r_{1})}{(r_{1}-\gamma)log(r_{1}-\gamma)}+\frac{\phi(r_{1})}{r_{1}log(r_{1})}\Bigr)\Bigr)\lambda r 
-3(2\pi+\lambda)r\left(1-\phi(r_{1})\right)\Bigr]\Bigr|,
\end{equation}

\[
\rho -|p_{t}|= \Bigl[\frac{1}{8\pi+3\lambda}\Bigl[\frac{1-\phi(r_{1})+r\left(-\frac{\phi(r_{1})}{(r_{1}-\gamma)log(r_{1}-\gamma)}+
\frac{\phi(r_{1})}{r_{1}log(r_{1})}\right)}{r^{2}} +\frac{1}{6r^{3}(4\pi+\lambda)(3\lambda^{2}+62\pi \lambda+24\pi^{2})}
\]
\[
\times \Bigl(3\lambda \Bigl((\lambda^{2}+66\pi\lambda+40\pi^{2})-(2\pi+\lambda)(4\pi+\lambda)r^{3}\Bigr)r(1-\phi(r_{1}))+
\lambda\Bigl(\lambda(4\pi+\lambda)r^{3}-(3\lambda^{2}+70\pi\lambda+24\pi^{2})
\]
\[
+2\lambda^{2}r(1-\phi(r_{1}))\Bigr) r \Bigl(1-\phi(r_{1})+r\Bigl(-\frac{\phi(r_{1})}{(r_{1}-\gamma)log(r_{1}-\gamma)}+
\frac{\phi(r_{1})}{r_{1}log(r_{1})}\Bigr)\Bigr)\Bigr)\Bigr]\Bigr] 
\]
\[
-\Bigl|\frac{1}{4}\frac{1}{r^{3}(4\pi+\lambda)\left(3\lambda^{2}+62\pi\lambda+24\pi^{2}\right)}
\Bigl[3\left(\lambda^{2}+66\pi\lambda+40\pi^{2}\right)r\left(1-\phi(r_{1})\right)+\Bigl(-\left(3\lambda^{2}+70\pi\lambda+24\pi^{2}\right)r
\]
\begin{equation}
\label{eq22}
+2\lambda^{2}r^{2}(1-\phi(r_{1}))\Bigr)\Bigl(1-\phi(r_{1}) +r\Bigl(-\frac{\phi(r_{1})}{(r_{1}-\gamma)log(r_{1}-\gamma)}+
\frac{\phi(r_{1})}{r_{1}log(r_{1})}\Bigr)\Bigr)\Bigr]\Bigr|,
\end{equation}

\[
\rho + p_{r}+2p_{t}= \Bigl[ \frac{1}{8\pi+3\lambda}\Bigl[\frac{1-\phi(r_{1})+r\left(-\frac{\phi(r_{1})}{(r_{1}-\gamma)log(r_{1}-\gamma)}+
\frac{\phi(r_{1})}{r_{1}log(r_{1})}\right)}{r^{2}} +\frac{1}{6r^{3}(4\pi+\lambda)(3\lambda^{2}+62\pi \lambda+24\pi^{2})}
\]
\[
 \times \Bigl(3\lambda \Bigl((\lambda^{2}+66\pi\lambda+40\pi^{2})-(2\pi+\lambda)(4\pi+\lambda)r^{3}\Bigr)r(1-\phi(r_{1}))+
\lambda\Bigl(\lambda(4\pi+\lambda)r^{3}-(3\lambda^{2}+70\pi\lambda+24\pi^{2})
\]
\[
+2\lambda^{2}r(1-\phi(r_{1}))\Bigr) r \Bigl(1-\phi(r_{1})+r\Bigl(-\frac{\phi(r_{1})}{(r_{1}-\gamma)log(r_{1}-\gamma)}+
\frac{\phi(r_{1})}{r_{1}log(r_{1})}\Bigr)\Bigr)\Bigr)\Bigr]\Bigr] 
\]
\[
+\Bigl[\frac{1}{6\lambda^{2}+124\pi\lambda+48\pi^{2}}\Bigl[\Bigl(1-\phi(r_{1})+r\Bigl(-\frac{\phi(r_{1})}{(r_{1}-\gamma)log(r_{1}-\gamma)}+
\frac{\phi(r_{1})}{r_{1}log(r_{1})}\Bigr)\Bigr)\lambda r
\]
\[
-3(2\pi+\lambda)r\left(1-\phi(r_{1})\right)\Bigr]\Bigr] +2\Bigl[\frac{1}{4}\frac{1}{r^{3}(4\pi+\lambda)\left(3\lambda^{2}+
62\pi\lambda+24\pi^{2}\right)}\Bigl[3\left(\lambda^{2}+66\pi\lambda+40\pi^{2}\right)r\left(1-\phi(r_{1})\right) 
\]
\begin{equation}
\label{eq23}
+\Bigl(-\left(3\lambda^{2}+70\pi\lambda+24\pi^{2}\right)r +2\lambda^{2}r^{2}(1-\phi(r_{1}))\Bigr)\Bigl(1-\phi(r_{1}) +
r\Bigl(-\frac{\phi(r_{1})}{(r_{1}-\gamma)log(r_{1}-\gamma)}+\frac{\phi(r_{1})}{r_{1}log(r_{1})}\Bigr)\Bigr)\Bigr]\Bigl],
\end{equation}
where in all the above relations,  $\phi(r_{1}) =\frac{log(r_{1}-\gamma)}{log(r_{1})}$, here $r_{1}=1+r$. Now, the anisotropy parameter 
($\triangle = p_{t}-p_{r}$) is computed as:

\[
\triangle = \Bigl[\frac{1}{4}\frac{1}{r^{3}(4\pi+\lambda)\left(3\lambda^{2}+62\pi\lambda+24\pi^{2}\right)}
\Bigl[3\left(\lambda^{2}+66\pi\lambda+40\pi^{2}\right)r\left(1-\phi(r_{1})\right)+\Bigl(-\left(3\lambda^{2}+70\pi\lambda+24\pi^{2}\right)r 
\]
\[
 +2\lambda^{2}r^{2}(1-\phi(r_{1}))\Bigr)\Bigl(1-\phi(r_{1}) +r\Bigl(-\frac{\phi(r_{1})}{(r_{1}-\gamma)log(r_{1}-\gamma)}+
\frac{\phi(r_{1})}{r_{1}log(r_{1})}\Bigr)\Bigr)\Bigr]\Bigr]
\]
\begin{equation}
\label{eq24}
- \Bigl[\frac{1}{6\lambda^{2}+124\pi\lambda+48\pi^{2}}\Bigl[\Bigl(1-\phi(r_{1})+r\Bigl(-\frac{\phi(r_{1})}{(r_{1}-\gamma)log(r_{1}-\gamma)}
+\frac{\phi(r_{1})}{r_{1}log(r_{1})}\Bigr)\Bigr)\lambda r -3(2\pi+\lambda)r\left(1-\phi(r_{1})\right)\Bigr]\Bigr].
\end{equation}

If $\triangle = 0$, then the geometry has an isotropic pressure.\\

If $\triangle < 0, $ then it would seen that the geometry is attractive. If $\triangle > 0 $ then it would seen that the geometry is 
repulsive. If $\triangle = 0 $ then isotropic pressure throughout the geometry.\\

Now, let us consider the two relations :
\[
p_{r} = \omega_{r} \rho, 
\]
\begin{equation}
\label{eq25}
p_{t} = \omega_{t}\rho,
\end{equation}
as the equations of state for matter inside the wormholes, where $\omega_{r}$ and $\omega_{t}$ are variables known as radial state 
parameter and tangential state parameter. The combinations of above two equations of state have been already invoked in the 
literature \cite {ref44,ref56,ref61,ref62}. From the above two relations, we can get the functions for $\omega_{r}$ and $\omega_{t}$ as

\[
\omega_{r}= \Bigl[\frac{1}{6\lambda^{2}+124\pi\lambda+48\pi^{2}}\Bigl[\Bigl(1-\phi(r_{1})+
r\Bigl(-\frac{\phi(r_{1})}{(r_{1}-\gamma)log(r_{1}-\gamma)}+\frac{\phi(r_{1})}{r_{1}log(r_{1})}\Bigr)\Bigr)\lambda r 
\]
\[
-3(2\pi+\lambda)r\left(1-\phi(r_{1})\right)\Bigr] \Bigr]\Bigl{/} \Bigl[ \frac{1}{8\pi+3\lambda}\Bigl[\frac{1-\phi(r_{1})+
r\left(-\frac{\phi(r_{1})}{(r_{1}-\gamma)log(r_{1}-\gamma)}+\frac{\phi(r_{1})}{r_{1}log(r_{1})}\right)}{r^{2}} 
\]
\[
 +\frac{1}{6r^{3}(4\pi+\lambda)(3\lambda^{2}+62\pi \lambda+24\pi^{2})} \Bigl(3\lambda \Bigl((\lambda^{2}+66\pi\lambda+
40\pi^{2})-(2\pi+\lambda)(4\pi+\lambda)r^{3}\Bigr)r(1-\phi(r_{1}))
\]
\[
+\lambda\Bigl(\lambda(4\pi+\lambda)r^{3}-(3\lambda^{2}+70\pi\lambda+24\pi^{2})+2\lambda^{2}r(1-\phi(r_{1}))\Bigr) 
r \Bigl(1-\phi(r_{1})+r\Bigl(-\frac{\phi(r_{1})}{(r_{1}-\gamma)log(r_{1}-\gamma)} 
\]
\begin{equation}
\label{eq26}
+\frac{\phi(r_{1})}{r_{1}log(r_{1})}\Bigr)\Bigr)\Bigr)\Bigr]\Bigr]  
\end{equation}

\[ 
\omega_{t}= \Bigl[\frac{1}{4}\frac{1}{r^{3}(4\pi+\lambda)\left(3\lambda^{2}+62\pi\lambda+24\pi^{2}\right)}
\Bigl[3\left(\lambda^{2}+66\pi\lambda+40\pi^{2}\right)r\left(1-\phi(r_{1})\right)+\Bigl(-\left(3\lambda^{2}+70\pi\lambda+24\pi^{2}\right)r 
\]
\[
+2\lambda^{2}r^{2}(1-\phi(r))\Bigr)\Bigl(1-\phi(r_{1}) +r\Bigl(-\frac{\phi(r_{1})}{(r_{1}-\gamma)log(r_{1}-\gamma)}+
\frac{\phi(r_{1})}{r_{1}log(r_{1})}\Bigr)\Bigr)\Bigr]\Bigr]
\]
\[
\Bigl{/} \Bigl[ \frac{1}{8\pi+3\lambda}\Bigl[\frac{1-\phi(r_{1})+r\left(-\frac{\phi(r_{1})}{(r_{1}-\gamma)log(r_{1}-\gamma)}+
\frac{\phi(r_{1})}{r_{1}log(r_{1})}\right)}{r^{2}} +\frac{1}{6r^{3}(4\pi+\lambda)(3\lambda^{2}+62\pi \lambda+24\pi^{2})}
\]
\[
 \times \Bigl(3\lambda \Bigl((\lambda^{2}+66\pi\lambda+40\pi^{2})-(2\pi+\lambda)(4\pi+\lambda)r^{3}\Bigr)r(1-\phi(r_{1}))+
\lambda\Bigl(\lambda(4\pi+\lambda)r^{3}-(3\lambda^{2}+70\pi\lambda+24\pi^{2})
\]
\begin{equation}
\label{eq27}
+2\lambda^{2}r(1-\phi(r_{1}))\Bigr) r \Bigl(1-\phi(r_{1})+r\Bigl(-\frac{\phi(r_{1})}{(r_{1}-\gamma)log(r_{1}-\gamma)}+
\frac{\phi(r_{1})}{(r_{1}log(r_{1})}\Bigr)\Bigr)\Bigr)\Bigr]\Bigr].
\end{equation}

\noindent Now, we discuss the four fundamental energy conditions :

 Null Energy Condition (NEC), Weak Energy Condition (WEC), Strong Energy Condition (SEC) and Dominant Energy Condition (DEC). Generally, these 
 energy conditions are in the form of inequalities for the contraction of time-like and null vector fields with respect to the matter 
 describing properties i.e. the Einstein tensor and the energy-momentum tensor appearing in Einstein's field equations. The mathematical 
 and the physical formulation of the energy conditions is discussed below:
\begin{itemize}
\item \textbf{NEC}: It is unaccented confinement and stands for the attractive nature of gravity. In terms of principle pressures, it is 
expressed as : $\rho +p_{r} \geq 0$, $\rho+p_{t}\geq 0$. 
\item \textbf{SEC}: It ensures the positivity of energy density along with the fulfillment of NEC. In terms of principle pressures, it is 
expressed as : $\rho \geq 0$, $\rho +p_{r} \geq 0$, $\rho+p_{t}\geq 0$. 
\item \textbf{WEC}: It also represents the attractive nature of gravity. In terms of principle pressures, it is expressed as : 
$\rho +p_{r} \geq 0$, $\rho+p_{t}\geq 0$, $\rho +p_{r}+2p_{t} \geq 0$.
\item \textbf{DEC}: It additions the positivity of energy density imposing limits to the velocity of energy transfer with the 
speed of light. In terms of principle pressures, it is expressed as : $\rho \geq 0$, $\rho -|p_{r}| \geq 0$, $\rho-|p_{t}|\geq 0$. 
\end{itemize}

%%%%%%%%%%%%%%%%%%%%%%%%%%%%%%%%%%%%%%%%%%%%%%%%%%%%%%%% Section 4 %%%%%%%%%%%%%%%%%%%%%%%%%%%%%%%%%%%%%%%%%%%%%%%%%%%%%%%%%
\section{Results and Discussions}
 
In the presenting paper, in the context of $f(R, T)$ gravity concept, the presence of wormhole solutions was investigated with a linear 
$f(R, T)$ function described by $f(R, T)=R+2\lambda T$, where $\lambda$ is an arbitrary coupling constant. 
In section $2$, the background of $f(R,T)$ gravity theory has been explained. In section $3$, the field equations are derived and their 
respective implicit solutions have been obtained with the help of our newly proposed shape function $b(r)=r\left[1-\frac{log(1+r-\gamma)}
{log(1+r)}\right]$. Moreover, terms related to the energy conditions, the anisotropy parameter $\triangle$ and the state parameters 
$\omega_{r}$, $\omega_{t}$ are computed with the help of implicit solutions for $\rho$, $p_{r}$ and $p_{t}$. In this section, we discuss 
the results and the analysis related to the absence and presence of exotic matter, has been presented from the respective validation and 
violation of four basic energy conditions with our new shape function for the different values of coupling constant 
$\lambda$ in $f(R,T)=R+2\lambda T$ gravity theory. For the mathematical simplicity of our analysis, we have chosen radius of the throat 
to be unity i.e. $\gamma =1$ and hence $1\leq r \leq \infty$.  For different values of $\lambda \in [-100,100]$, a detailed study of the 
results have been summarized in the tables $1$, $2$ and $3$. It may be easily observed from these tables that for some particular range of 
$\lambda$, all the energy conditions are validated pointing the non-existence of exotic matter, while for some ranges of $\lambda$, all 
the conditions are violated showing the presence of exotic matter. Moreover, there is a particular range of $\lambda$, for which some of 
the energy conditions are validated and some are violated. For more precise results, we have chosen particular values of $\lambda$ from 
these ranges and a clear picture of the behaviour of energy conditions, state parameters ($\omega_{r}$, $\omega_{t}$) and anisotropy 
parameter ($\triangle$) w.r.t. radial coordinate $r$ has been presented in the figures $2$-$9$. \\

In fig. $2$, we have taken $\lambda=-75$, it can be easily seen from fig. $2(a)$ and fig. $2(b)$ that the NEC, WEC and SEC are validated 
for the wormhole geometry while the terms for DEC ($\rho-|p_{r}| \geq 0$, $\rho-|p_{t}| \geq 0$) are violated. In fig. $3$, for $\lambda = -13$, 
it has been observed from fig. $3(a)$ and fig. $3(b)$ that all the energy conditions are violated near the throat of the wormhole showing 
the presence of exotic matter. From fig. $4$ for $\lambda=-9.5$, it can be clearly observed from fig. $4(a)$ and fig. $4(b)$ that all the 
energy conditions being positive decreasing are validated for the wormhole geometry. Therefore, $\lambda=-9.5$ is the value that 
predicts the non-existence of exotic matter near the throat of the wormhole. In fig. $5$ for $\lambda=-5$, fig. $5(a)$ depicts negative 
energy density $\rho$ which clearly depicts the presence of exotic matter and $\rho \rightarrow 0$ as $r \rightarrow 4$. Moreover, the 
term corresponding to NEC, WEC and SEC i.e. $\rho+p_{r}$ is a negative increasing function of $r$ near the throat of a wormhole, as $r$ 
increases firstly it approaches to 0 and then attains a fixed positive value. Now, the other term common to NEC, WEC and SEC i.e. 
$\rho+p_{t}$ is a positive decreasing function of $r$ near the throat of the wormhole, as $r$ increases, it approaches to zero and then 
attains a fixed negative value. 
The term $\rho+p_{r}+2p_{t}$ corresponding to WEC is positive decreasing function of $r$ approaching to 0. Fig. $5(b)$ shows that the 
terms of DEC $\rho-|p_{r}|$ and $\rho-|p_{t}|$ are negative and increasing functions of $r$. Combining the behaviour of all these energy 
condition terms for $\lambda=-5$, the presence of exotic matter near the throat of the wormhole can be confirmed. Now, for $\lambda= -1.5$, 
it has been observed from fig. $6(a)$ and fig. $6(b)$ that energy density $\rho$ is negative increasing function of $r$, the term $\rho+p_{r}$ 
is positive decreasing function of $r$, the term $\rho+p_{t}$ is negative increasing function of $r$, the term $\rho+p_{r}+2p_{t}$ is shifting 
from negative increasing to positive decreasing and the terms $\rho-|p_{r}|$, $\rho-|p_{t}|$ are negative increasing functions of $r$. Combining 
the behaviour of all, we conclude that for $\lambda=-1.5$, there is a violation of energy conditions and hence, exotic matter is present near 
the throat of the wormhole. Now, for $\lambda=-1$ from fig. $7(a)$ and $7(b)$, it is easily observed that $rho$ becomes positive for $r > 2$, 
but the rest of the terms for the energy conditions such as $\rho+p_{r}$, $\rho-|p_{r}|$, $\rho-|p_{t}|$ are negative increasing, $\rho+p_{t}$ 
is positive decreasing and $\rho+p_{r}+2p_{t}$ is changing its behaviour from positive decreasing to negative increasing. The final result shows 
a violation near the throat of the wormhole affirming the presence of exotic matter. \\

Now, we discuss the case for general relativity for which the coupling constant $\lambda$ is considered to be zero i.e. $f(R)=R$. From 
fig. $8(a)$, the energy density $\rho$ is found to be a negative function of radial coordinate $r$, which indicates the presence of 
exotic matter in wormhole geometry. Moreover from $8(a)$, the term $\rho+p_{r}$ is negative increasing, the term $\rho+p_{t}$ is positive 
decreasing, the term $\rho+p_{r}+2p_{t}$ has its behaviour change from positive decreasing to negative increasing and from fig. $8(b)$, 
the DEC terms $\rho-|p_{r}|$, $\rho-|p_{t}|$ are found to be negative increasing. We may conclude that for the case of general relativity, 
there is net violation of the energy conditions and the presence of exotic matter is verified. At last, we discuss the positive values 
of $\lambda$. For reference, we have chosen $\lambda=1$ and the respective behaviours of energy condition terms have been shown in 
fig. $9(a)$ and $9(b)$. It may be observed that the behaviour of all the terms is the same as we have obtained for $\lambda=0$ i.e. there is 
the presence of exotic matter. We may say that for $\lambda \geq 0$, the presence of exotic matter is confirmed for the wormhole geometry. 
The final analysis regarding the violation/validation of the energy conditions for different chosen values of $\lambda$ in $[-100,100]$ has 
been presented in Table $4$.\\

Moreover, we have also plotted the state parameters ($\omega_{r}$, $\omega_{t}$) in figs. $2-9$(c) and figs. $2-9$(d) and the anisotropy 
parameter ($\triangle$) in figs. $2-9$ (e) w.r.t. the radial co-ordinate $r$ for different chosen values of $\lambda$ in $[-100,100]$. 
The combined analysis of their behavior and the nature of the geometry of the wormholes has been presented in Tables $5$ and $6$.

%%%%%%%%%%%%%%%%%%%%%%%%%%%%%%%%%%%%%%%%%%%%%%%%%%%%%%%%%%%%%%%%%%%%%%%%%%%%%%%%%%%%%%%%%%%%%%%%%%%%%%%%%%%%%%%%%%%%%%%%%%%%%%%%%%%%
\begin{figure}[H]
    \centering
    	(a)\includegraphics[width=7cm,height=7cm,angle=0]{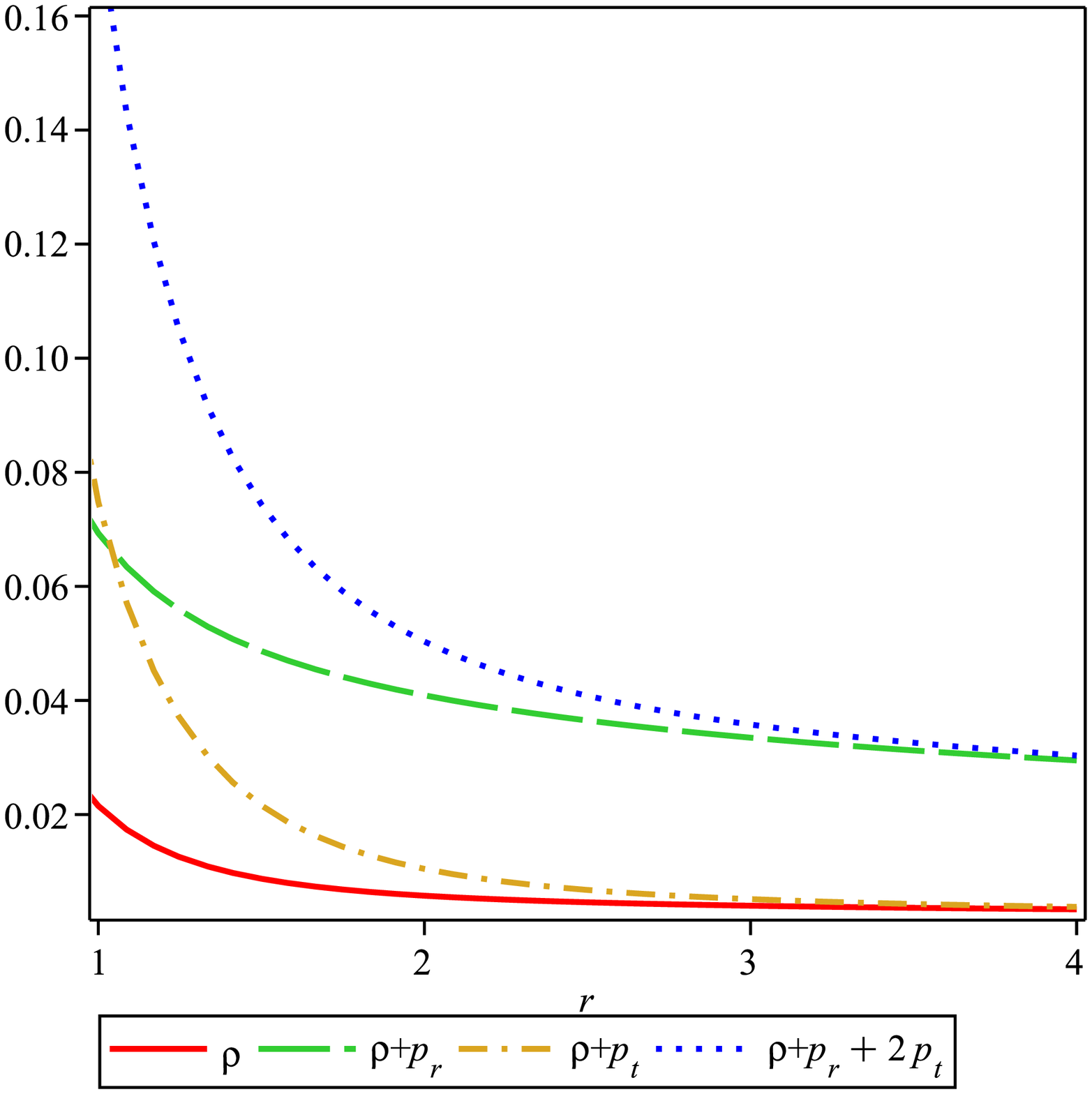}
    [b]	\includegraphics[width=7cm,height=7cm,angle=0]{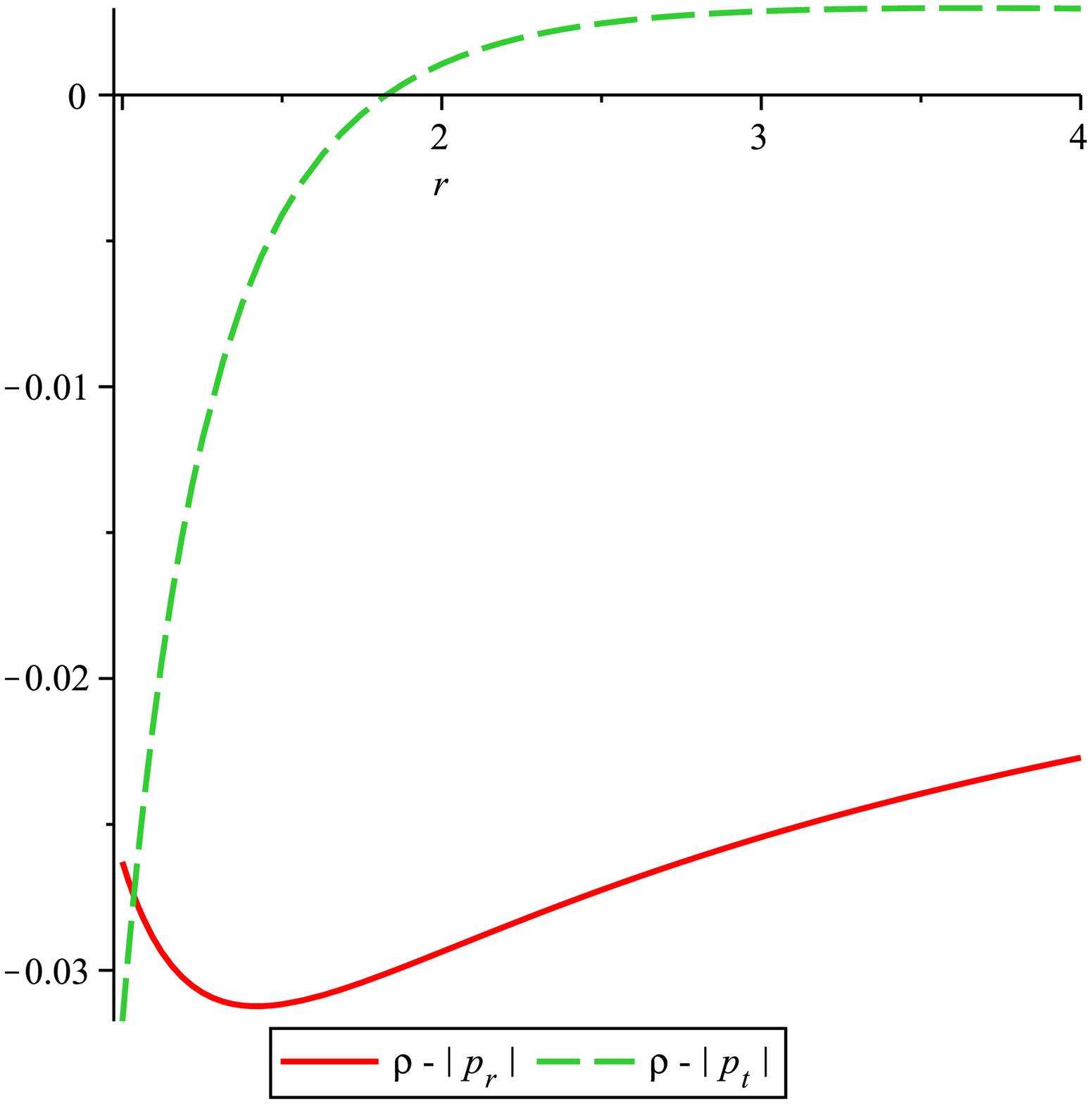} 
    [c]{\includegraphics[width=0.4\textwidth]{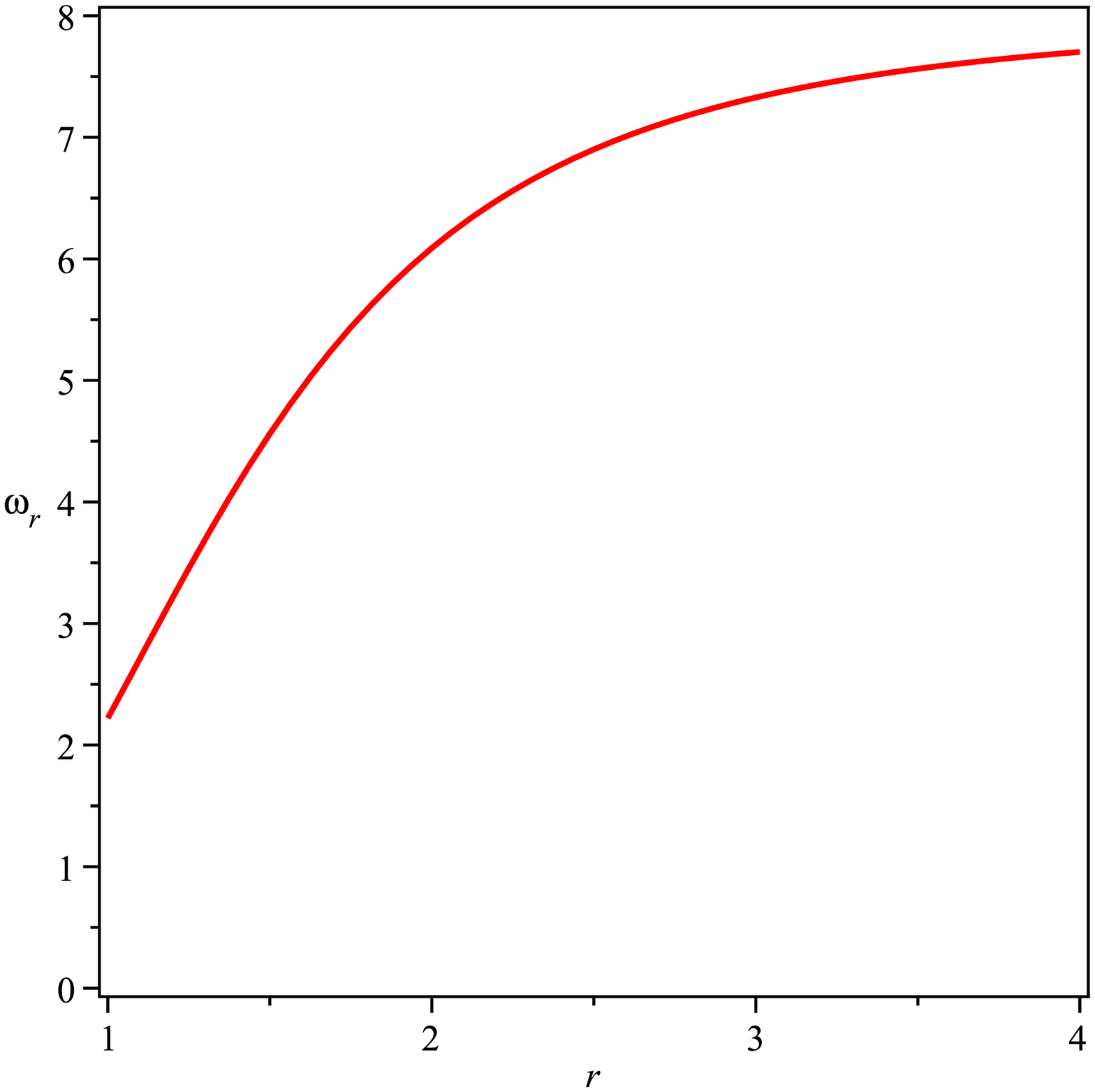}}
    [d]{\includegraphics[width=0.4\textwidth]{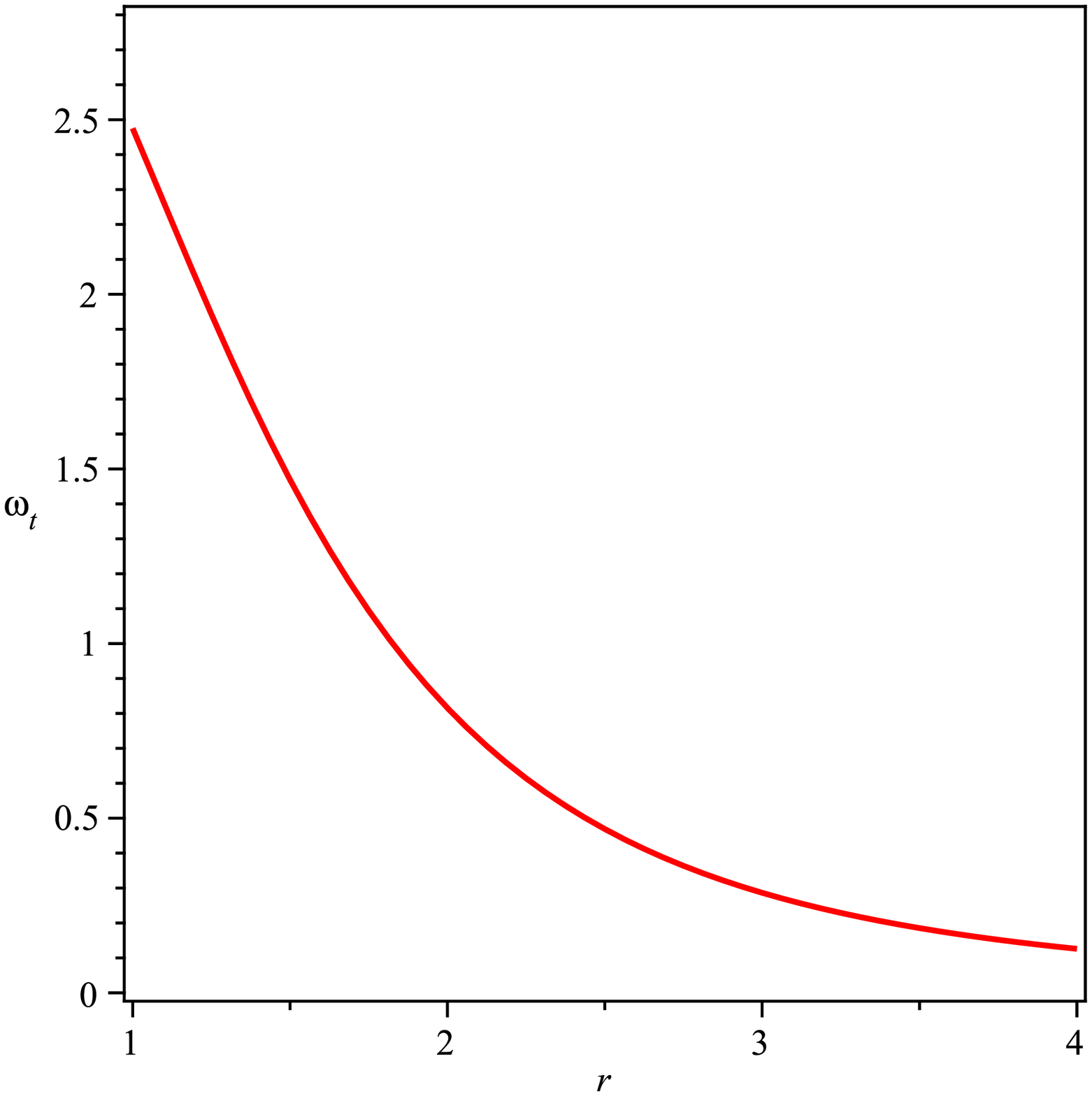}}
    [e]{\includegraphics[width=0.4\textwidth]{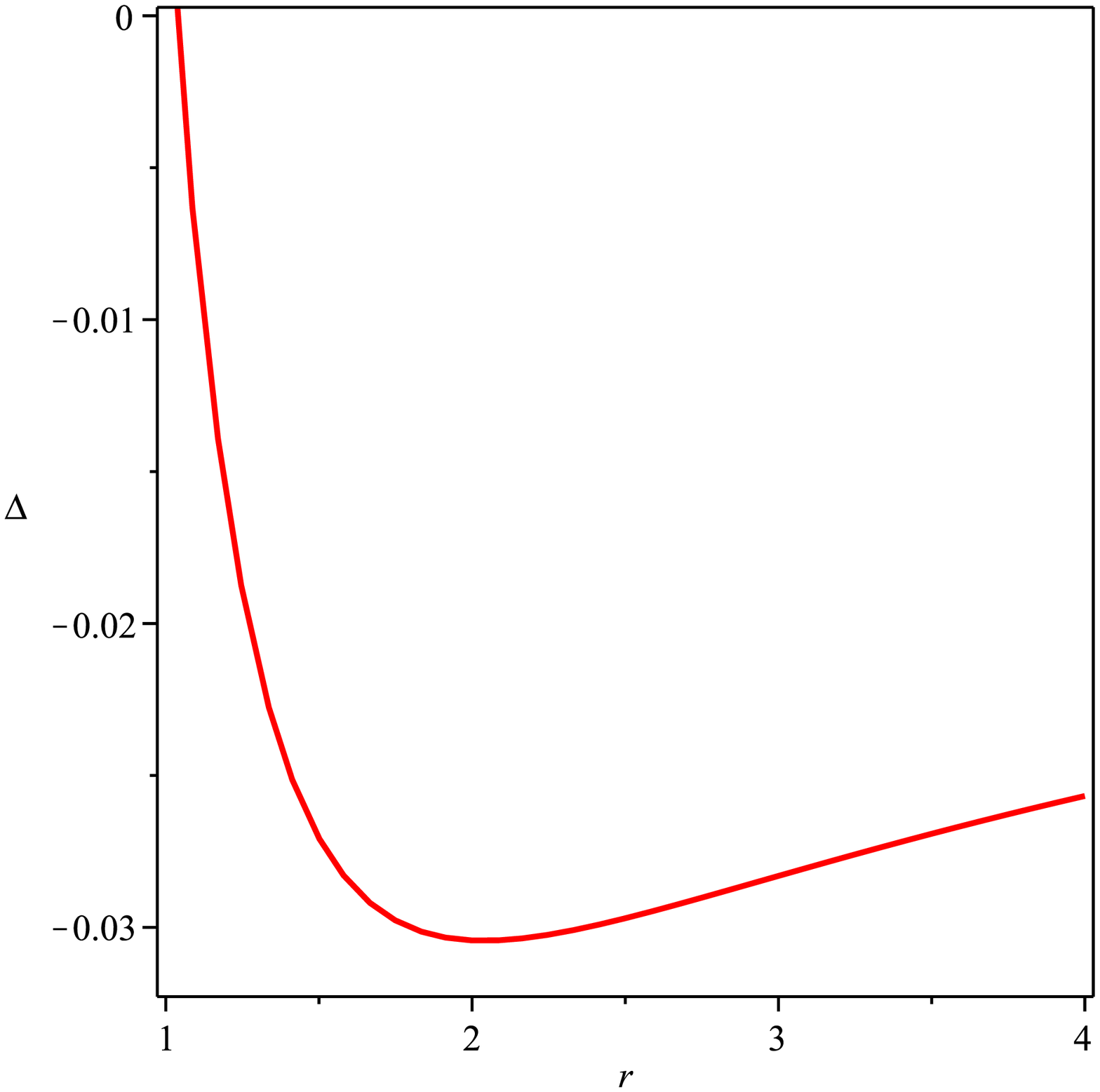}}
    \caption{The plot of [a] NEC, SEC, WEC, [b] DEC, [c] Radial EoS Parameter ($\omega_{r}$), [d] Tangential EoS parameter ($\omega_{t}$), 
    [e] Anisotropy Parameter ($\triangle$), vs. Radial Co-ordinate ($r$). Here $\gamma =1$, $\lambda=-75$.}
    %\label{fig:foobar}
    \label{fig:2}
\end{figure}
%%%%%%%%%%%%%%%%%%%%%%%%%%%%%%%%%%%%%%%%%%%%%%%%%%%%%%%%%%%%%%%%%%%%%%%%%%%%%%%%%%%%%%%%%%%%%%%%%%%%%%%%%%%%%%%%%%%%%%%%%%%%%%%%%%%%%
%%%%%%%%%%%%%%%%%%%%%%%%%%%%%%%%%%%%%%%%%%%%%%%%%%%%%%%%%%%%%%%%%%%%%%%%%%%%%%%%%%%%%%%%%%%%%%%%%%%%%%%%%%%%%%%%%%%%%%%%%%%%%%%
\begin{figure}[H]
    \centering
    [a]{\includegraphics[width=0.4\textwidth]{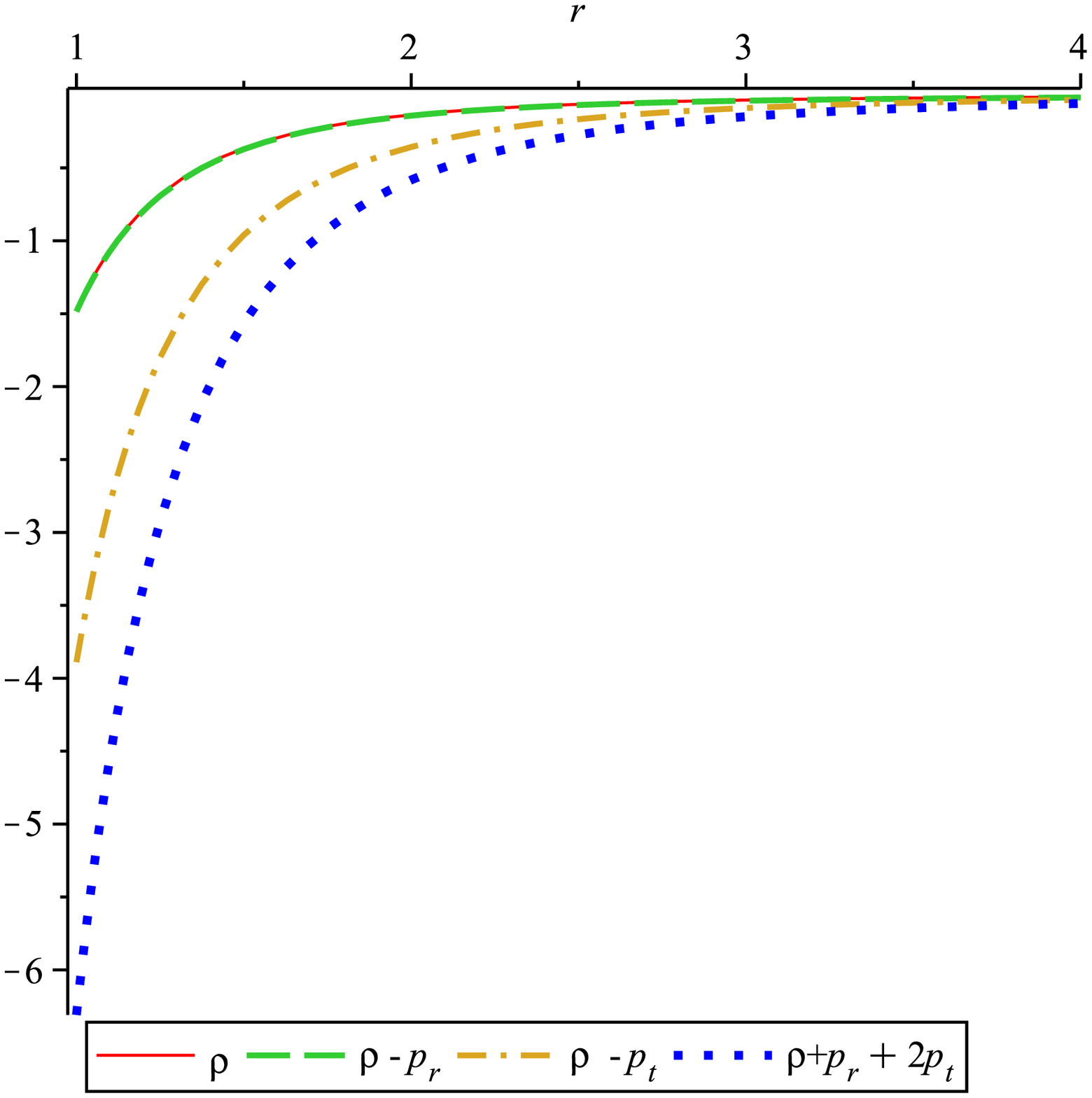}} 
    [b]{\includegraphics[width=0.4\textwidth]{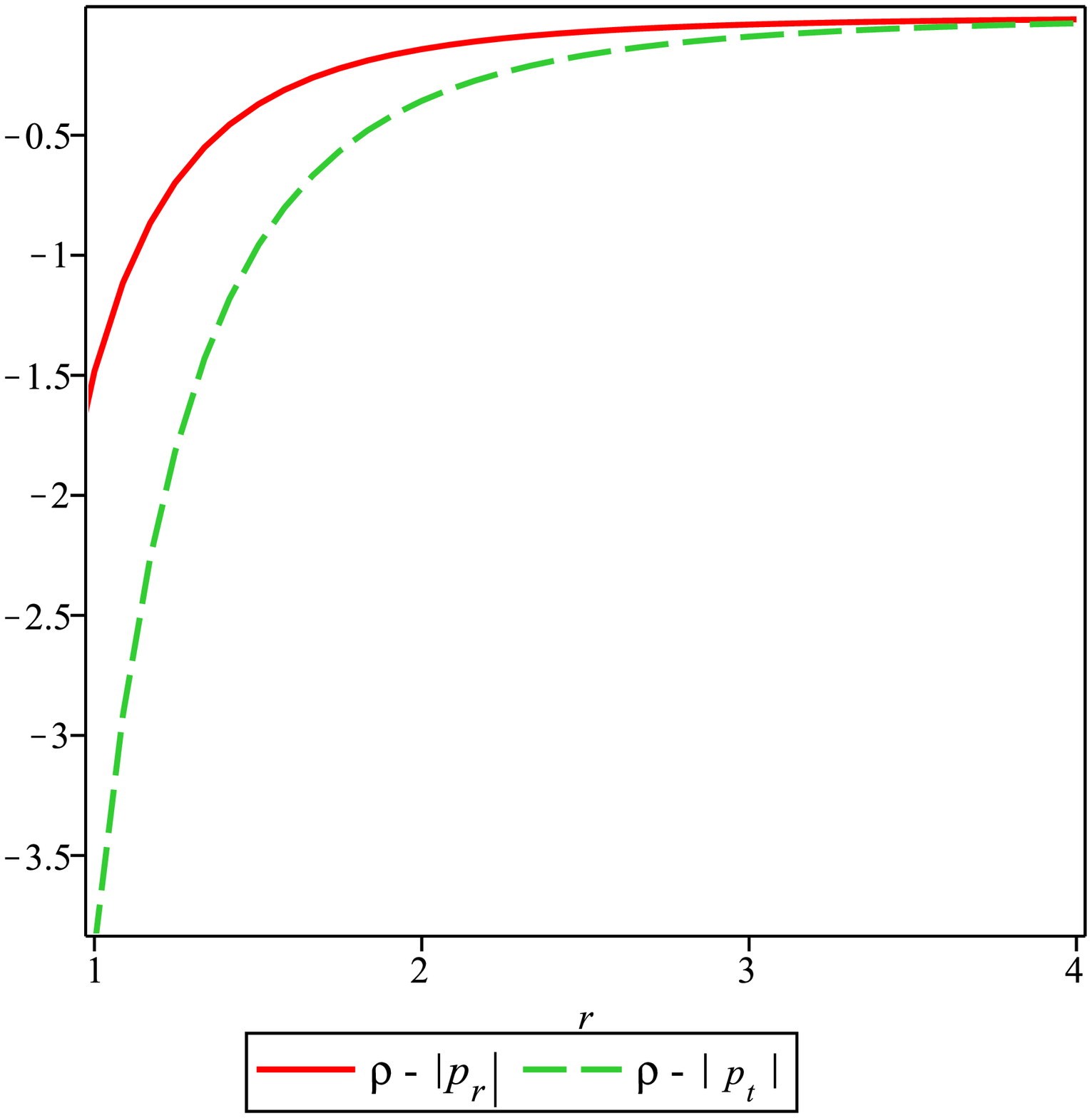}} 
    [c]{\includegraphics[width=0.4\textwidth]{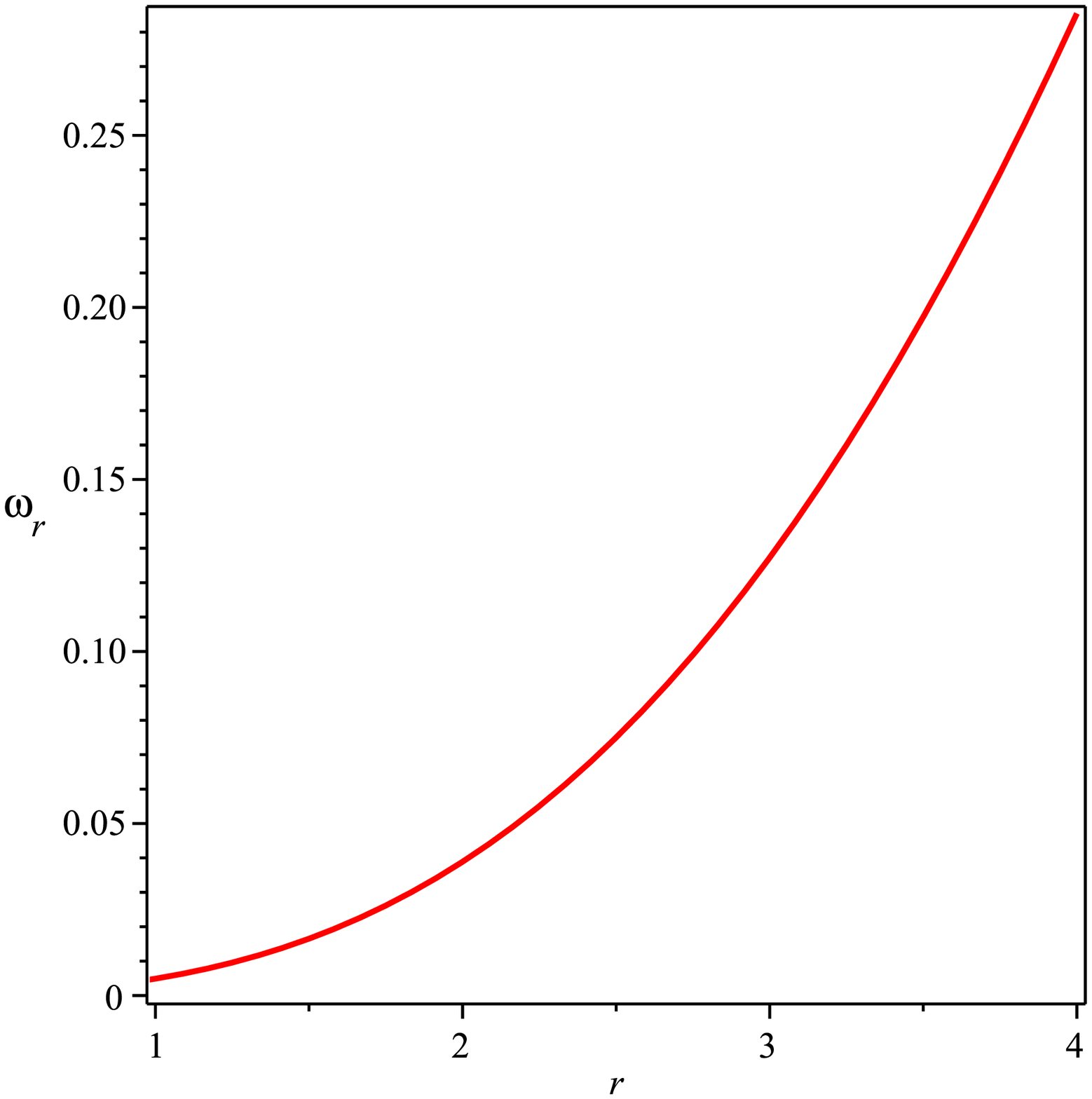}}
    [d]{\includegraphics[width=0.4\textwidth]{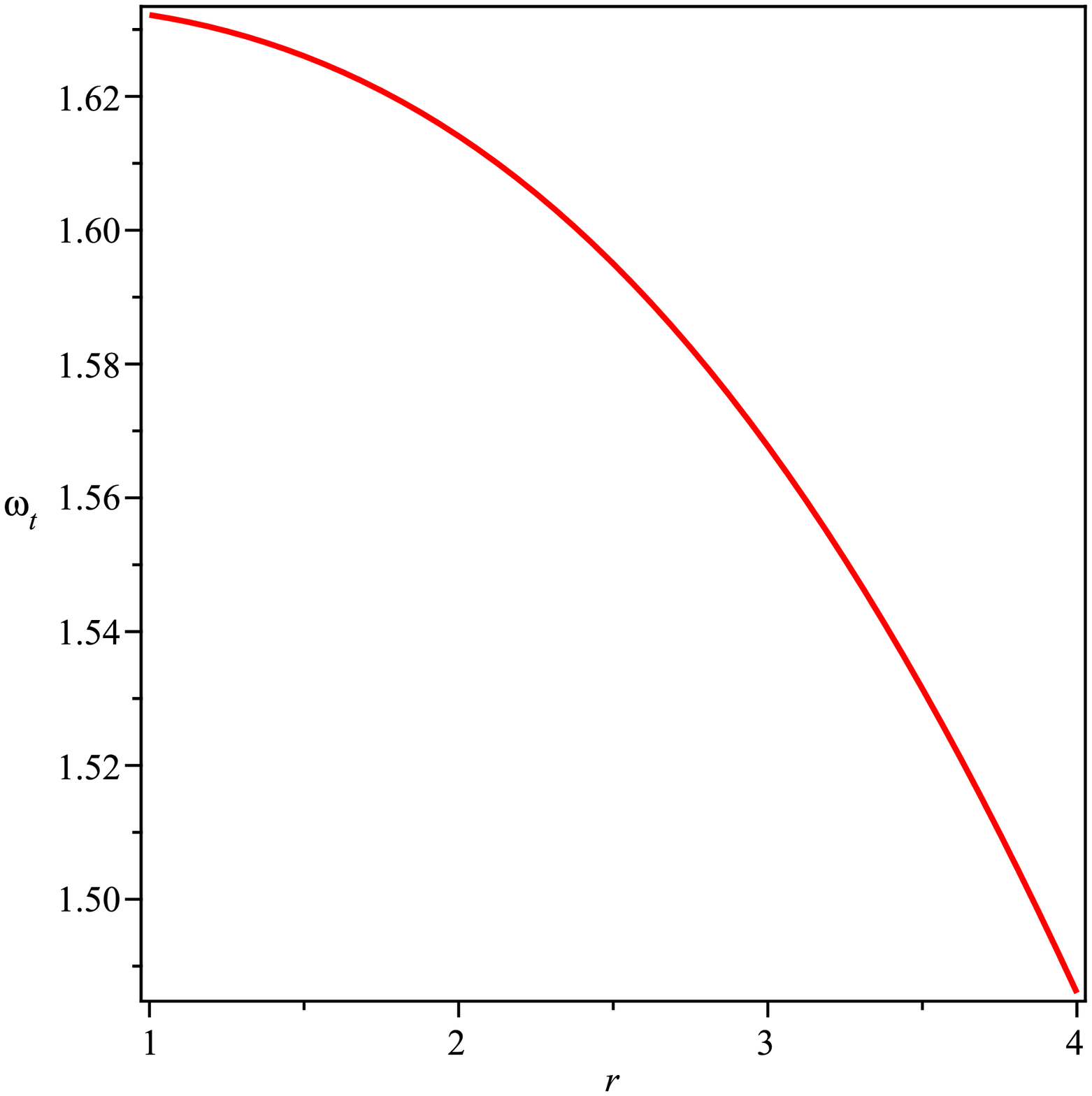}}
    [e]{\includegraphics[width=0.4\textwidth]{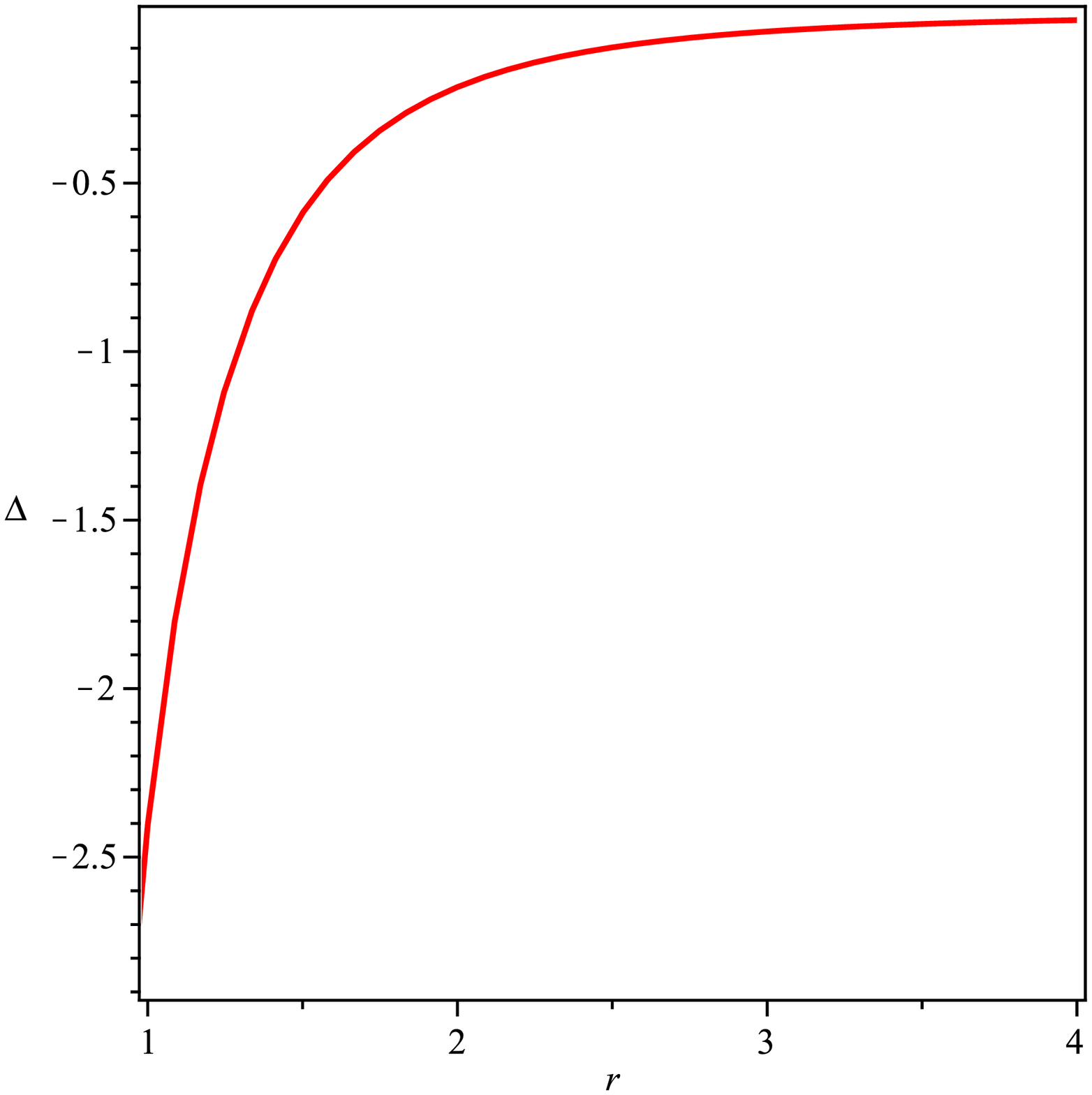}}
    \caption{The plot of [a] NEC, SEC, WEC, [b] DEC, [c] Radial EoS Parameter ($\omega_{r}$), [d]  Tangential EoS Parameter ($\omega_{t}$), 
    [e] Anisotropy Parameter ($\triangle$), vs. Radial Co-ordinate ($r$). Here $\gamma=1$, $\lambda=-13$.}
    %\label{fig:foobar}
    \label{fig:3}
\end{figure}
%%%%%%%%%%%%%%%%%%%%%%%%%%%%%%%%%%%%%%%%%%%%%%%%%%%%%%%%%%%%%%%%%%%%%%%%%%%%%%%%%%%%%%%%%%%%%%%%%%%%%%%%%%%%%%%%%%%%%%%%%%%%%%%%%%%%%%%%%%%%%%%%%
%%%%%%%%%%%%%%%%%%%%%%%%%%%%%%%%%%%%%%%%%%%%%%%%%%%%%%%%%%%%%%%%%%%%%%%%%%%%%%%%%%%%%%%%%%%%%%%%%%%%%%%%%%%%%%%%%%%%%%%%%%%%%%%%%%%%
\begin{figure}[H]
    \centering
    [a]{\includegraphics[width=0.38\textwidth]{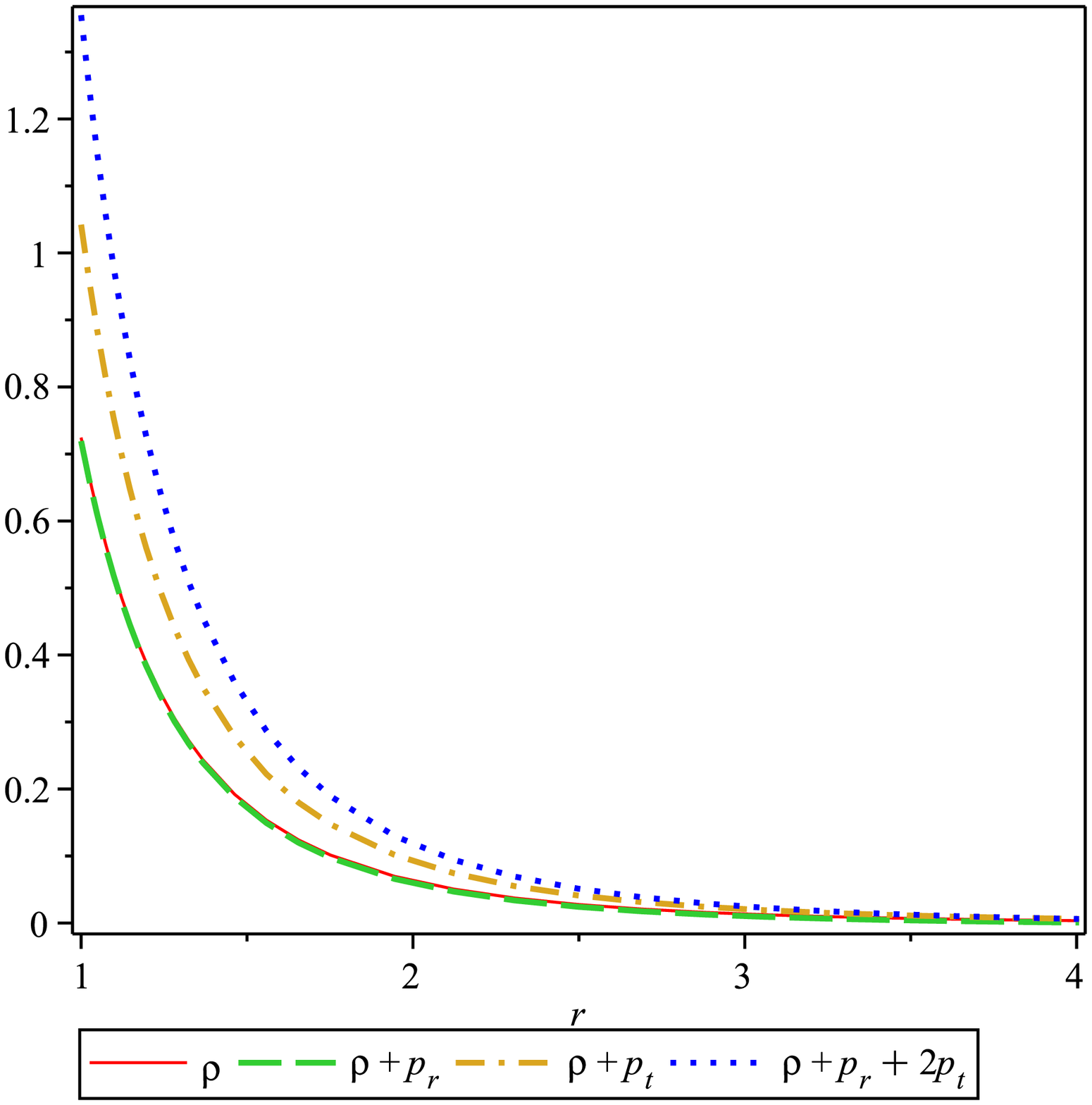}} 
    [b]{\includegraphics[width=0.38\textwidth]{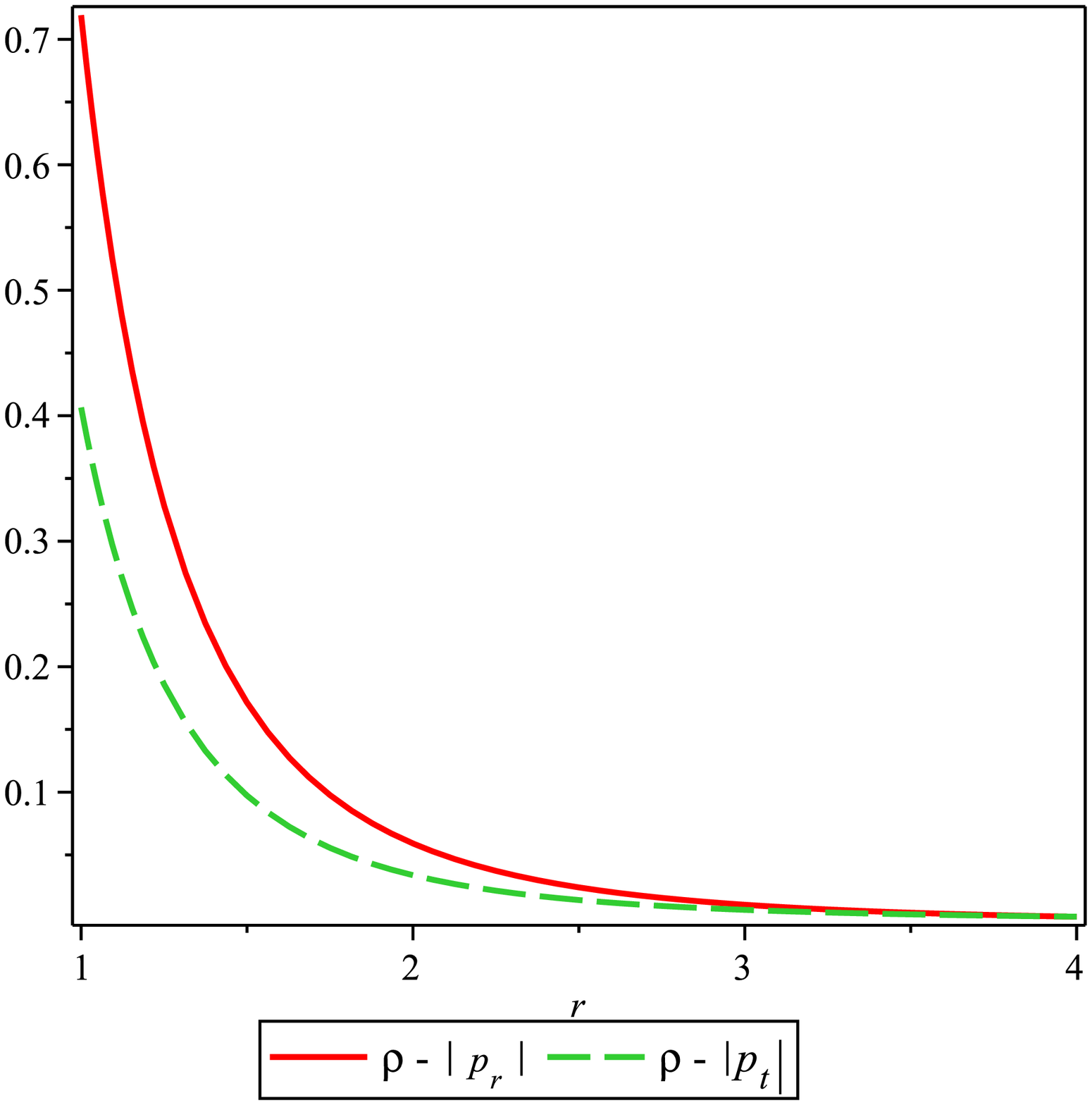}} 
    [c]{\includegraphics[width=0.38\textwidth]{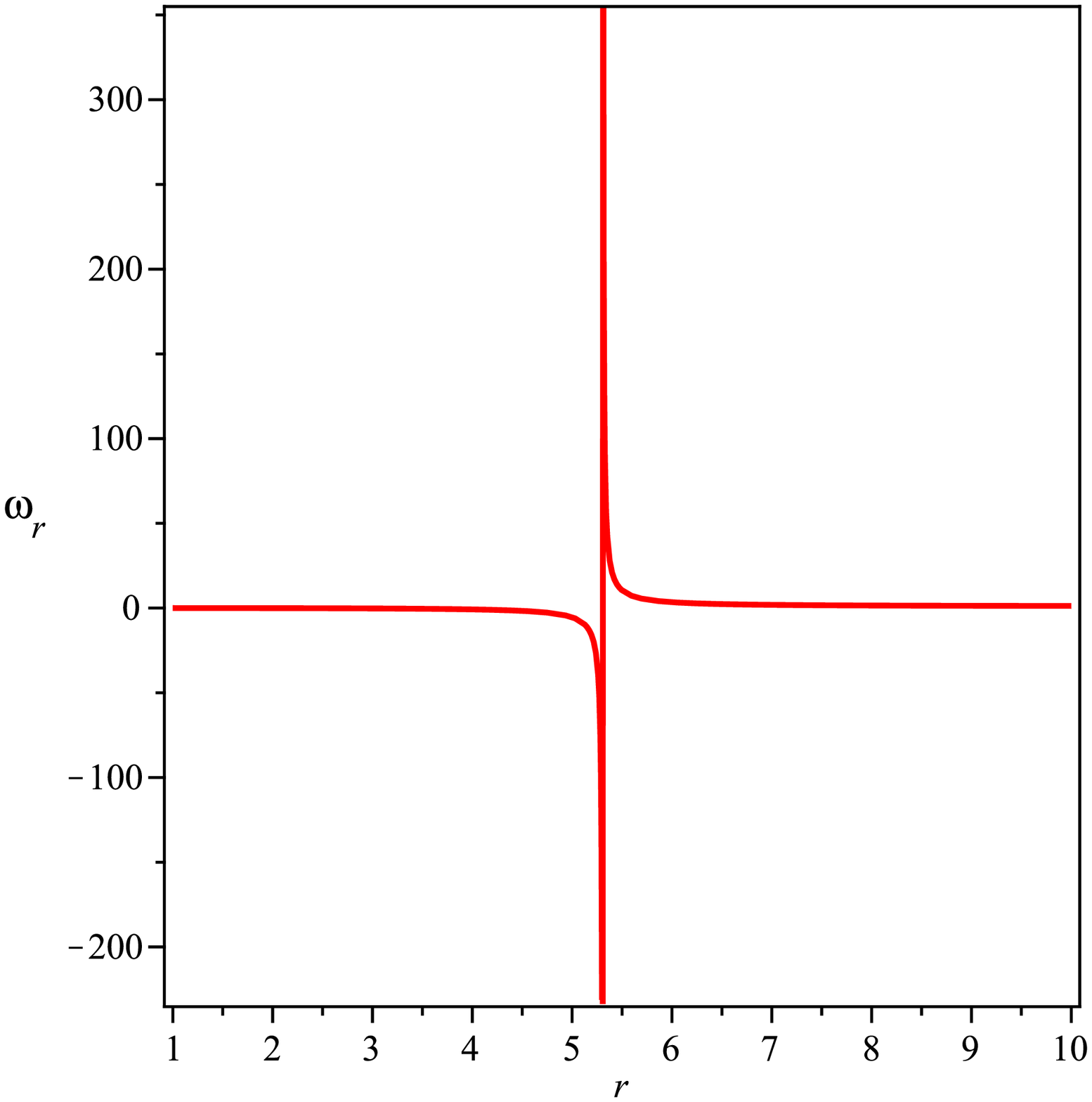}}
    [d]{\includegraphics[width=0.38\textwidth]{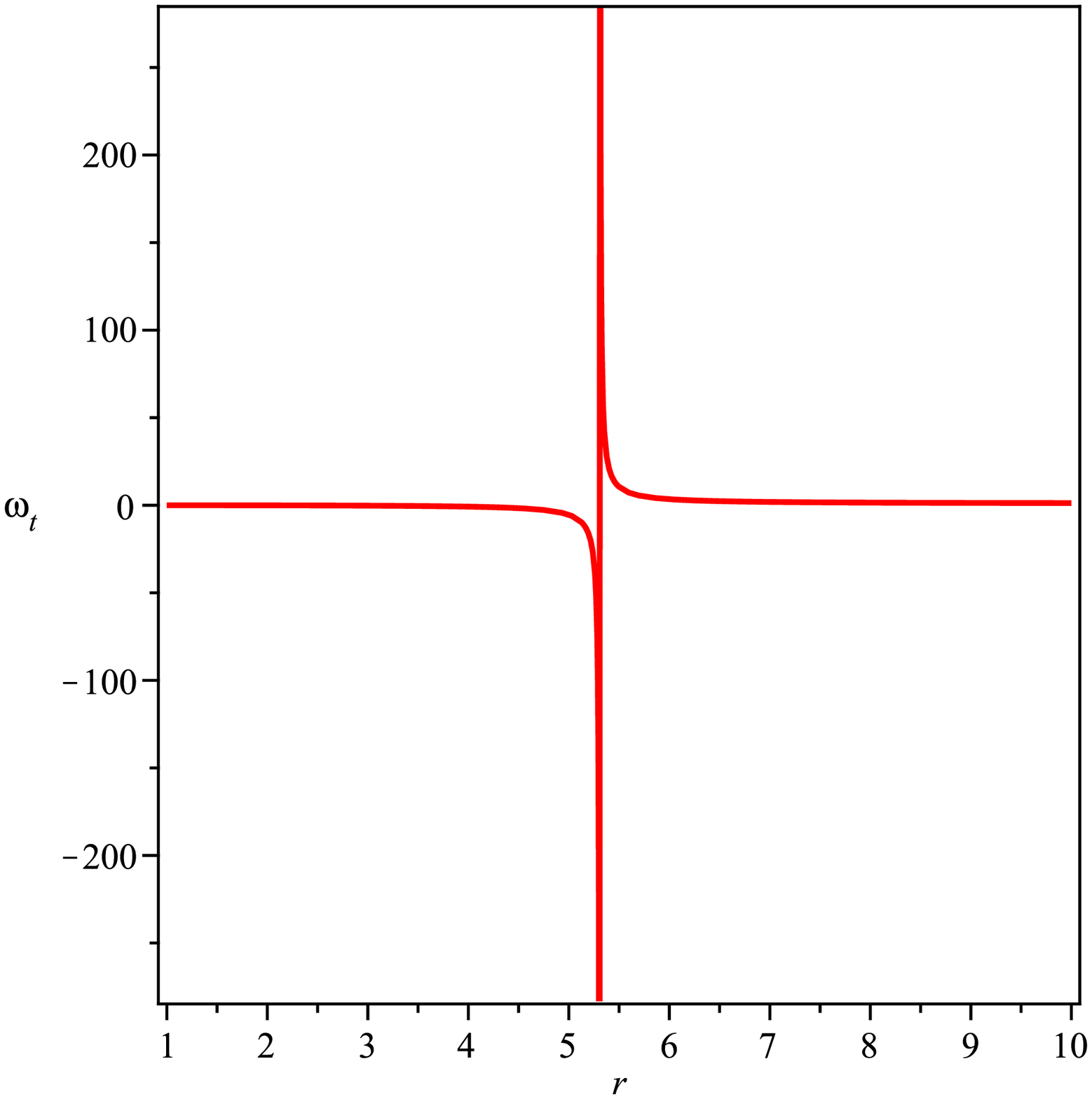}}
    [e]{\includegraphics[width=0.3\textwidth]{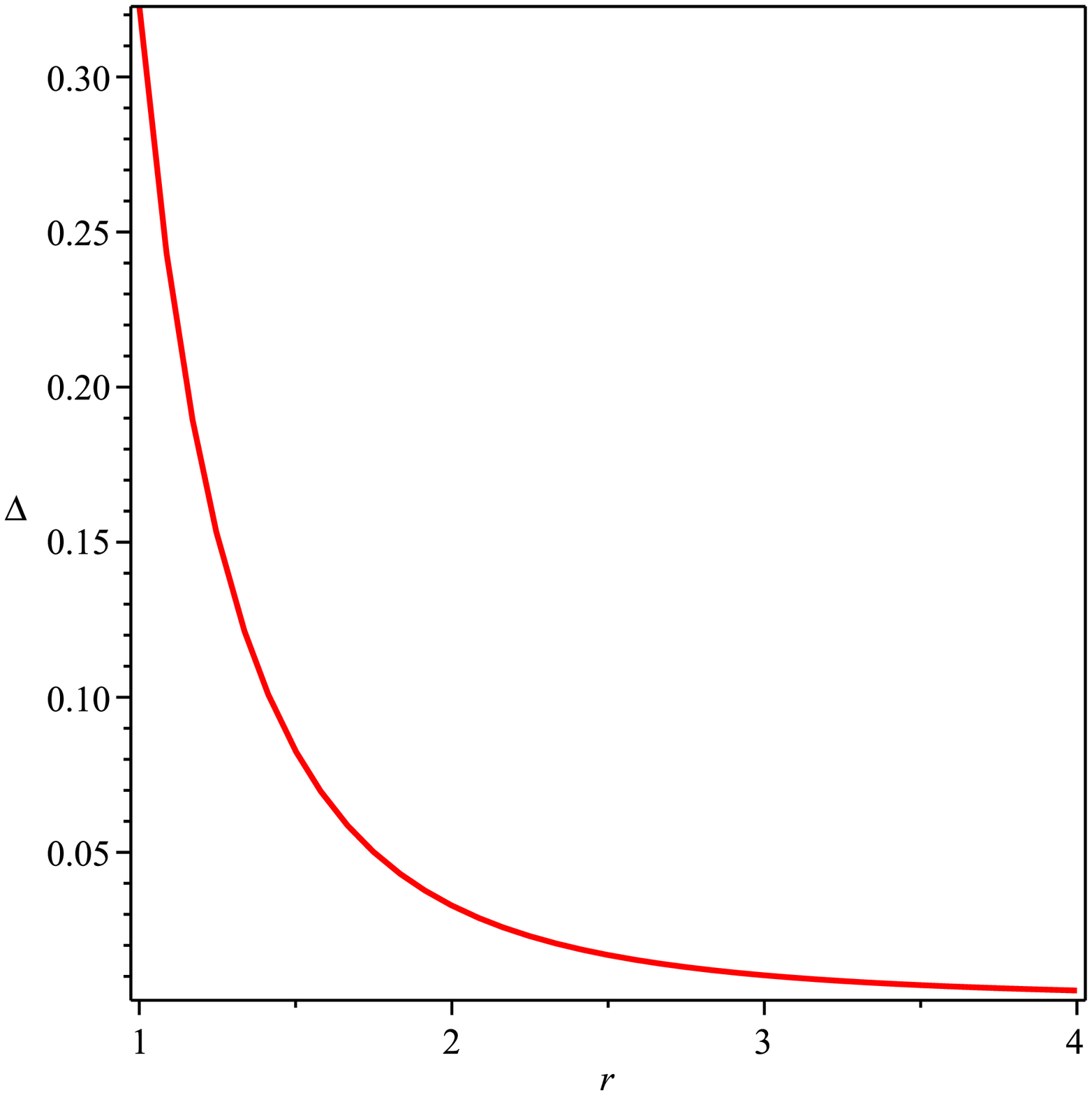}}
    \caption{The plot of [a] NEC, SEC, WEC, [b] DEC, [c] Radial EoS Parameter ($\omega_{r}$), [d] Tangential EoS parameter ($\omega_{t}$), [e] Anisotropy Parameter ($\triangle$), vs. Radial Co-ordinate ($r$). Here $\gamma=1$, $\lambda=-9.5$.}
   % \label{fig:foobar}
   \label{fig:4}
\end{figure}
%%%%%%%%%%%%%%%%%%%%%%%%%%%%%%%%%%%%%%%%%%%%%%%%%%%%%%%%%%%%%%%%%%%%%%%%%%%%%%%%%%%%%%%%%%%%%%%%%%%%%%%%%%%%%%%%%%%%%%%%%%%%%%%%%%%%%%%%%%%%%%%%%%
%%%%%%%%%%%%%%%%%%%%%%%%%%%%%%%%%%%%%%%%%%%%%%%%%%%%%%%%%%%%%%%%%%%%%%%%%%%%%%%%%%%%%%%%%%%%%%%%%%%%%%%%%%%%%%%%%%%%%%%%%%%%%%%%%%%%%%%%%%%
\begin{figure}[H]
    \centering
   [a]{\includegraphics[width=0.4\textwidth]{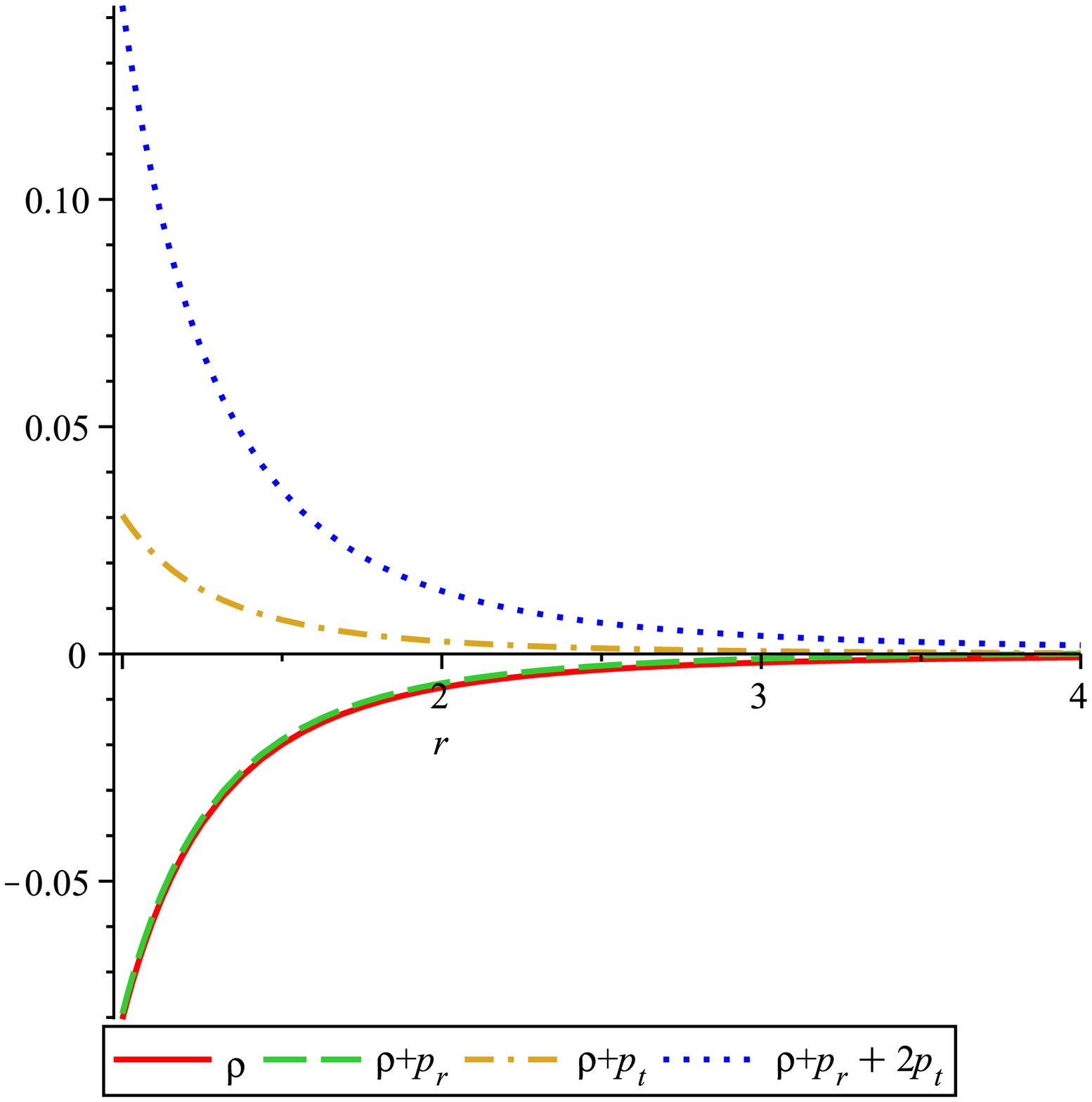}} 
   [b]{\includegraphics[width=0.4\textwidth]{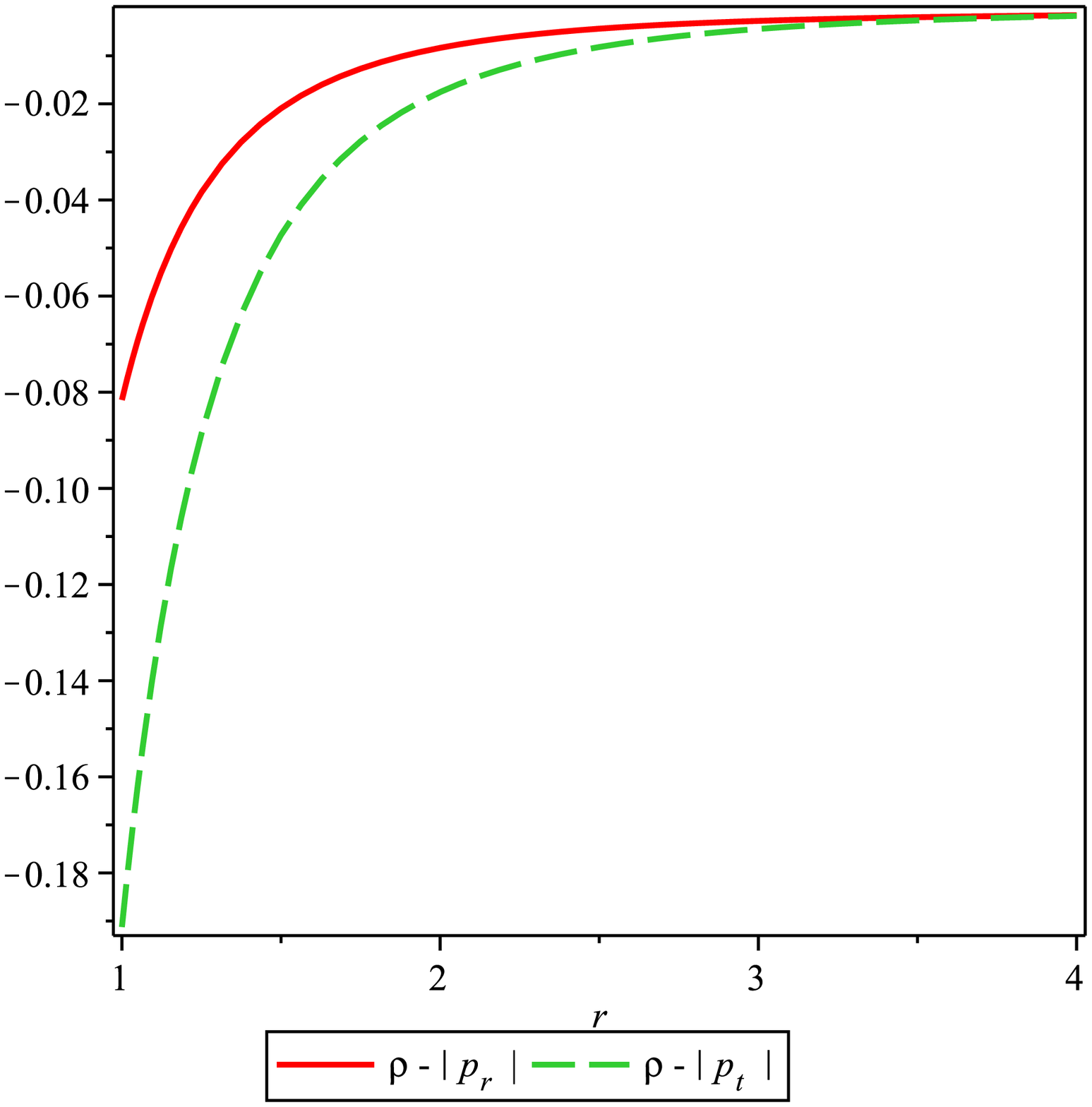}} 
    [c]{\includegraphics[width=0.4\textwidth]{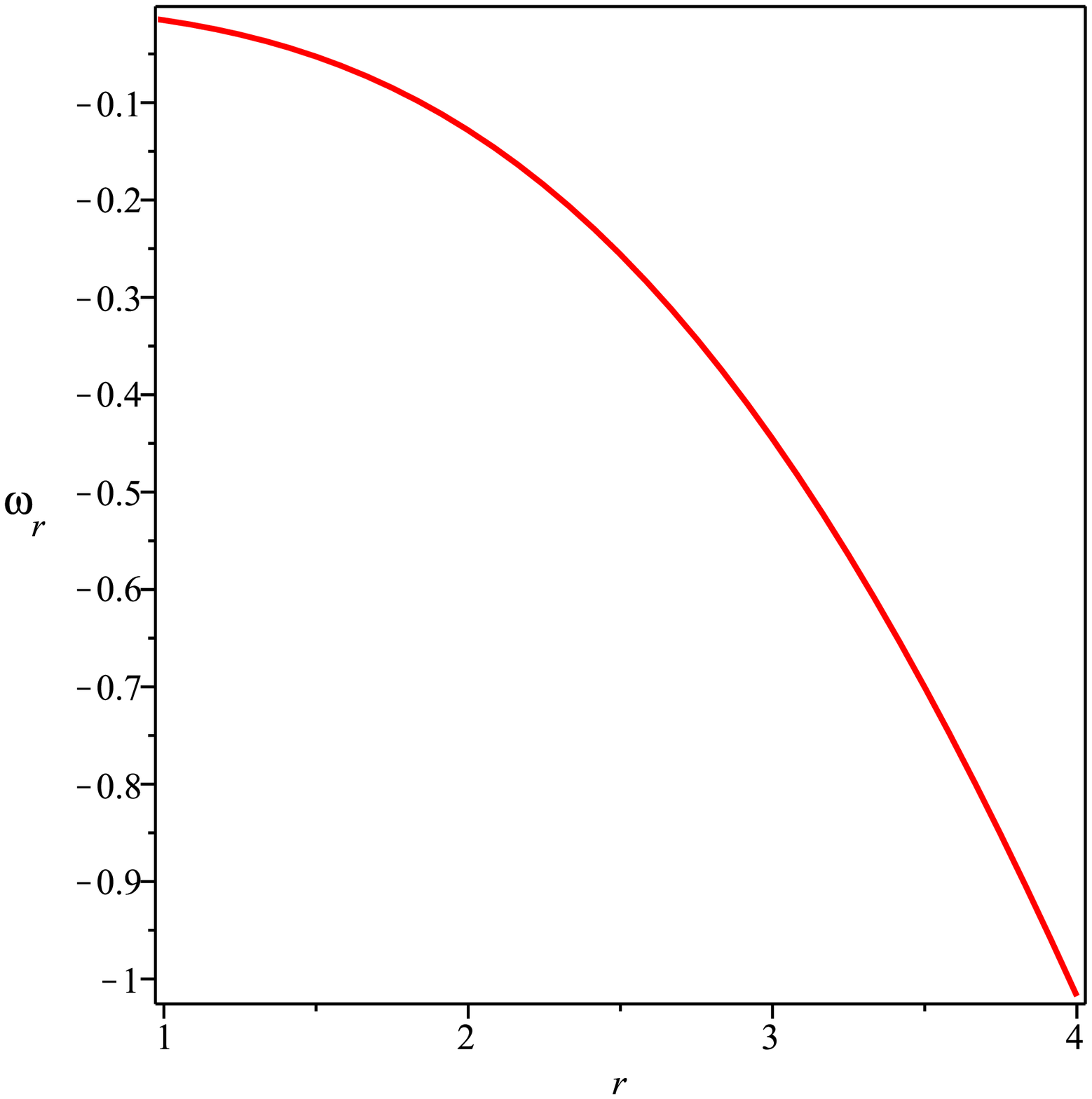}}
    [d]{\includegraphics[width=0.4\textwidth]{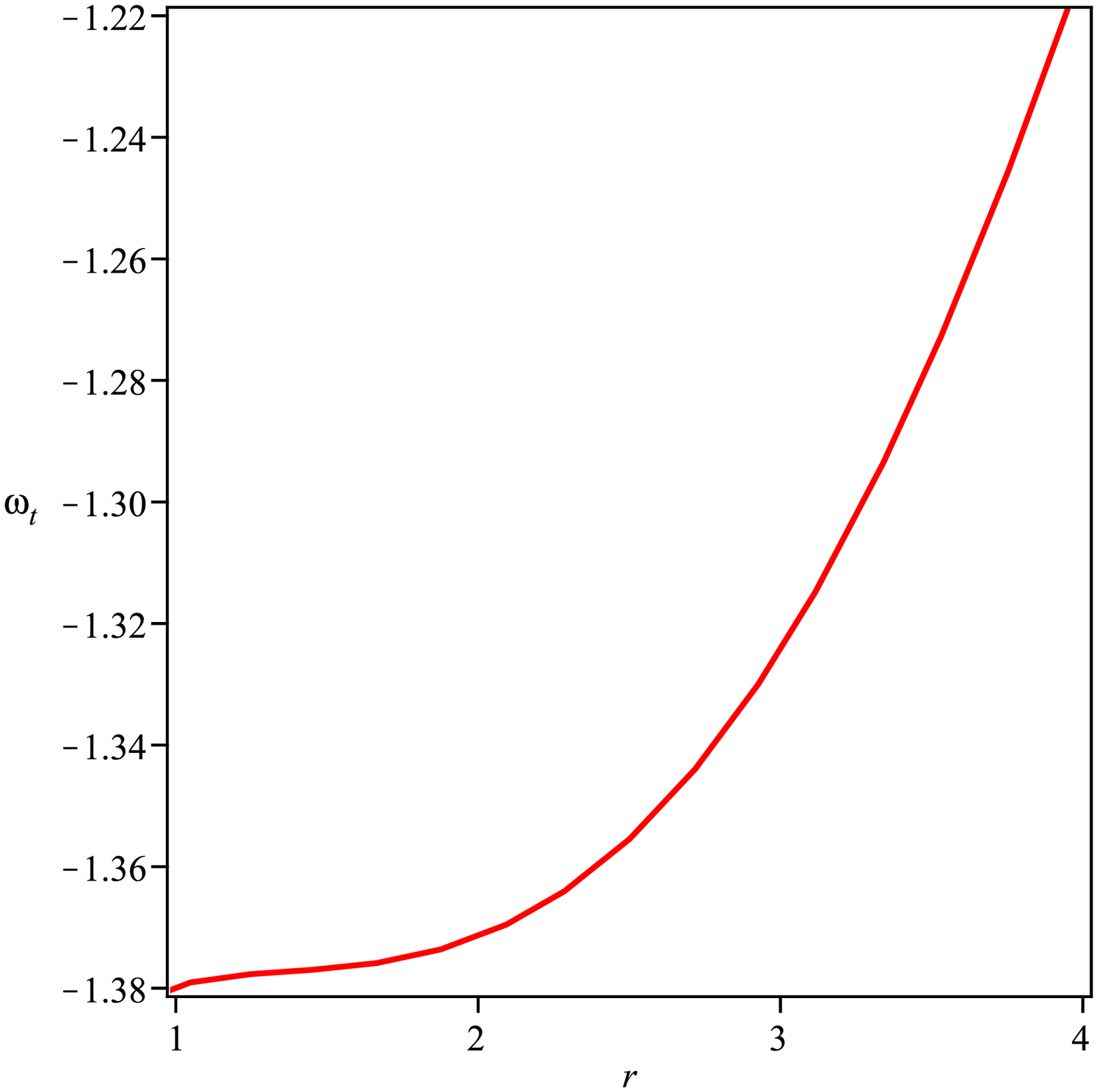}}
   [e]{\includegraphics[width=0.4\textwidth]{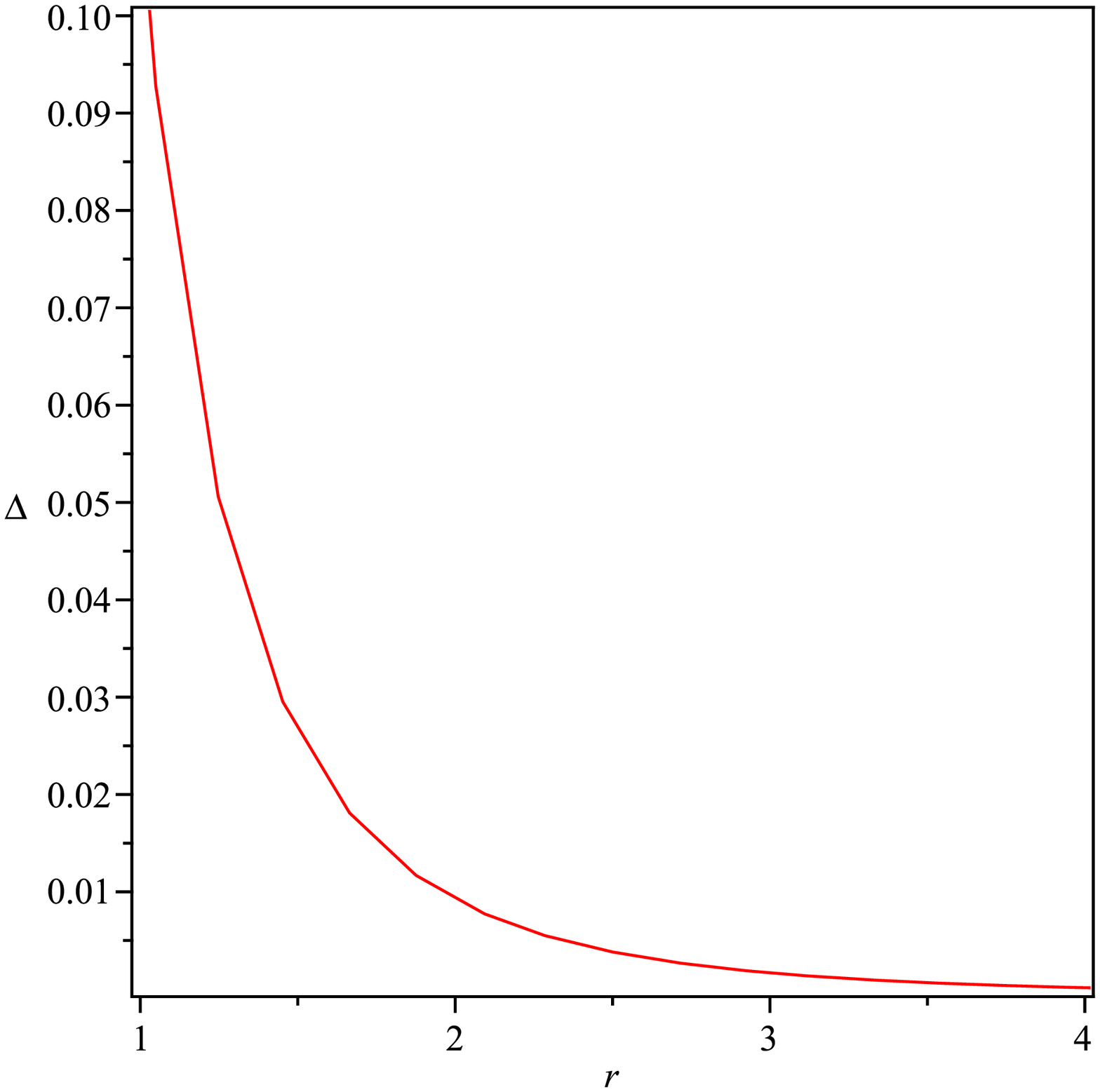}}
   \caption{The plot of [a] NEC, SEC, WEC, [b] DEC, [c] Radial EoS Parameter ($\omega_{r}$), [d] Tangential EoS parameter ($\omega_{t}$), 
    [e] Anisotropy Parameter ($\triangle$), vs. Radial Co-ordinate ($r$). Here $\gamma=1$, $\lambda=-5$.}
   % \label{fig:foobar}
   \label{fig:5}
\end{figure}
%%%%%%%%%%%%%%%%%%%%%%%%%%%%%%%%%%%%%%%%%%%%%%%%%%%%%%%%%%%%%%%%%%%%%%%%%%%%%%%%%%%%%%%%%%%%%%%%%%%%%%%%%%%%%%%%%%%%%%%%%%%%%%%%%%%%%%%%%%%%%
%%%%%%%%%%%%%%%%%%%%%%%%%%%%%%%%%%%%%%%%%%%%%%%%%%%%%%%%%%%%%%%%%%%%%%%%%%%%%%%%%%%%%%%%%%%%%%%%%%%%%%%%%%%%%%%%%%%%%%%%%%%%%%%%%%
\begin{figure}[H]
    \centering
    [a]{\includegraphics[width=0.4\textwidth]{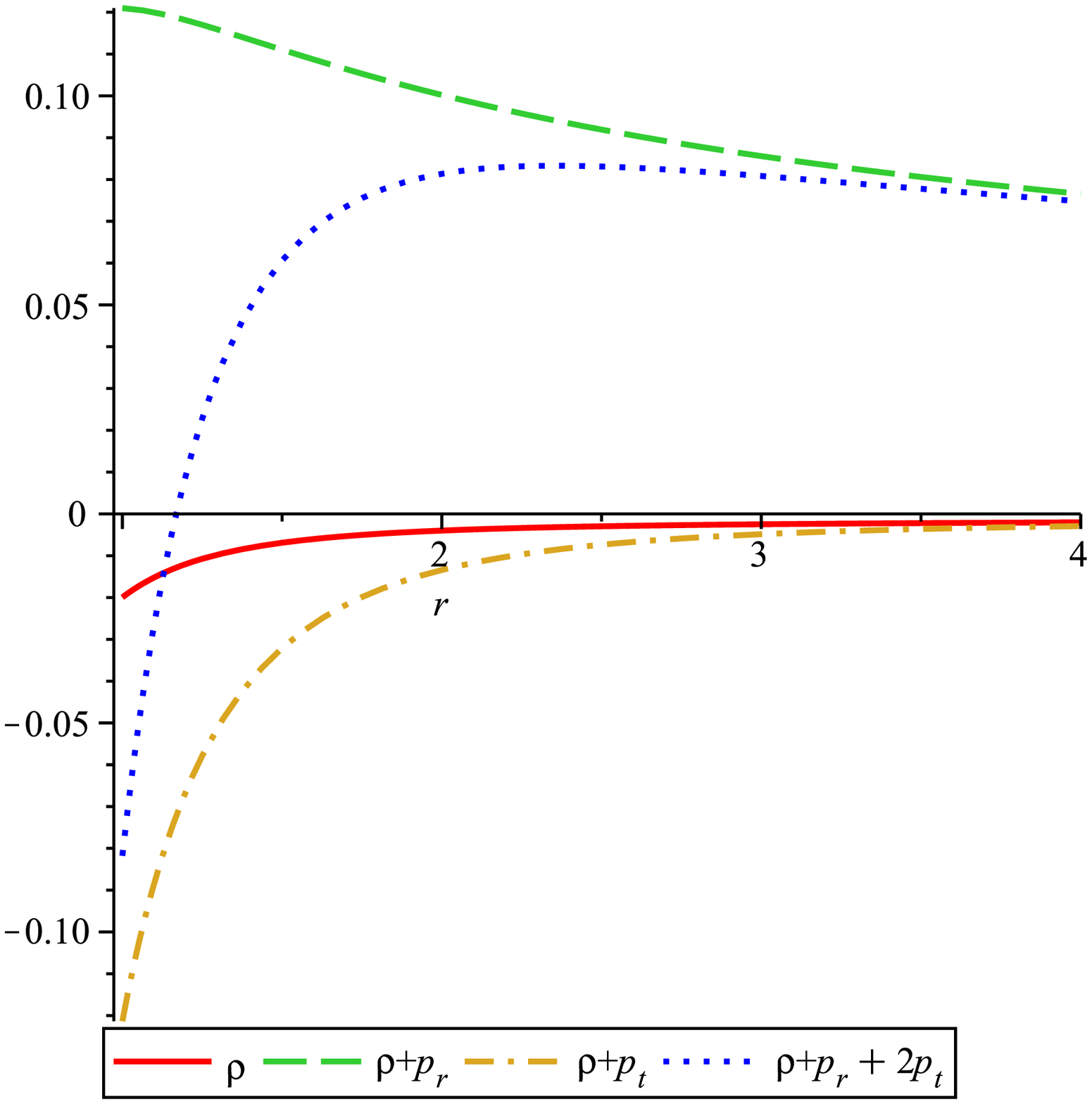}} 
    [b]{\includegraphics[width=0.4\textwidth]{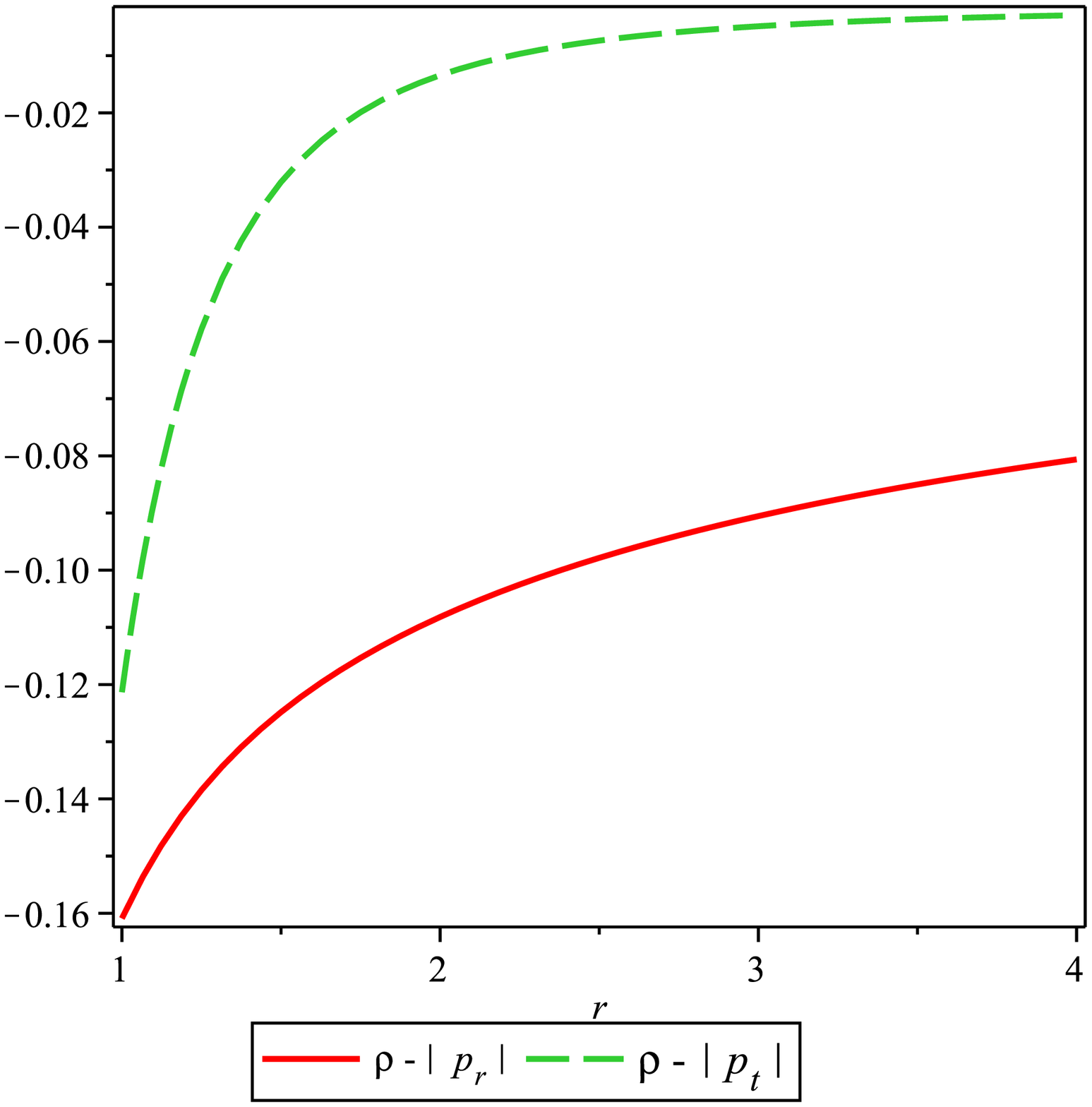}} 
    [c]{\includegraphics[width=0.4\textwidth]{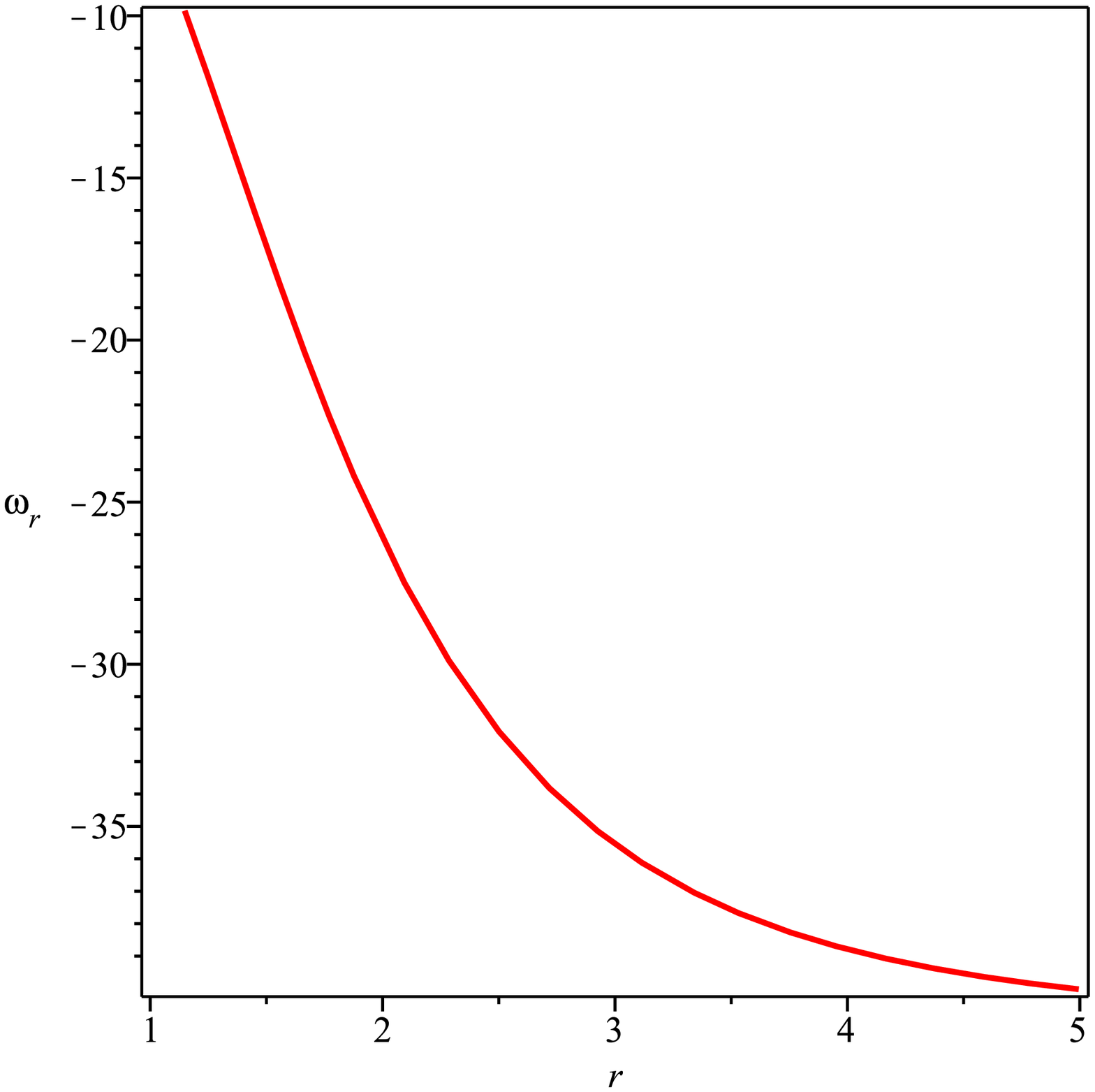}}
    [d]{\includegraphics[width=0.4\textwidth]{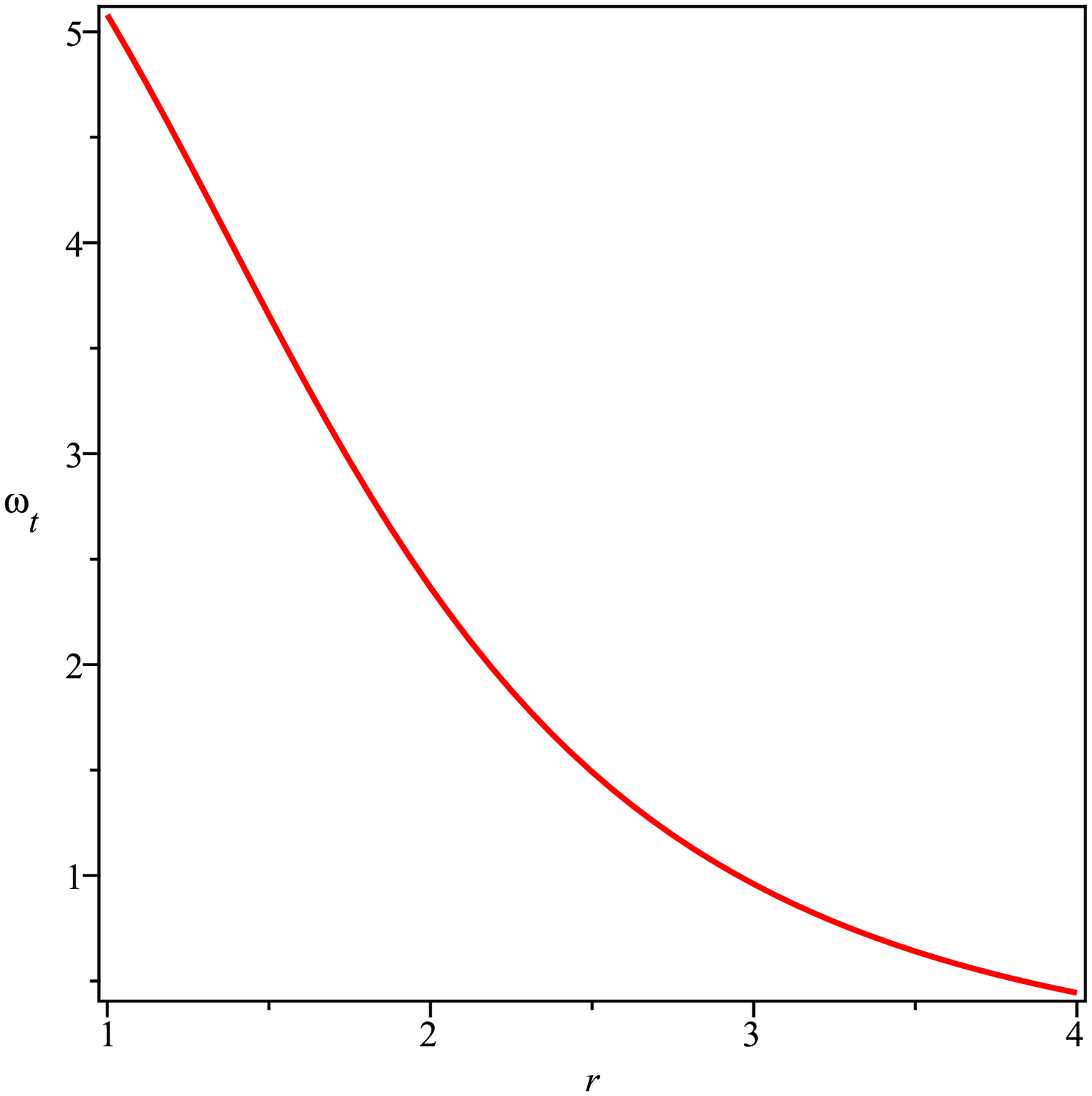}}
    [e]{\includegraphics[width=0.4\textwidth]{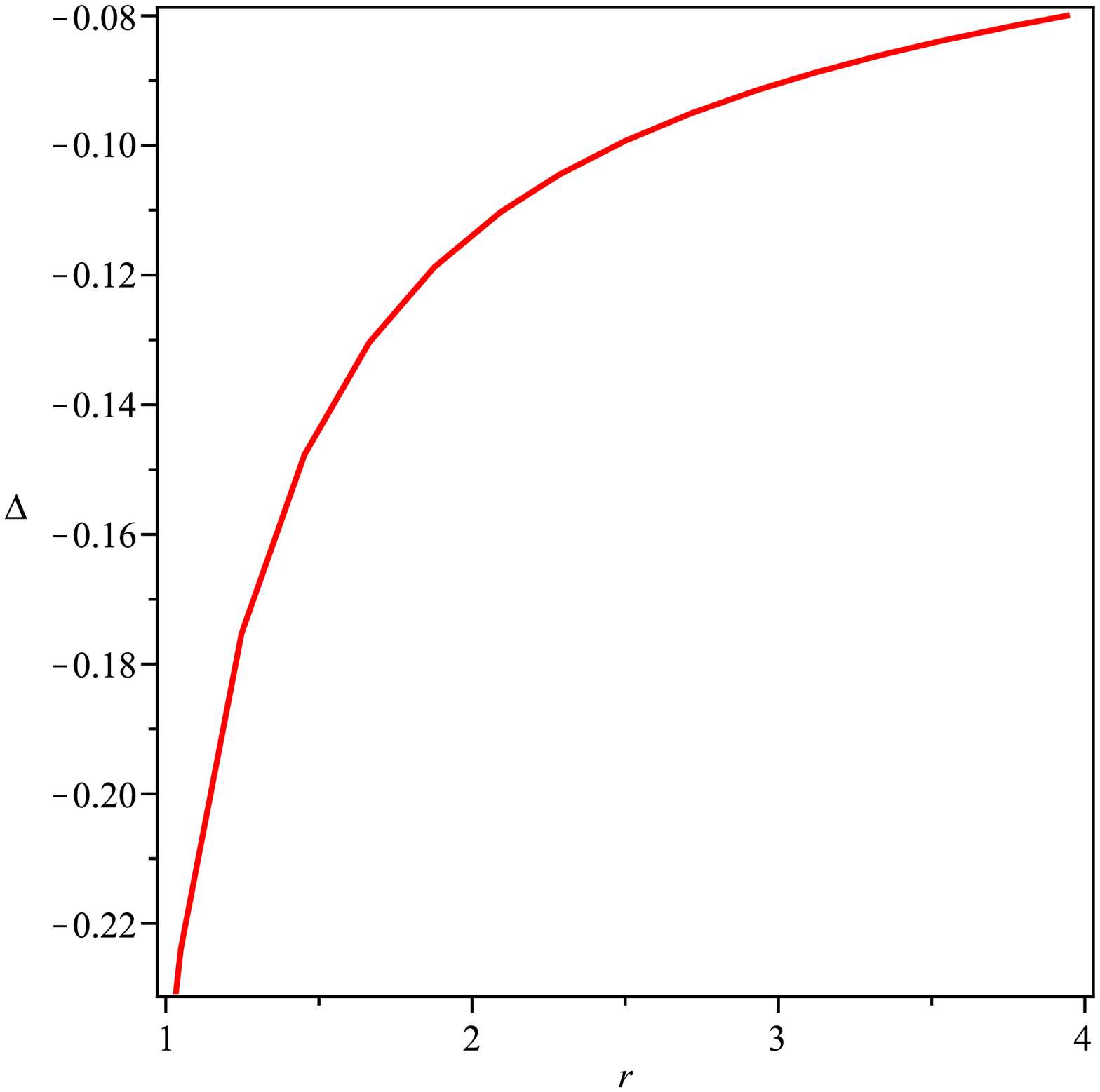}}
    \caption{The plot of [a] NEC, SEC, WEC, [b] DEC, [c] Radial EoS Parameter ($\omega_{r}$), [d] Tangential EoS parameter ($\omega_{t}$), 
    [e] Anisotropy Parameter ($\triangle$), vs. Radial Co-ordinate ($r$). Here $\gamma =1$, $\lambda=-1.5$.}
   % \label{fig:foobar}
   \label{fig:6}
\end{figure}
%%%%%%%%%%%%%%%%%%%%%%%%%%%%%%%%%%%%%%%%%%%%%%%%%%%%%%%%%%%%%%%%%%%%%%%%%%%%%%%%%%%%%%%%%%%%%%%%%%%%%%%%%%%%%%%%%%%%%%%%%%%%%%%%%%%%%%%%%%%%
%%%%%%%%%%%%%%%%%%%%%%%%%%%%%%%%%%%%%%%%%%%%%%%%%%%%%%%%%%%%%%%%%%%%%%%%%%%%%%%%%%%%%%%%%%%%%%%%%%%%%%%%%%%%%%%%%%%%
\begin{figure}[H]
    \centering
    	(a)\includegraphics[width=7cm,height=7cm,angle=0]{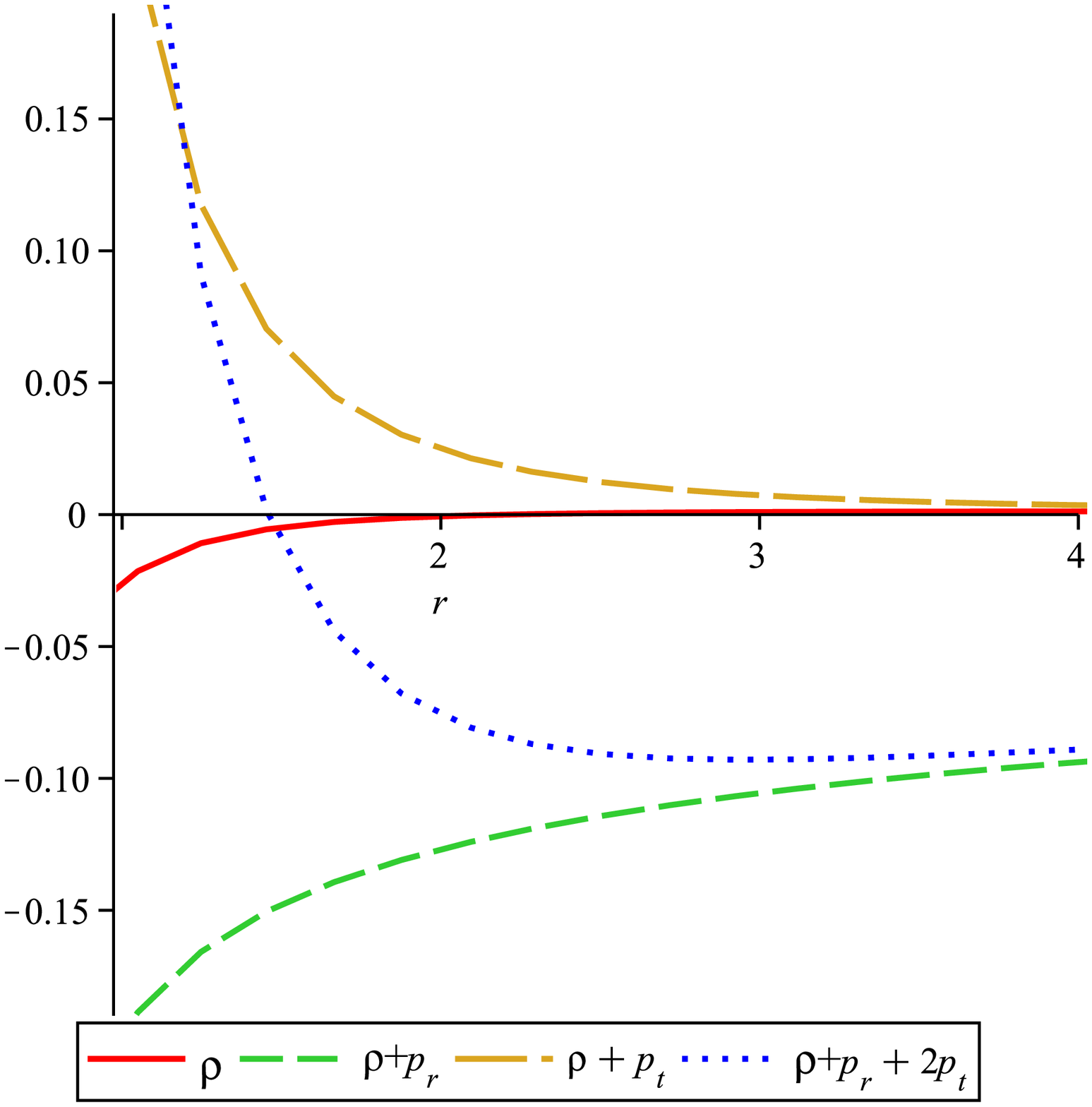}
    [b]{\includegraphics[width=0.4\textwidth]{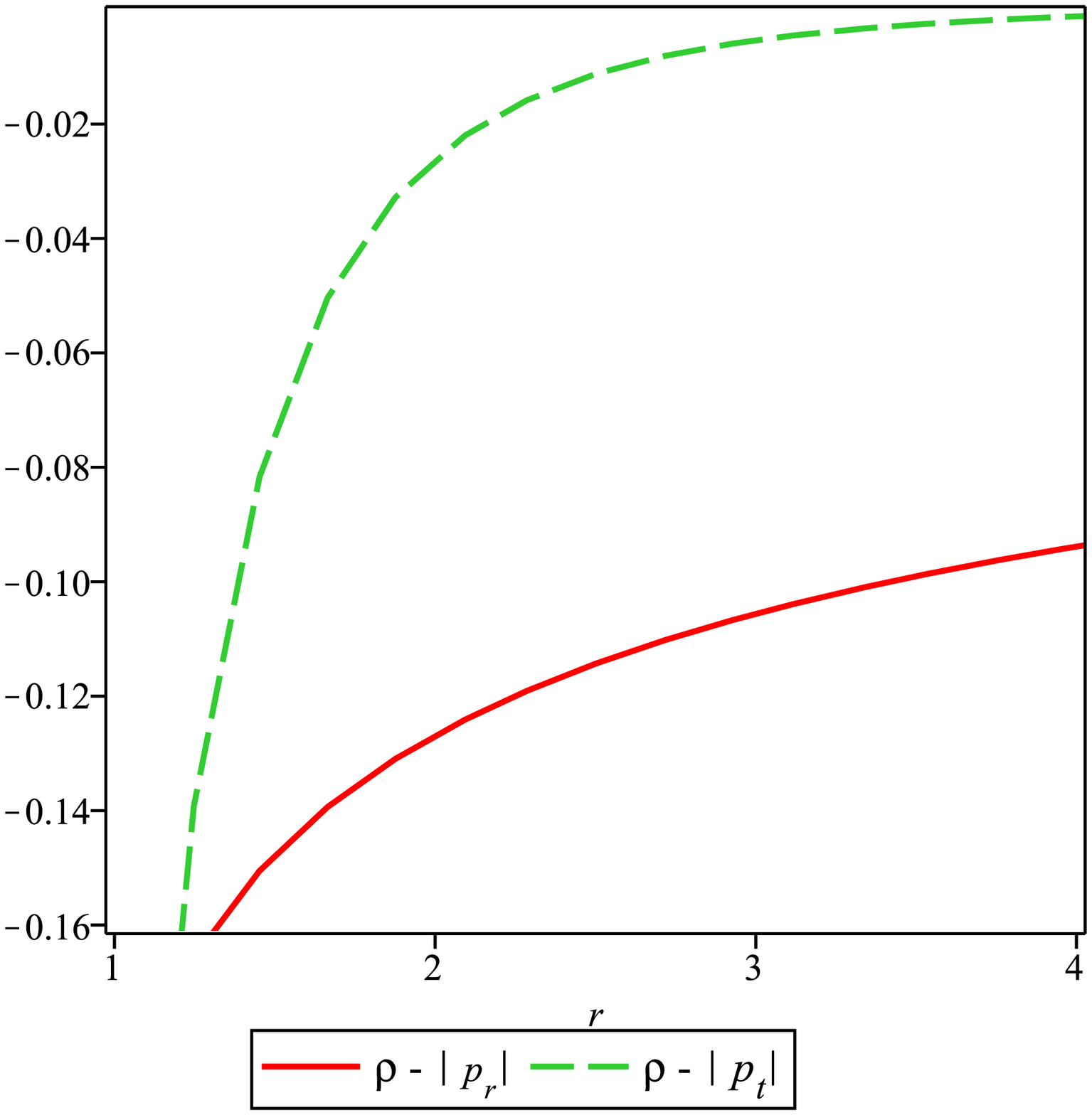}} 
    [c]{\includegraphics[width=0.4\textwidth]{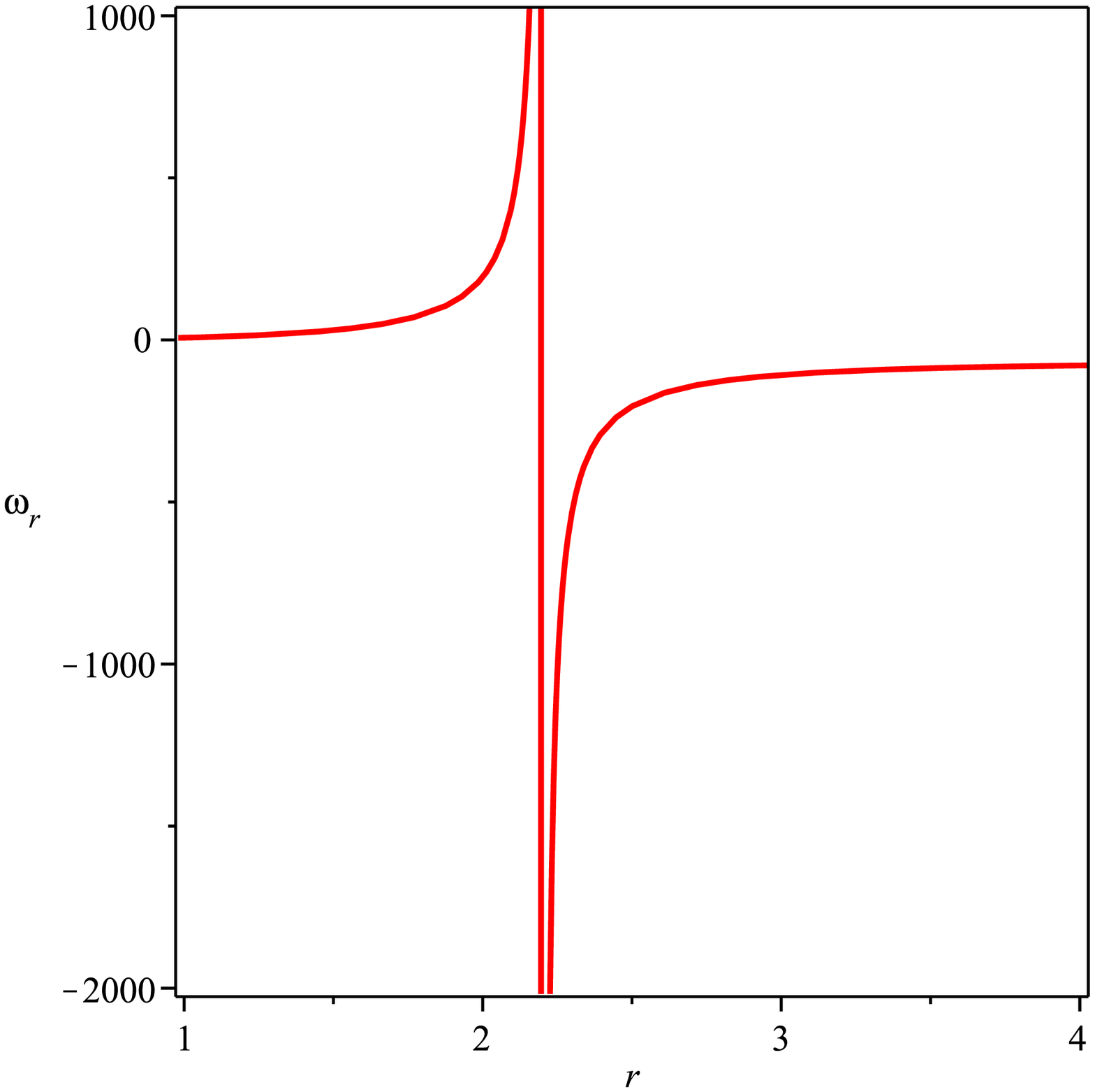}}
    [d]{\includegraphics[width=0.4\textwidth]{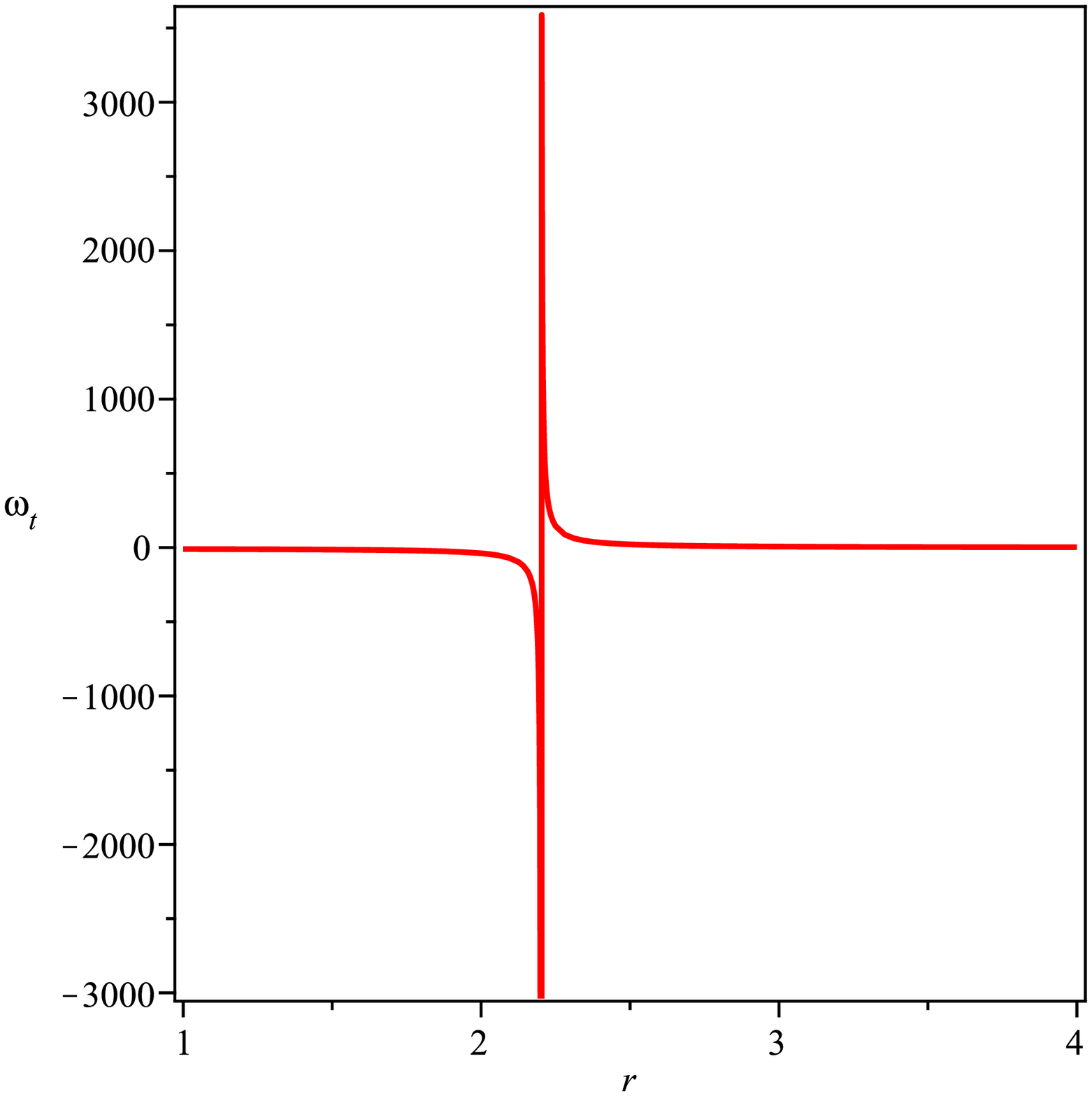}}
    [e]{\includegraphics[width=0.4\textwidth]{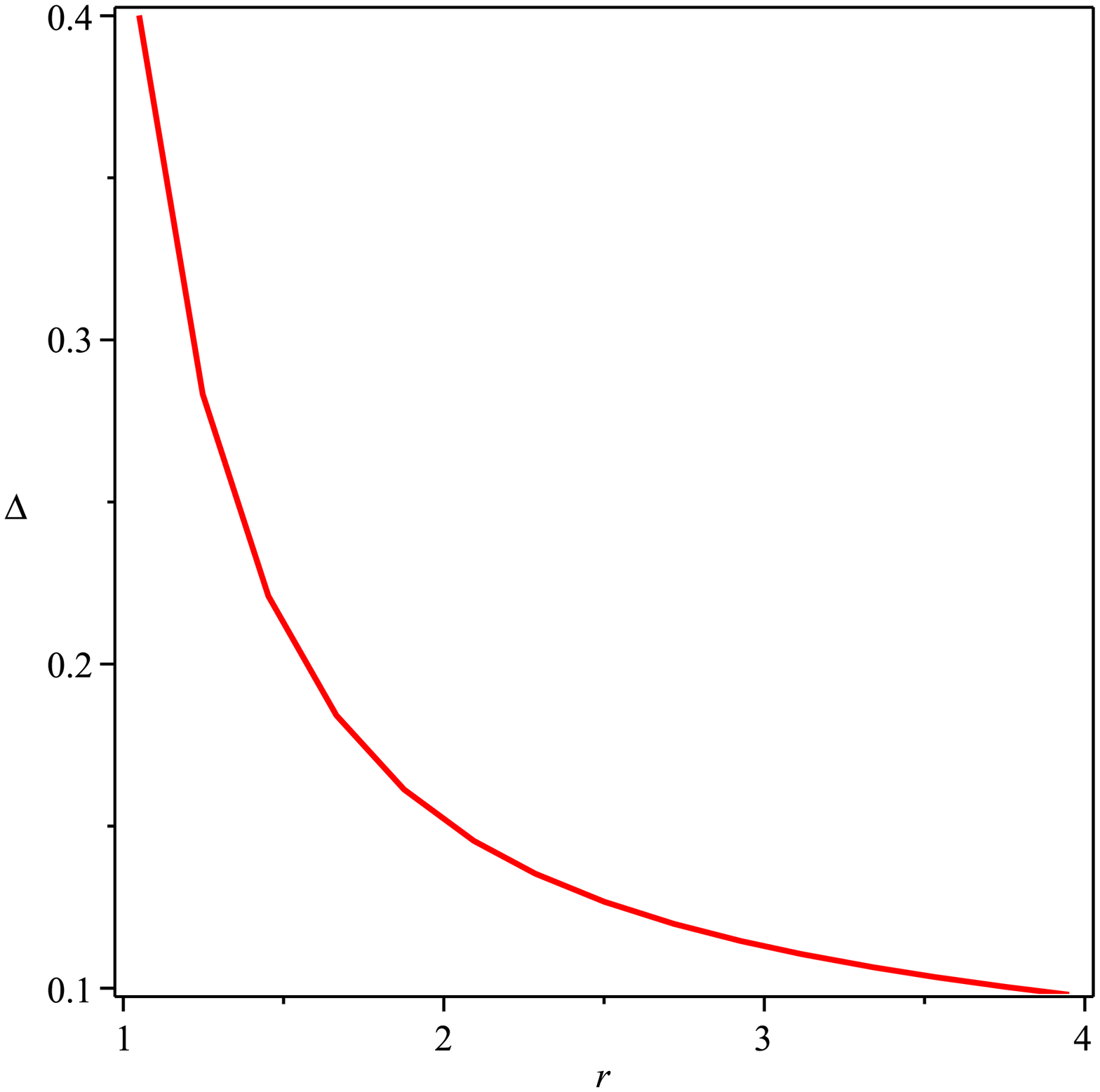}}
    
    \caption{The plot of [a] NEC, SEC, WEC, [b] DEC, [c] Radial EoS Parameter ($\omega_{r}$), [d] Tangential EoS parameter ($\omega_{t}$), 
    [e] Anisotropy Parameter ($\triangle$), vs. Radial Co-ordinate ($r$). Here $\gamma=1$, $\lambda=-1$.}
    %\label{fig:foobar}
    \label{fig:7}
\end{figure}
%%%%%%%%%%%%%%%%%%%%%%%%%%%%%%%%%%%%%%%%%%%%%%%%%%%%%%%%%%%%%%%%%%%%%%%%%%%%%%%%%%%%%%%%%%%%%%%%%%%%%%%%%%%%%%%%%%%%%%%%%%%%%%%%%%%%%%%%%
\begin{figure}[H]
    \centering
   [a]{\includegraphics[width=0.4\textwidth]{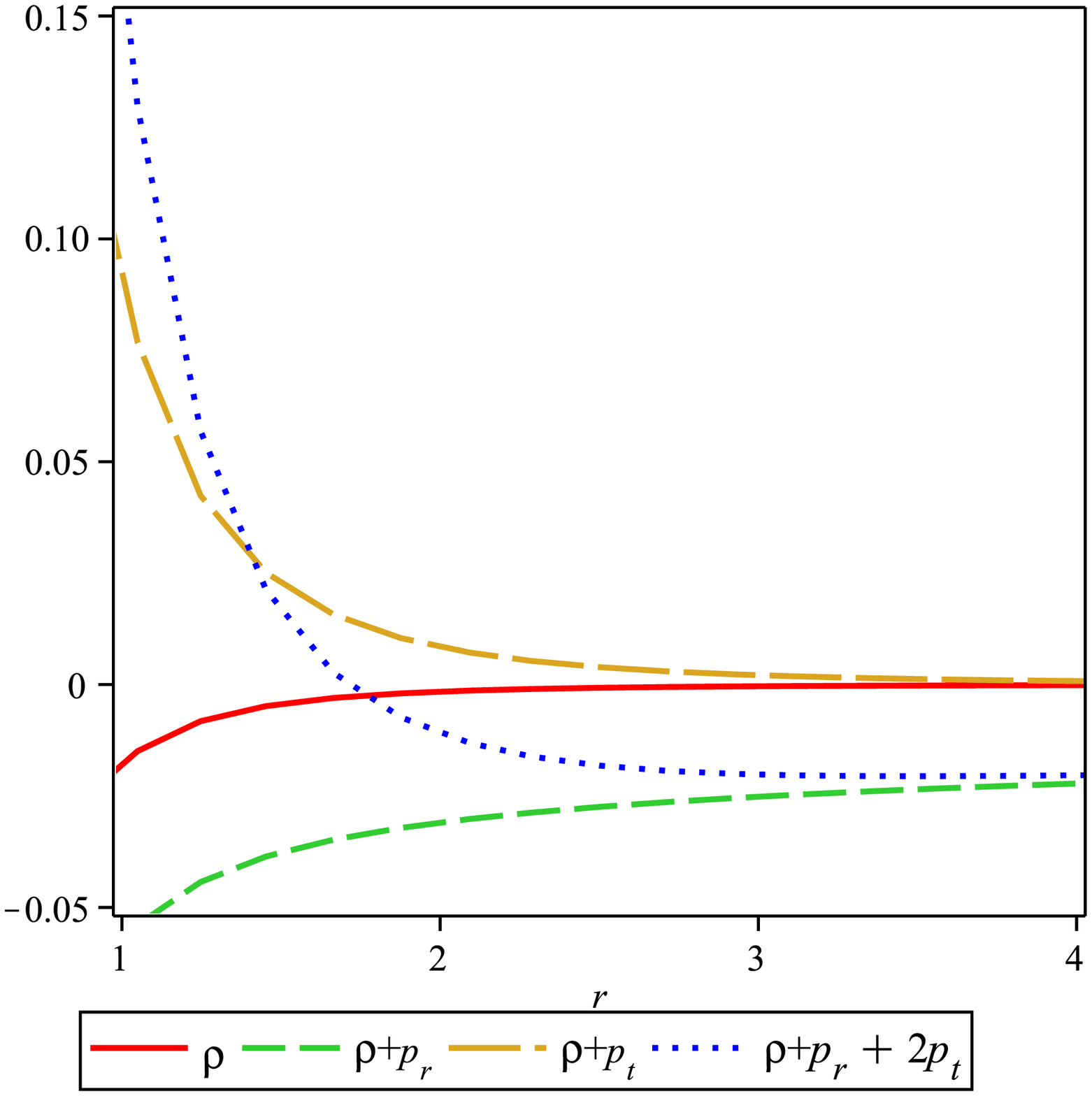}} 
   [b]{\includegraphics[width=0.4\textwidth]{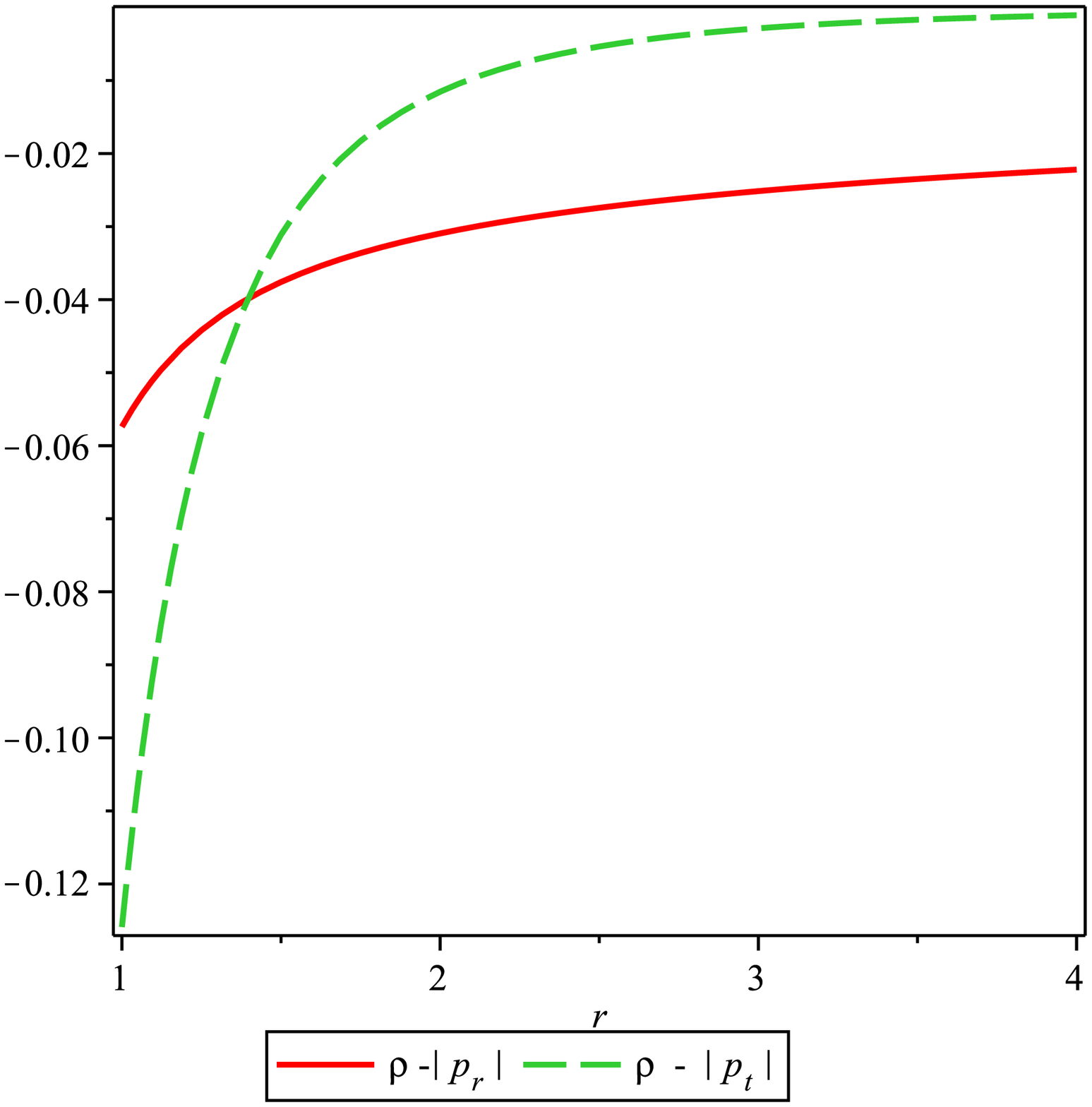}} 
   [c]{\includegraphics[width=0.4\textwidth]{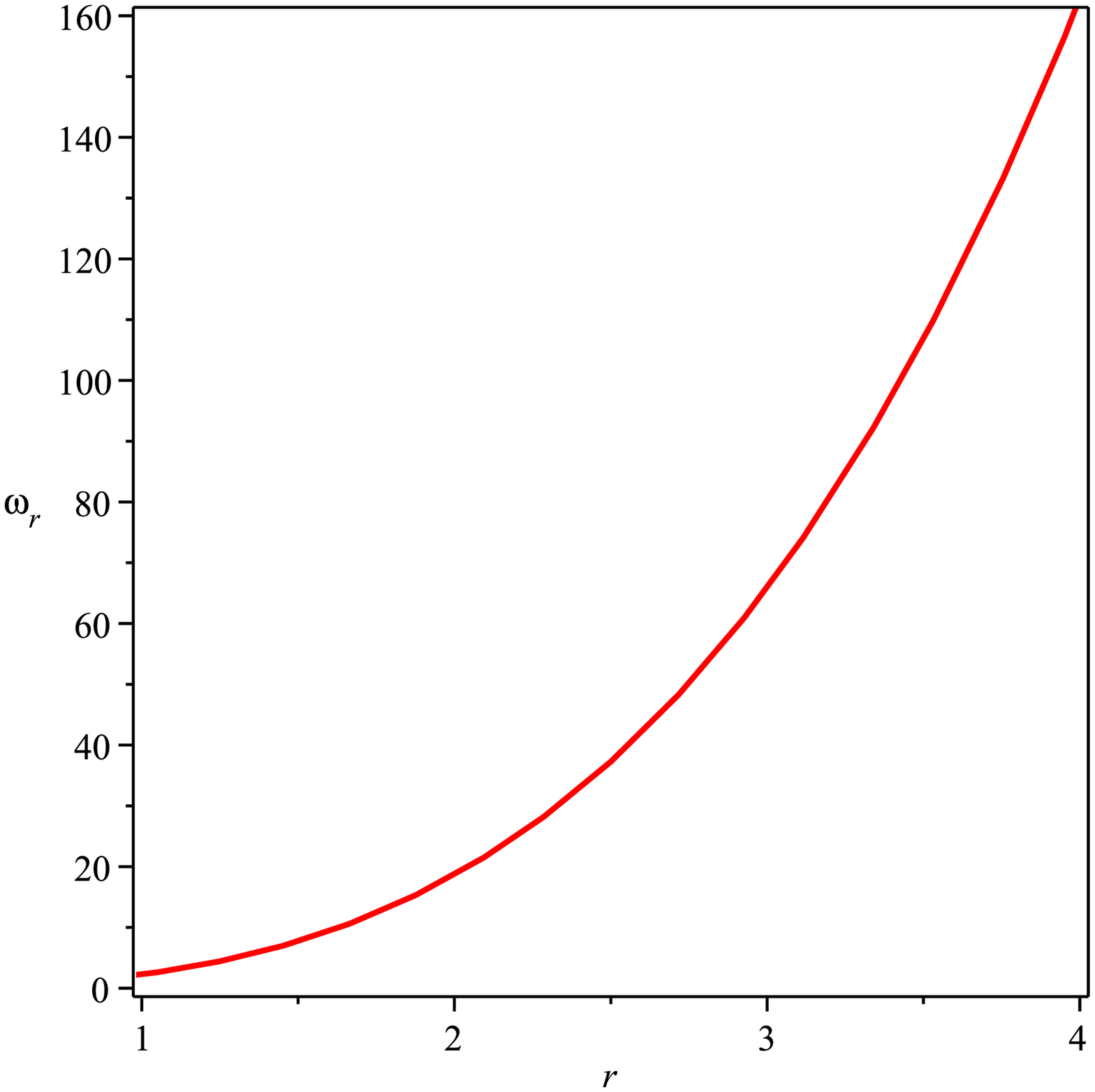}}
   [d]{\includegraphics[width=0.4\textwidth]{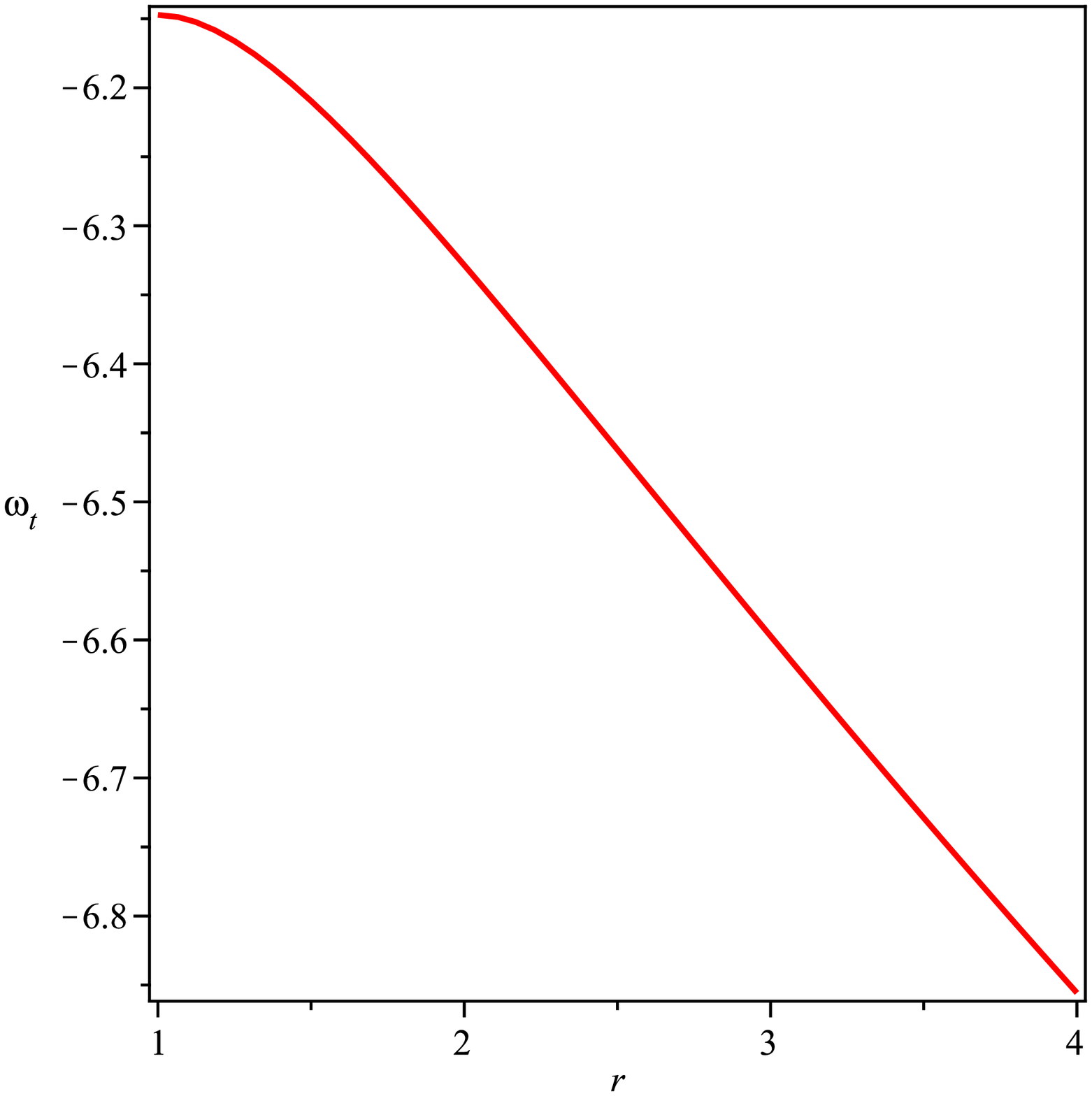}}
   [e]{\includegraphics[width=0.4\textwidth]{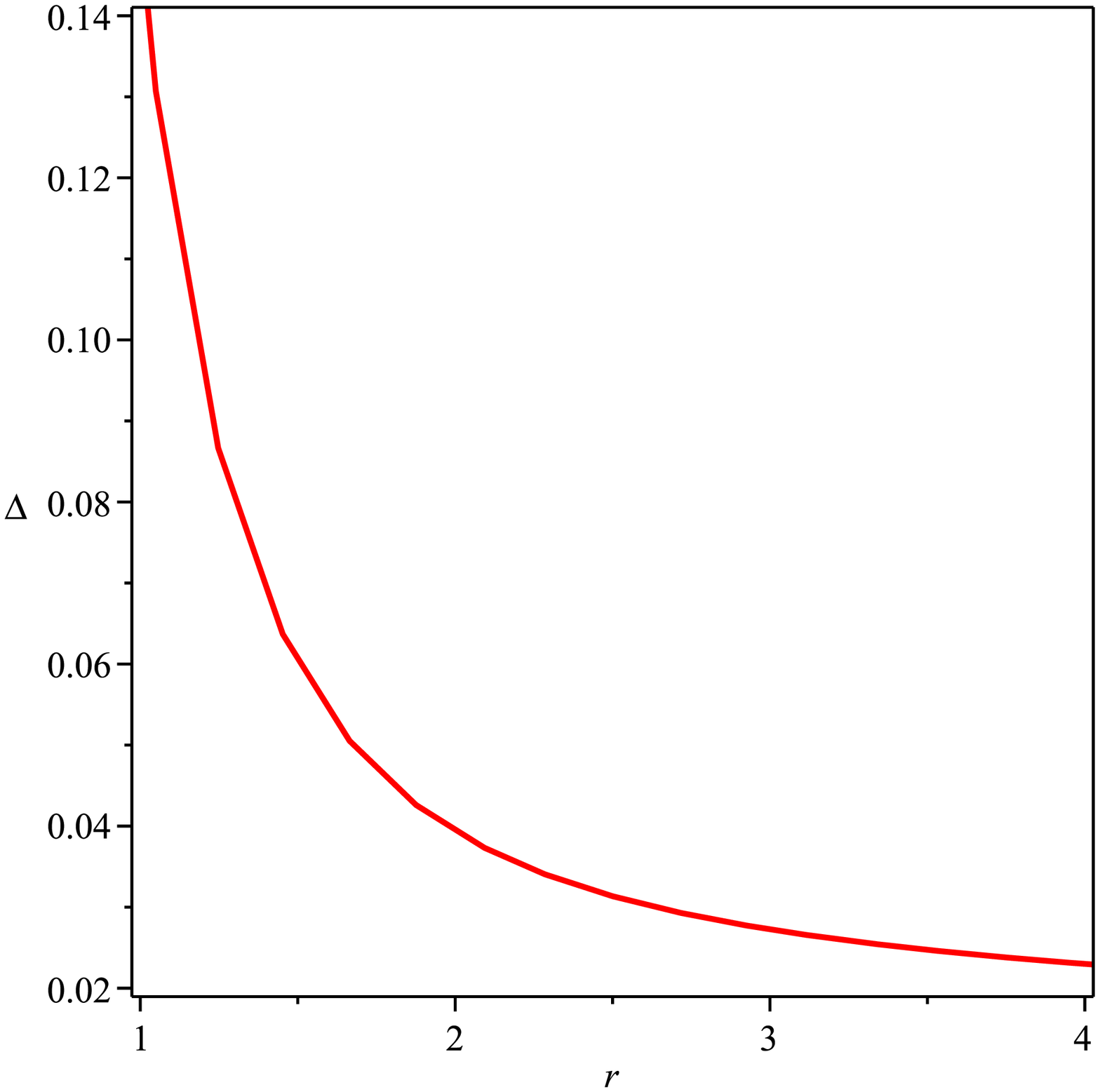}}
    \caption{The plot of [a] NEC, SEC, WEC, [b] DEC, [c] Radial EoS Parameter ($\omega_{r}$), [d] Tangential EoS parameter ($\omega_{t}$), 
    [e] Anisotropy Parameter ($\triangle$), vs. Radial Co-ordinate ($r$). Here $\gamma=1$, $\lambda=0$.}
    %\label{fig:foobar}
    \label{fig:8}
\end{figure}
%%%%%%%%%%%%%%%%%%%%%%%%%%%%%%%%%%%%%%%%%%%%%%%%%%%%%%%%%%%%%%%%%%%%%%%%%%%%%%%%%%%%%%%%%%%%%%%%%%%%%%%%%%%%%%%%%%%%%%%%%%%%%%%%%%%%%%%
%%%%%%%%%%%%%%%%%%%%%%%%%%%%%%%%%%%%%%%%%%%%%%%%%%%%%%%%%%%%%%%%%%%%%%%%%%%%%%%%%%%%%%%%%%%%%%%%%%%%%%%%%%%%%%%%%%%%%%%%%%%%%%%%%%
\begin{figure}[H]
    \centering
    [a]{\includegraphics[width=0.4\textwidth]{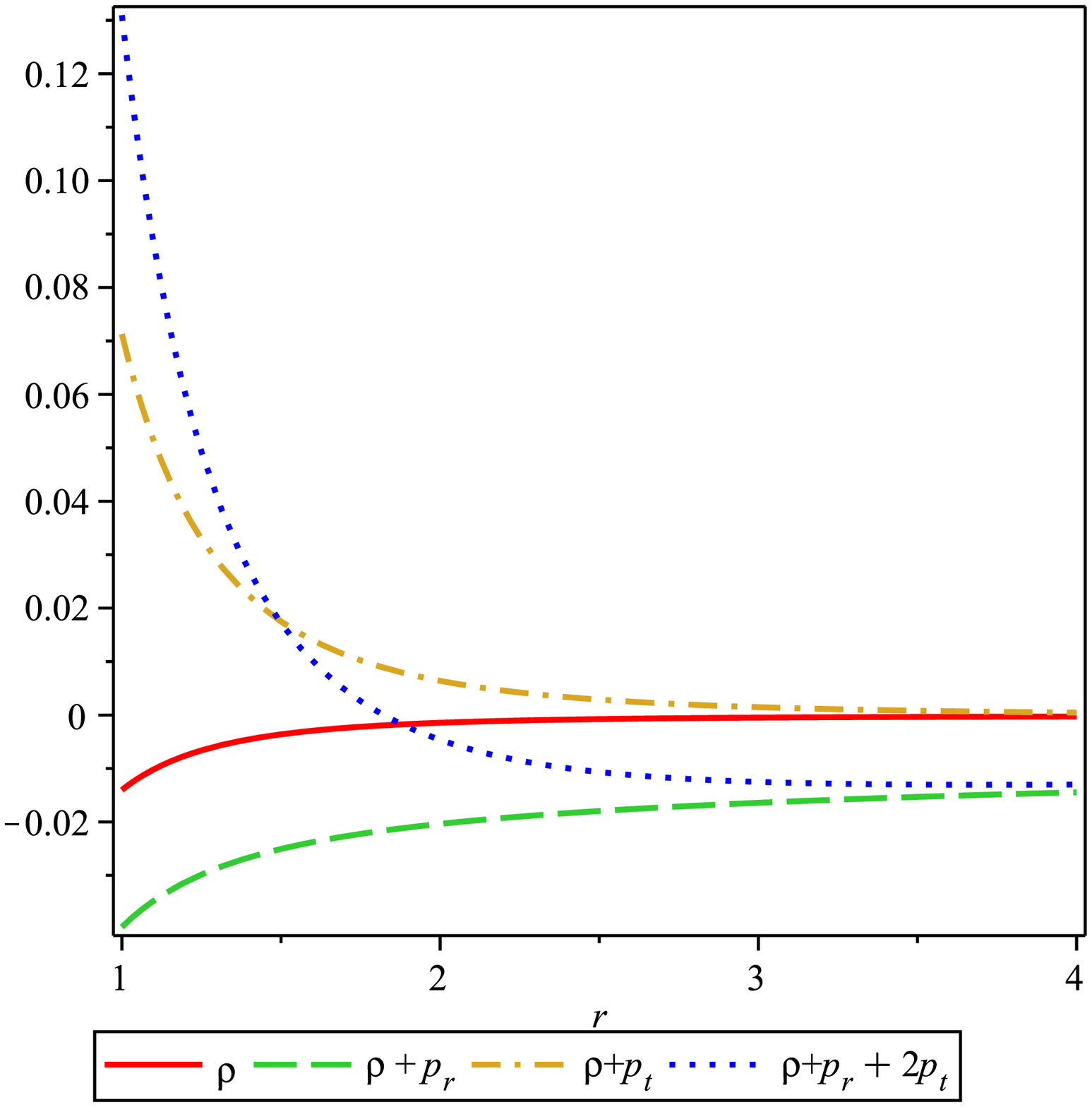}} 
    [b]{\includegraphics[width=0.4\textwidth]{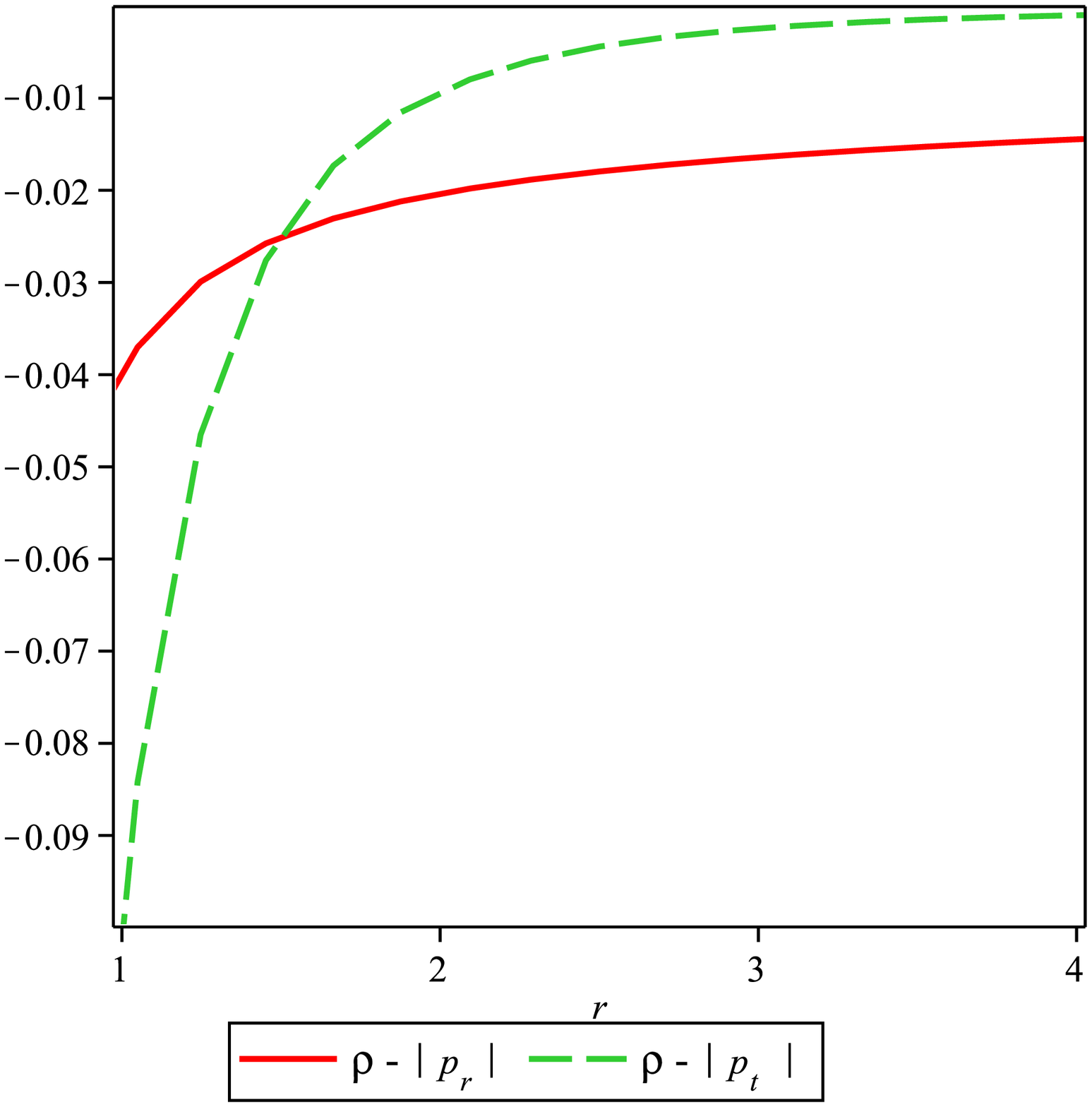}} 
    [c]{\includegraphics[width=0.4\textwidth]{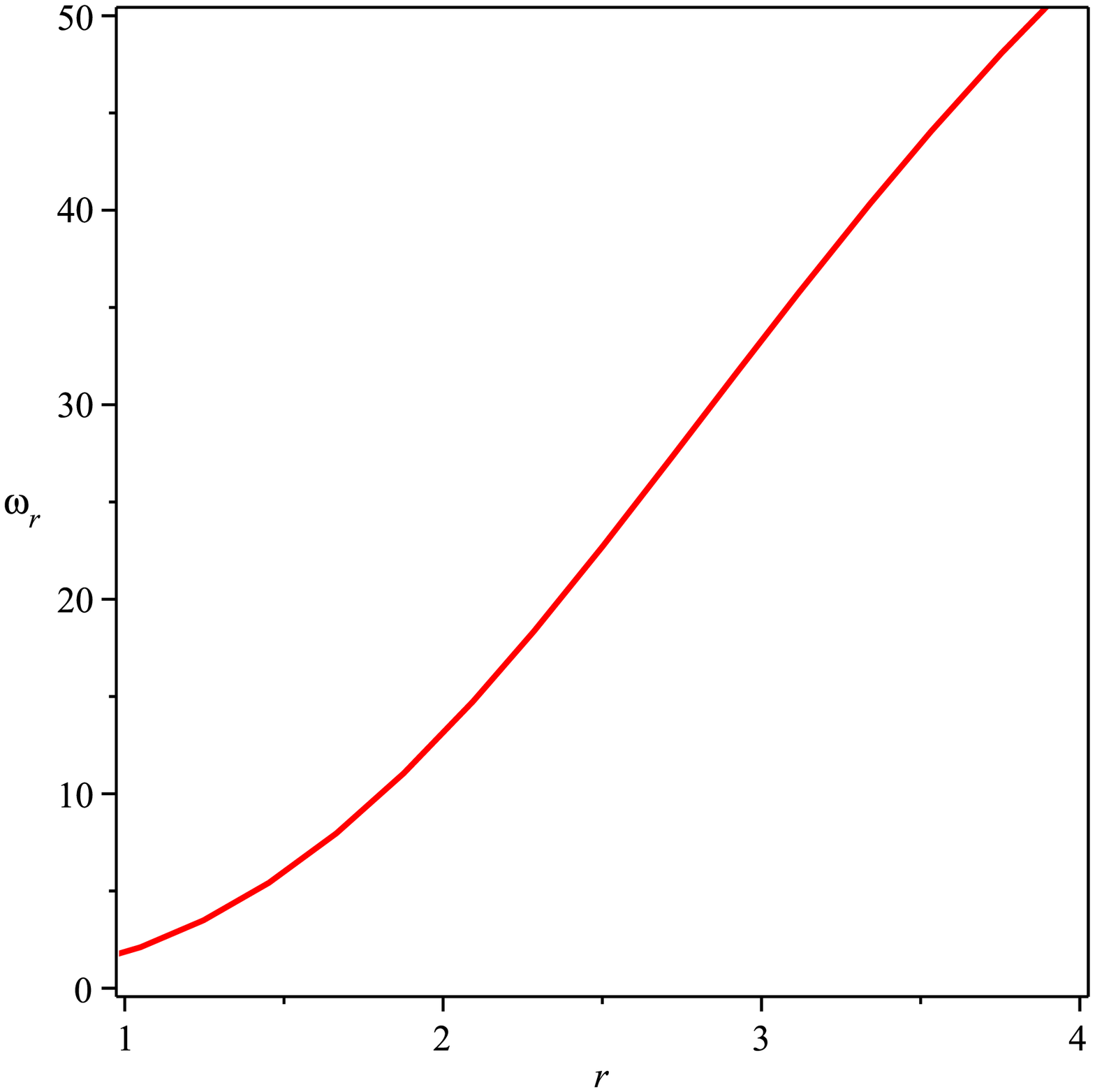}}
    [d]{\includegraphics[width=0.4\textwidth]{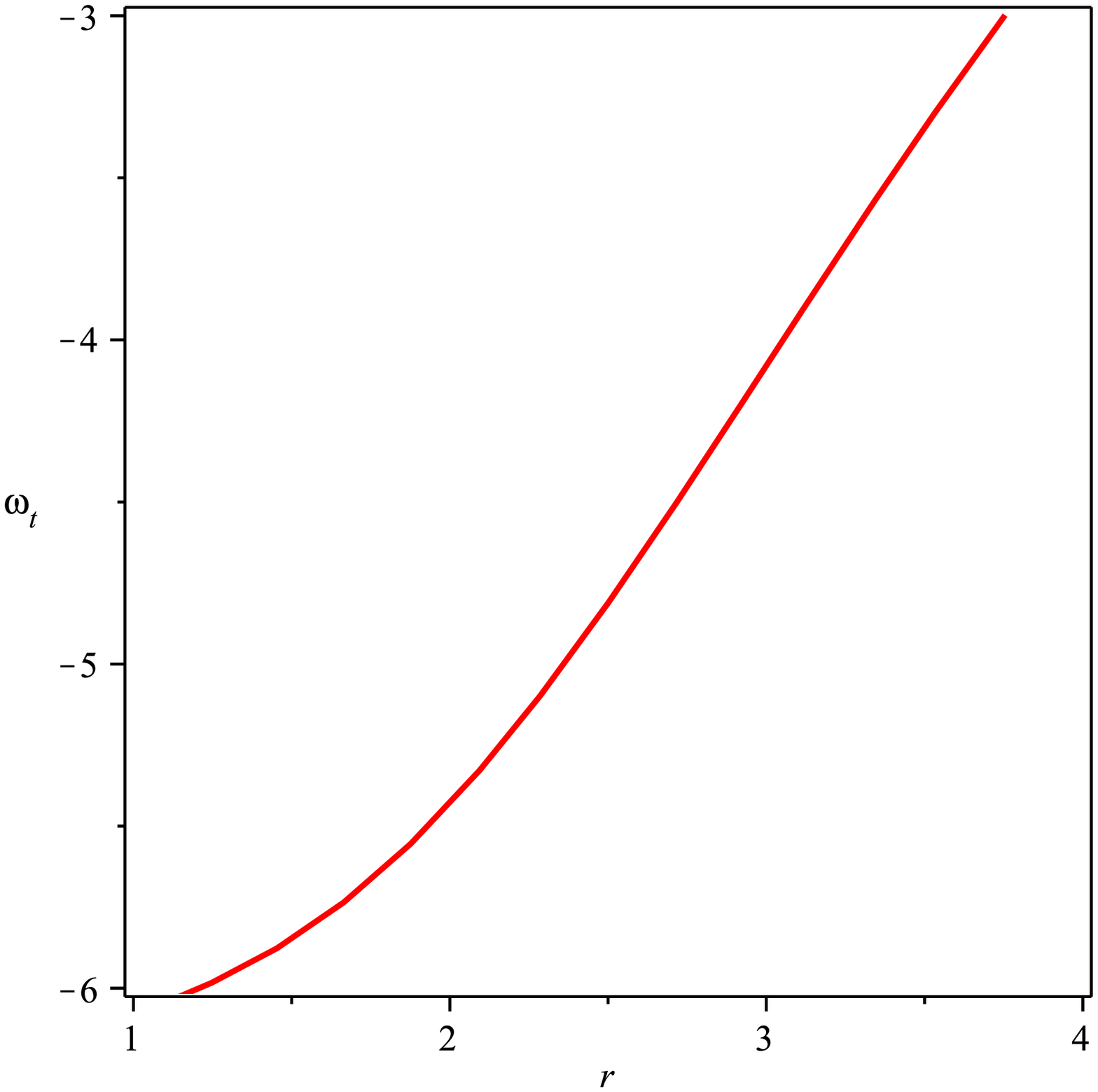}}
    [e]{\includegraphics[width=0.4\textwidth]{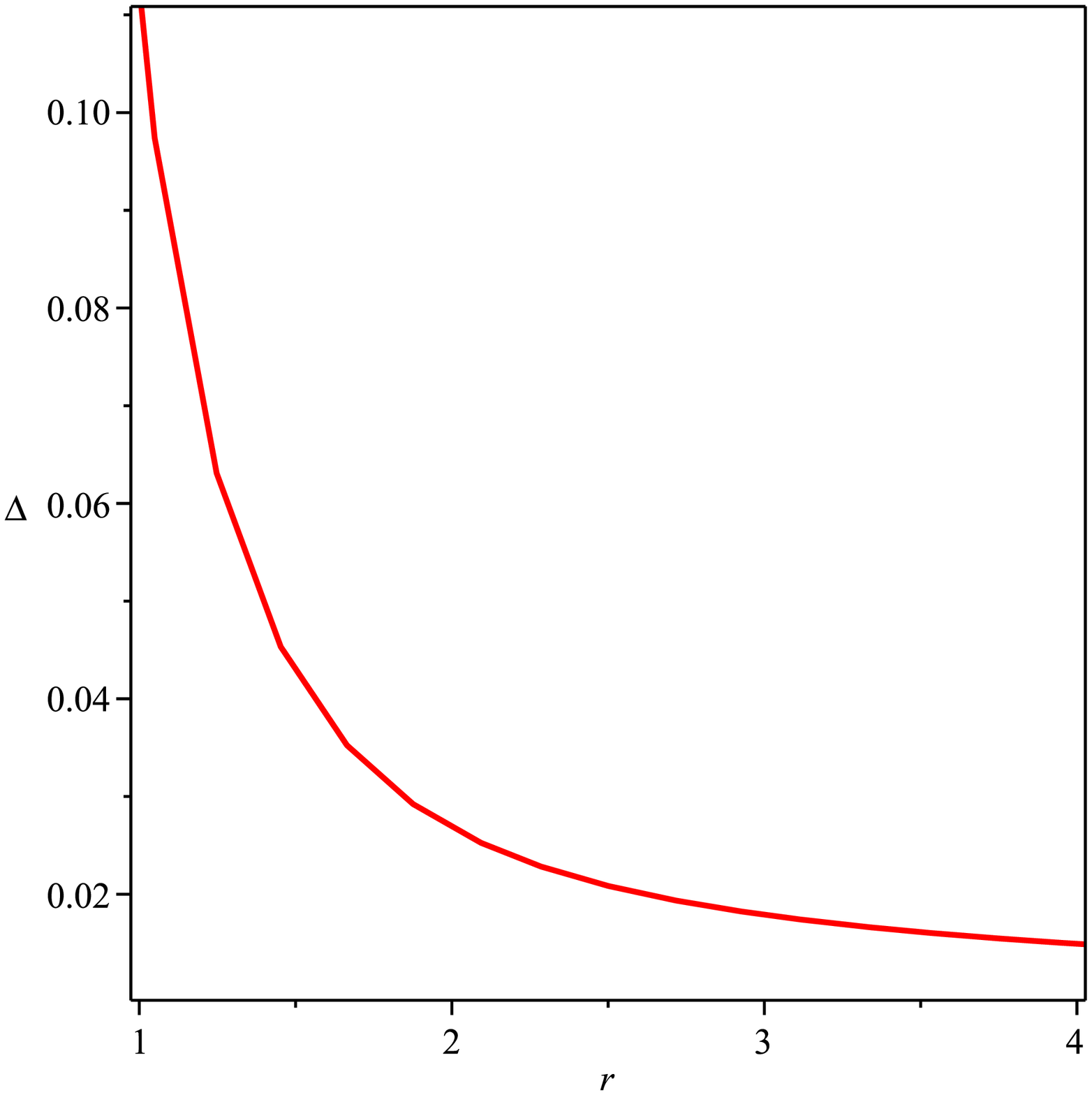}}
    \caption{The plot of [a] NEC, SEC, WEC, [b] DEC, [c] Radial EoS Parameter ($\omega_{r}$), [d] Tangential EoS parameter ($\omega_{t}$), 
    [e] Anisotropy Parameter ($\triangle$), vs. Radial Co-ordinate ($r$). Here $\gamma =1$, $\lambda=1$.}
    %\label{fig:foobar}
    \label{fig:9}
\end{figure}
%%%%%%%%%%%%%%%%%%%%%%%%%%%%%%%%%%%%%%%%%%%%%%%%%%%%%%%%%%%%%%%%%%%%%%%%%%%%%%%%%%%%%%%%%%%%%%%%%%%%%%%%%%%%%%%%%%%%%%%%%%%%%%%%%%%%%%%%%
%%%%%%%%%%%%%%%%%%%%%%%%%%%%%%%%%%%%%%%%%%%%%%%%%%%%%%%%%%%%%%%%%%%%%%%%%%%%%%Table %%%%%%%%%%%%%%%%%%%%%%%%%%%%%%%%%%%%%%%%%%%%%%%%%%
\renewcommand{\arraystretch}{1.3}
\setlength{\tabcolsep}{8.5pt}
\begin{table}[H]
	\caption{Prediction of physical nature of different combinations of $\rho$, $p_{r}$ and $p_{t}$ for $\lambda \in [-100,-12)$ }
	\centering
	\begin{tabular}{c c c c c  }
		\hline\hline
		Parameter & $[-100,-64]$ & $(-64,-63)$ & $[-63,-13]$ & $(-13,-12)$\\
		[0.5ex]
		\hline
		\multirow{2}{*}{$\rho$}   &  $+$ve & $(-64,\approx-63.68)$ $+$ve & $-$ve & $(-13, \approx -12.56) -$ve \\
		& &  $[\approx -63.68,-63)$ $-$ve & & $[\approx-12.56, -12)$ $+$ve      \\
		\hline
		\multirow{2}{*}{$\rho+p_{r}$}   &  $+$ve & $(-64,\approx-63.68)$ $+$ve & $-$ve & $(-13, \approx -12.56) -$ve \\
		& &  $[\approx -63.68,-63)$ $-$ve & & $[\approx-12.56, -12)$ $+$ve      \\
		\hline
		\multirow{2}{*}{$\rho+p_{t}$}   &  $+$ve & $(-64,\approx-63.68)$ $+$ve & $-$ve & $(-13, \approx -12.56) -$ve \\
		& &  $[\approx -63.68,-63)$ $-$ve & & $[\approx-12.56, -12)$ $+$ve      \\
		\hline
		\multirow{2}{*}{$\rho-|p_{r}|$}   &  $-$ve & $-$ve & $-$ve & $(-13, \approx -12.56) -$ve \\
		&  & & & $[\approx-12.56, -12)$ $+$ve     \\
		\hline
		\multirow{2}{*}{$\rho-|p_{t}|$}  &  $-$ve to $+$ve & $(-64,\approx-63.68)$ $-$ve to $+$ve & $-$ve & $-$ve  \\
		& &  $[\approx-63.68,-63)$ $-$ve & &    \\
		& & & &  \\
		\hline
		\multirow{2}{*}{$\rho+p_{r}+2p_{t}$}   &  $+$ve & $(-64,\approx-63.68)$ $+$ve & $-$ve & $(-13, \approx -12.56) -$ve  \\
		& &  $[\approx -63.68,-63)$ $-$ve & & $[\approx-12.56, -12)$ $+$ve     \\[1ex]
		\hline
	\end{tabular}
	\label{table:1}
\end{table}
%%%%%%%%%%%%%%%%%%%%%%%%%%%%%%%%%%%%%%%%%%%%%%%%%%%%%%%%%%%%%%%%%%%%%%%%%%%%%%%%%%%%%%%%%%%%%%%%%%%%%%%%%%%%%%%%%%%%%%%%%%%%%%%%%%%%%%%%%%%%%%%%%%
%%%%%%%%%%%%%%%%%%%%%%%%%%%%%%%%%%%%%%%%%%%%%%%%%%%%%%%%%%%%%% Table %%%%%%%%%%%%%%%%%%%%%%%%%%%%%%%%%%%%%%%%%%%%%%%%%%%%%%%%%%%%%%%%%%%%%%%
\renewcommand{\arraystretch}{1.3}
\setlength{\tabcolsep}{8.5pt}
\begin{table}[H]
	\caption{Prediction of physical nature of different combinations of $\rho$, $p_{r}$ and $p_{t}$ for $\lambda \in [-12,-6]$ }
	\centering
	\begin{tabular}{c c c}
		\hline\hline
		Parameter & $[-12,-9]$ & $(-9,-6]$ \\
		[0.5ex]
		\hline
		\multirow{2}{*}{$\rho$}    & $+$ve & $(-9,\approx-8.37) +$ve \\
		&  & $[\approx -8.37,-6] -$ve   \\
		\hline
		\multirow{2}{*}{$\rho +p_{r}$}           & $+$ve & $(-9,\approx-8.37) +$ve \\
		&  & $[\approx -8.37,-6] -$ve   \\
		\hline
		\multirow{3}{*}{$\rho +p_{t}$}           & $+$ve & $(-9,\approx-8.37) +$ve \\
		& & $[\approx -8.37, \approx -6.66) -$ve\\
		& & $[\approx-6.66, -6] -$ve to $+$ ve\\ 
		\hline
		\multirow{2}{*}{$\rho -|p_{r}|$}           & $+$ve & $(-9,\approx-8.37) +$ve \\
		&  & $[\approx -8.37,-6] -$ve   \\
		\hline
		\multirow{3}{*}{$\rho -|p_{t}|$} & $[-12,\approx-11.04) -$ve & $(-9,\approx-8.37) +$ve \\
		& $[\approx -11.04, \approx-9.91) +$ve to $-$ve & $[\approx -8.37, -6] -$ve\\
		&  $[\approx-9.91,-9] +$ve &\\
		\hline
		\multirow{4}{*}{$\rho+p_{r}+2p_{t}$}  & $+$ve &  $(-9, \approx -8.37) +$ve \\
		& & [$\approx-8.37, \approx-7.14) -$ve\\
		& & [$\approx-7.14,\approx-6.82) +$ve to $-$ve\\
		& & [$\approx-6.82,-6] +$ve \\ [1ex]
		\hline
	\end{tabular}
	\label{table:2}
\end{table}
%%%%%%%%%%%%%%%%%%%%%%%%%%%%%%%%%%%%%%%%%%%%%%%%%%%%%%%%%%%%%%%%%%%%%%%%%%%%%%%%%%%%%%%%%%%%%%%%%%%%%%%%%%%%%%%%%%%%%%%%%%%%%%%%%%%%%%
%%%%%%%%%%%%%%%%%%%%%%%%%%%%%%%%%%%%%%%%%%%%%%%%%%%%%%%%%%%%%%%%%%%%% Table %%%%%%%%%%%%%%%%%%%%%%%%%%%%%%%%%%%%%%%%%%%%%%%%%%%%%%%%%%%%%%%
\renewcommand{\arraystretch}{1.3}
\setlength{\tabcolsep}{8.5pt}
\begin{table}[H]
	\caption{Prediction of physical nature of different combinations of $\rho$, $p_{r}$ and $p_{t}$ for $\lambda \in$ $\{-5,-4,-3,-2,-1.5\} 
	\bigcup [-1,100]$ }
	\centering
	\begin{tabular}{c c c c c c c}
		\hline\hline
		Parameter & $\lambda=-5$ & $\lambda=-4$ & $\lambda=-3$ & $\lambda=-2$ & $\lambda=-1.5$ &  $\lambda \geq -1$ \\
		[0.5ex]
		\hline
		$\rho$    & $-$ve & $-$ve & $-$ve & $-$ve &  $-$ve  to $+$ve & $-$ve  \\
		\hline
		$\rho+p_{r}$  &  $-$ve & $-$ve to $+$ve & $-$ve to $+$ve & $+$ve & $+$ve &$-$ve   \\
		\hline
		\multirow{2}{*}{$\rho+p_{t}$}  & $+$ve &  $+$ve & $+$ve to $-$ve  & $-$ve & $-$ve & $[-1,\approx8.18) +$ve  \\
		& & & & & & $[\approx 8.18,100]$ $+$ve to $-$ve \\
		\hline
		$\rho-|p_{r}|$ & $-$ve &  $-$ve & $-$ve &  $-$ve & $-$ve & $-$ve  \\
		\hline
		$\rho-|p_{t}|$  & $-$ve &  $-$ve & $-$ve &  $-$ve & $-$ve & $-$ve \\
		\hline
		$\rho+p_{r}+2p_{t}$  & $+$ve & $+$ve & $+$ve & $+$ve  & $-$ve to $+$ve & $+$ve to $-$ve \\
		[1ex]
		\hline
	\end{tabular}
	\label{table:3}
\end{table}
%%%%%%%%%%%%%%%%%%%%%%%%%%%%%%%%%%%%%%%%%%%%%%%%%%%%%%%%%%%%%%%%%%%%%%%%%%%%%%%%%%%%%%%%%%%%%%%%%%%%%%%%%%%%%%%%%%%%%%%%%%%%%%%%%%%%%%%%%%%%
%%%%%%%%%%%%%%%%%%%%%%%%%%%%%%%%%%%%%%%%%%%%%%%%%%%%%%%%%%%%%%%%%% Table  %%%%%%%%%%%%%%%%%%%%%%%%%%%%%%%%%%%%%%%%%%%%%%%%%%%%%%%%%%%%%%%5555
\renewcommand{\arraystretch}{1.3}
\setlength{\tabcolsep}{8.5pt}
\begin{table}[H]
	\caption{Violation/Validation of Energy Conditions for different values of $\lambda \in [-100,100]$ }
	\centering
	\begin{tabular}{c c c c c c}
		\hline\hline
		$\lambda$ & NEC & SEC &  WEC & DEC & Matter Content\\
		[0.5ex]
		\hline
		-75   & Validated  & Validated &  Validated & Violated & Exotic\\
		-13  & Violated & Violated & Violated  & Violated & Exotic \\
		-9.5 &Validated  & Validated &  Validated & Validated & Non-Exotic\\
		-5 & -Violated & Violated & Violated  & Violated & Exotic  \\
		-1.5 & Violated & Violated & Violated  & Violated & Exotic \\
		-1 & Violated & Violated & Violated  & Violated & Exotic \\
		0 & Violated & Violated & Violated  & Violated & Exotic \\
		1 & Violated & Violated & Violated  & Violated & Exotic \\
		%$p_{t}-p_{r}$  & &  &   \\
		[1ex]
		\hline
	\end{tabular}
	\label{table:4}
\end{table}
%%%%%%%%%%%%%%%%%%%%%%%%%%%%%%%%%%%%%%%%%%%%%%%%%%%%%%%%%%%%%%%%%%%%%%%%%%%%%%%%%%%%%%%%%%%%%%%%%%%%%%%%%%%%%%%%%%%%%%%%%%%%%%%%%%%%%%%%%%%
%%%%%%%%%%%%%%%%%%%%%%%%%%%%%%%%%%%%%%%%%%%%%%%%%%%%%%%%%%%%%%%% Table %%%%%%%%%%%%%%%%%%%%%%%%%%%%%%%%%%%%%%%%%%%%%%%%%%%%%%%%%%%%%%%%%
\renewcommand{\arraystretch}{1.3}
\setlength{\tabcolsep}{8.5pt}
\begin{table}[H]
	\caption{Prediction of physical nature of state Parameters $\omega_{r}$, $\omega_{t}$ for different values of $\lambda \in [-100,100]$ }
	\centering
	\begin{tabular}{c c c c c}
		\hline\hline
		$\lambda$ & $\omega_{r}$ & $\omega_{r}(\gamma)$ &  $\omega_{t}$ & $\omega_{t}(\gamma)$\\
		[0.5ex]
		\hline
		-75   & +$\uparrow$  & 2.221 &  +$\downarrow$ $\rightarrow 0$& 2.475 \\
		-13  &  +$\uparrow$ & 0.005 & +$\downarrow$ $\rightarrow 0$   & 1.632  \\
		-9.5 & 0 $\rightarrow$ signature flipping ($5<r<6$) $\rightarrow$ 0 & 0 & 0 $\rightarrow$ signature flipping ($5<r<6$) 
		$\rightarrow$ 0 &0 \\
		-5 & - $\downarrow$ &  -0.015 &-$\uparrow$ & -1.38  \\
		-1.5 & - $\downarrow$  & -7.062 & +$\downarrow$ $\rightarrow$ 0 & 5.08 \\
		-1 & 0 $\rightarrow$ signature flipping ($2<r<2.5$) $\rightarrow$ 0 & 0 &0 $\rightarrow$ signature flipping ($2<r<2.5$) $\rightarrow$ 0& 0 \\
		0 & +$\uparrow$  & 2.259 & -$\downarrow$ & -6.147  \\
		1 & +$\uparrow$  & 1.829 & -$\uparrow$ $\rightarrow$ 0 & -6.085   \\
		%$p_{t}-p_{r}$  & &  &   \\
		[1ex]
		\hline
	\end{tabular}
	\label{table:5}
\end{table}
%%%%%%%%%%%%%%%%%%%%%%%%%%%%%%%%%%%%%%%%%%%%%%%%%%%%%%%%%%%%%%%%%%%%%%%%%%%%%%%%%%%%%%%%%%%%%%%%%%%%%%%%%%%%%%%%%%%%%%%%%%%%%%%%%%%%%%%%%%%%
%%%%%%%%%%%%%%%%%%%%%%%%%%%%%%%%%%%%%%%%%%%%%%%%%%%%%%%%%%%%%%%%%%%%%%% Table %%%%%%%%%%%%%%%%%%%%%%%%%%%%%%%%%%%%%%%%%%%%%%%%%%%%%%%%%%%%%%%
\renewcommand{\arraystretch}{1.3}
\setlength{\tabcolsep}{8.5pt}
\begin{table}[H]
	\caption{Prediction of physical nature of Anisotropy Parameter $\triangle$ for different values of $\lambda \in [-100,100]$ }
	\centering
	\begin{tabular}{c c c}
		\hline\hline
		$\lambda$ &  $\triangle$ & Geometry\\
		[0.5ex]
		\hline
		-75   &  +$\downarrow$ to -$\uparrow$ & Repulsive to Attractive\\
		-13  &   + $\downarrow$ & Repulsive\\
		-5 &  +$\downarrow$ to -$\uparrow$ & Repulsive to Attractive \\
		-1.5 &  -$\uparrow$ & Attractive\\
		-1 &  + $\downarrow$ & Repulsive\\
		0 &  +$\downarrow$ & Repulsive \\
		1 &  +$\downarrow$ & Repulsive  \\
		%$p_{t}-p_{r}$  & &  &   \\
		[1ex]
		\hline
	\end{tabular}
	\label{table:6}
\end{table}
%%%%%%%%%%%%%%%%%%%%%%%%%%%%%%%%%%%%%%%%%%%%%%%%%%%%%%%%%%%%%%%%%%%%%%%%%%%%%%%%%%%%%%%%%%%%%%%%%%%%%%%%%%%%%%%%%%%%%%%%%%%%
%%%%%%%%%%%%%%%%%%%%%%%%%%%%%%%%%%%%%%%%%%%%%%%% Section 5 %%%%%%%%%%%%%%%%%%%%%%%%%%%%%%%%%%%%%%%%%%%%%%%%%%%%%%%%%%%
\section{Conclusion}

In this work, we have investigated the possibility that the existence of wormholes could be explored in the context of modified $f(R,T)$ 
gravity theories. For this, we have chosen a new mathematical form of the shape function which varies with the logarithm of the 
radial coordinate, both directly and inversely. This shape function is being used for obtaining the implicit solution to the 
field equations in f(R, T) gravity theory. In continuation to that, the analysis of physical parameters and inequalities such as 
the energy density $\rho$, the energy conditions (NEC, WEC, SEC, DEC), the radial state parameter $\omega_{r}$, the tangential 
state parameter $\omega_{t}$ and the anisotropy parameters $\triangle$, has been presented in detail in tables 1$-$6. For the 
purpose to highlight the traversable and asymptotically flat WH structure information, we worked out with constant redshift 
function and variable shape function satisfying the basic requirements. Further, violation of NEC verifying the presence of 
exotic matter is a key component explaining the existence of stable traversable wormholes. In light of this, we noticed that 
for some specific intervals of $\lambda \in [-100,100]$ say $D=[-100,\approx-63.68)\bigcup[-12,\approx-8.37)$, the null energy 
condition is validated whilst for the rest of intervals i.e. $[-100,100]-D$, it remains violated. A detailed study showing the 
results of all energy conditions for $\lambda \in [-100.100]$ have been summarized in the tables 1$-$3.\\

Finally, we may conclude that $\lambda$ plays a significant role in the development of wormhole solutions in the $f(R,T)$ gravity theory.
 Tables clearly present the picture that for a very tiny spectrum of $\lambda$ say $D_{1}=[\approx-9.91,-9]$, all the energy conditions 
 are validated pointing towards the non-existence of exotic matter while for a broader spectrum i.e. $[-100,100]-D_{1}$, the violation 
 of energy conditions and presence of exotic matter has been confirmed. For the precision of results in a
discrete form, we have chosen particular values of $\lambda$ from these ranges and a clear picture of the behaviour of energy conditions, 
state parameters ($\omega_{r}$, $\omega_{t}$) and anisotropy parameter ($\triangle$) w.r.t. radial co-ordinate $r$ has been presented 
in the figures 2$-$9. Moreover, final analysis regarding the violation/validation of the energy conditions for different chosen values 
of $\lambda$ in $[-100,100]$ has been presented in table $4$ and the combined analysis of the behaviours of  state parameters 
($\omega_{r}$, $\omega_{t}$) and the nature of the geometry of the wormholes predicted from the anisotropy parameter ($\triangle$) has 
been presented in tables $5$ and $6$. Now, we wish to highlight the particular case for general relativity for which the coupling 
constant $\lambda$ is considered to be zero i.e. $f(R)=R$. For this scenario, the energy density $\rho$ is found to be a negative 
function of radial coordinate $r$ which indicates the presence of exotic matter in wormhole geometry. We may conclude that for the 
case of general relativity, there is a net violation of the energy conditions and the presence of exotic matter is verified. For the 
presented work, we, here conclude that the f(R,T) theory of gravity with our newly proposed logarithmic shape function is found to 
be a suitable choice to describe the existence of traversable and asymptotically flat wormholes with the majority of exotic matter 
presence near the throat of the wormhole. 

 %%%%%%%%%%%%%%%%%%%%%%%%%%%%%%%%%%%%%%%%%%%%%%%%%%%%%%%%%%%%%%%%%%%%%%%%%%%%%%%%%%%%%%%%%%%%%%%%%%%%%%%%%%%%%%%%%%%%%%%%%%%%%%%%
\section*{Acknowledgments}
The author (AP) thank the IUCAA, Pune, India for providing the facility during a visit where a part of this work was completed.
%%%%%%%%%%%%%%%%%%%%%%%%%%%%%%%%%%%%%%%%%%%%%%%%%%%%%%%%%%%%%%%%%%%%%%%%%%%%%%%%%%%%%%%%%%%%%%%%%%%%%%%%%%%%%%%%%%%%%%%%

\end{document}